\documentclass[aps,prb,reprint,longbibliography,superscriptaddress]{revtex4-1}

\usepackage{graphicx}
\usepackage{color}
\usepackage{ifsym}
\usepackage{amssymb}
\usepackage{bm}
\usepackage{upgreek}

\begin{document}


\title{Domain Wall Motion and Interfacial Dzyaloshinskii-Moriya Interactions\\ in Pt/Co/Ir$(t_\mathrm{Ir})$/Ta Multilayers}



\author{Kowsar Shahbazi}
\email[Email:]{pyks@leeds.ac.uk}
\affiliation{School of Physics and Astronomy, University of Leeds, Leeds LS2 9JT, United Kingdom}

\author{Joo-Von Kim}
\affiliation{Centre for Nanoscience and Nanotechnology (C2N), CNRS, Universit\'{e} Paris-Sud, Universit\'{e} Paris-Saclay, 91120 Palaiseau, France}

\author{Hans T. Nembach}
\affiliation{Electromagnetics Division, National Institute of Standards and Technology, Boulder, Colorado, 80305}

\author{Justin M. Shaw}
\affiliation{Electromagnetics Division, National Institute of Standards and Technology, Boulder, Colorado, 80305}

\author{Andreas Bischof}
\affiliation{IBM Research-Zurich, S\"{a}umerstrasse 4, CH-8803 R\"{u}schlikon, Switzerland}

\author{Marta D. Rossell}
\affiliation{IBM Research-Zurich, S\"{a}umerstrasse 4, CH-8803 R\"{u}schlikon, Switzerland}
\affiliation{Electron Microscopy Center, EMPA, Swiss Federal Laboratories for Materials Science and Technology, \"{U}berlandstrasse 129, 8600 D\"{u}bendorf, Switzerland}

\author{Vincent Jeudy}
\affiliation{Laboratoire de Physique des Solides, CNRS, Universit\'{e} Paris-Sud, Universit\'{e} Paris-Saclay, 91405 Orsay Cedex, France}

\author{Thomas A. Moore}
\affiliation{School of Physics and Astronomy, University of Leeds, Leeds LS2 9JT, United Kingdom}

\author{Christopher H. Marrows}
\email[Email:]{c.h.marrows@leeds.ac.uk}
\affiliation{School of Physics and Astronomy, University of Leeds, Leeds LS2 9JT, United Kingdom}


\date{\today}

\begin{abstract}
The interfacial Dzyaloshinskii-Moriya interaction (DMI) is important for chiral domain walls (DWs) and for stabilizing magnetic skyrmions. We study the effects of introducing increasing thicknesses of Ir, from zero to 2~nm, into a Pt/Co/Ta multilayer between the Co and Ta. We observe a marked increase in magnetic moment, due to the suppression of the dead layer at the interface with Ta, but the perpendicular anisotropy is hardly affected. All samples show a universal scaling of the field-driven domain wall velocity across the creep and depinning regimes. Asymmetric bubble expansion shows that DWs in all of the samples have the left-handed N\'{e}el form. The value of in-plane field at which the creep velocity shows a minimum drops markedly on the introduction of Ir, as does the frequency shift of the Stokes and anti-Stokes peaks in Brillouin light scattering measurements. Despite this qualitative similarity, there are quantitative differences in the DMI strength given by the two measurements, with BLS often returning higher values. Many features in bubble expansion velocity curves do not fit simple models commonly used to date, namely a lack of symmetry about the velocity minimum and no difference in velocities at high in-plane field. These features are explained by the use of a model in which the depinning field is allowed to vary with in-plane field in a way determined from micromagnetic simulations. This theory shows that velocity minimum underestimates the DMI field, consistent with BLS returning higher values. Our results suggest that the DMI at an Ir/Co interface has the same sign as the DMI at a Pt/Co interface.
\end{abstract}


\maketitle


\section{Introduction}

In today's society, the amount of data needing to be stored and/or processed is growing rapidly \cite{Hilbert2011}. Whilst hard disks still dominate the data storage landscape in terms of volume of data stored, they feature moving parts that increase energy consumption and decrease reliability. Magnetic domain walls (DWs) and skyrmions are the smallest magnetic components that can be used in a new generation of magnetic recording media/processing devices (so-called race-track memories) to overcome these obstacles\cite{Parkin2015,Fert2013}. To be able to make use of them efficiently, one of the most important parameters to optimize and control is the interfacial Dzyaloshinskii-Moriya interaction (DMI). The DMI changes the magnetostatically favorable Bloch wall to N{\'e}el walls with a fixed chirality \cite{thiaville12EPL} in multilayers with perpendicular magnetic anisotropy (PMA) which in-turn makes them sensitive to spin-orbit torques \cite{khvalkovskiy2013matching} so that they can be moved by current pulses.

The first step towards optimization of any parameter is to be able to measure it easily and reliably. Several different methods have been used to evaluate the strength of the DMI, $D$. Current driven domain wall motion under in-plane (InP) applied field has been widely investigated\cite{ryu2013chiral,Emori13NatMat,Martinez13APL,Torrejon14NatCom}. But using current to study DMI complicates the situation as usually spin Hall effect, Rashba, and DMI are present simultaneously\cite{Garello2013}. Microscopy measurements such as spin-polarized scanning tunnelling microscopy\cite{Meckler09PRL}, spin-polarized low-energy electron microscopy\cite{Chen13NatCom}, and photoemission electron microscopy combined with x-ray magnetic circular dichroism \cite{Boulle16NatNano} can also be used, but are only suitable for particular kinds of sample. Brillouin light scattering (BLS) uses non-reciprocal propagation of the spin waves in materials with DMI to measure $D$, but is time-consuming and so not suitable for routine measurements of large numbers of samples \cite{nembach15NatPhys,Belmeguenai15PRB}.

One widely-adopted technique to estimate DMI is asymmetric bubble expansion, since it requires minimal sample preparation and relatively inexpensive equipment to implement. The concept was introduced by Je \textit{et al.}\cite{Je13PRB} and extended by Hrabec \textit{et al.}\cite{Hrabec14PRB}. It was known that the DMI in systems with broken inversion symmetry splits the degeneracy between right-handed and left-handed twists in the magnetisation, and enforces homochiral N{\'e}el walls in layers with perpendicular magnetic anisotropy if sufficiently strong. As a result, the DMI can be represented as an intrinsic field across the DW \cite{thiaville12EPL}. The central idea of the asymmetric bubble expansion method is that this intrinsic field may be enhanced or (partially) cancelled by an externally applied in-plane (InP) field. This affects the wall energy and hence its creep velocity under an out-of-plane (OoP) field. Elongation of domains in an InP field was observed decades ago in garnet bubble domain materials\cite{Gallagher79JAP}, and was described in detail by De Leeuw, Van Den Doel and Enz\cite{deleeuw80RPP}. Still, the very first time DMI was suggested as one of the probable causes of this elongation in 2010 was by Kabanov \textit{et al.} \cite{Kabanov10IEEE}. They noticed changes of DW velocity with InP applied field and variation of elongation direction with field sign.

Likewise, Je \textit{et al.} \cite{Je13PRB} attributed the asymmetrical growth of bubble domains to breaking of DMI related N\'{e}el wall radial symmetry on either side of an expanding bubble with InP field. Typically curves for the DW creep velocity $v$ as a function of in-plane applied field $H_\mathrm{InP}$ are measured and fitted to a simple creep model to reveal the DMI field $H_\mathrm{DMI}$ and hence $D$. Whilst experimentally straightforward, interpretation of the results has not always been easy. Some literature reported excellent matches for this model \cite{Petit15APL, Khan16APL, Pham16EPL, kim15APL}, but other experimental investigations revealed basic assumptions of the model are not necessarily correct for all PMA materials. For instance, Soucaille \textit{et al.} mentioned a change of their domain wall roughness with InP field \cite{Soucaille16PRB}. Nevertheless, a common issue is simply that the $v(H_\mathrm{InP})$ curves do not have the simple form expected \cite{Lavrijsen15PRB, Vanatka15JP,lau16PRB,Soucaille16PRB,Ajejas17APL,Shepley18PRB}. To overcome this problem, some researchers went to the extent of applying fast pulses of high fields to work in the flow regime \cite{Vanatka15JP} or doing complex analytical calculations of the DW energy for the whole bubble \cite{pellegren17PRL}. All in all, using a simple creep model to evaluate DMI from asymmetrical bubble expansion is not always as straightforward as first thought, and there are anomalies that require further study.

Here we investigate asymmetric bubble expansion to evaluate DMI in a heavy metal (HM)/ferromagnet (FM) multilayer by systematic change of one parameter in a series of the samples, highlighting some anomalies that cannot be described by the simple creep model. Moreover, since DW dynamics will effect the behaviour of the bubble propagation, we also performed an extensive investigation of the details of DW dynamics in the creep and depinning regimes. Combining these results in a model in which the wall creep velocity depends on the depinning field that separates these regimes, which in turn depends on $H_\mathrm{InP}$, we are able to demonstrate the origin of two of these anomalies, namely the lack of symmetry of $v(H_\mathrm{InP})$ curves about their minima, and the tendency for these curves to join together at high fields. This model also shows that the field at which the minimum in $v(H_\mathrm{InP})$ occurs underestimates $H_\mathrm{DMI}$. We also compare our results for $D$ from asymmetric bubble expansion with those from BLS measurements on the same set of samples.

The multilayers we chose to study had the form Pt/Co/Ir($t_\mathrm{Ir}$)/Ta, in which the only quantity varied was the thickness $t_\mathrm{Ir}$ of the Ir layer. The presence of Ir brings another aspect to this work. Pt/Co/Ta multilayers show a high net spin Hall angle \cite{woo14eAPL} and DMI-stabilized skyrmion structures have been reported in them \cite{woo16NatMat}. Both the net spin Hall angle and the net DMI $D$ arises from differences between the effects arising from the heavy metal layers above and below the Co. Whilst Pt and Ta have large and opposite spin Hall angles, giving a large overall difference \cite{woo14eAPL}, the same may not be true of the DMI. The Pt/Co interface has been already shown to exhibit a sizeable DMI \cite{kashid14PRB, Belmeguenai15PRB}. On the other hand, a Co/Ta interface is expected to have a low DMI with the same sign as Co/Pt interface \cite{kashid14PRB, Yang15PRL}, so the resulting DMI of such multilayers is less than what one can get with a single Pt/Co interface. On the other hand, the DMI at an Ir/Co interface is predicted to be large and have the opposite sign to that for a Pt/Co interface \cite{Yang15PRL}, motivating Pt/Co/Ir as the basis for skyrmion-bearing multilayers \cite{Moreau16NatNan,Soumyanarayanan2017}. Considering the fact that the spin Hall effect takes place throughout the bulk of a heavy metal layer, while the DMI is generated at an interface, inserting Ir at the interface whilst retaining Ta for the bulk of the layer appears attractive to combine the two effects. Nevertheless, doubts have been raised about the actual sign of the DMI for an Ir/FM interface \cite{Kim16APL, Ajejas17APL}. Here we also go into the detail of the effect of inserting Ir and compare our results with other reports.

\section{Experiment}

\subsection{Sample Growth and Measurement Methods}

\begin{figure*}
  \includegraphics[width=12cm]{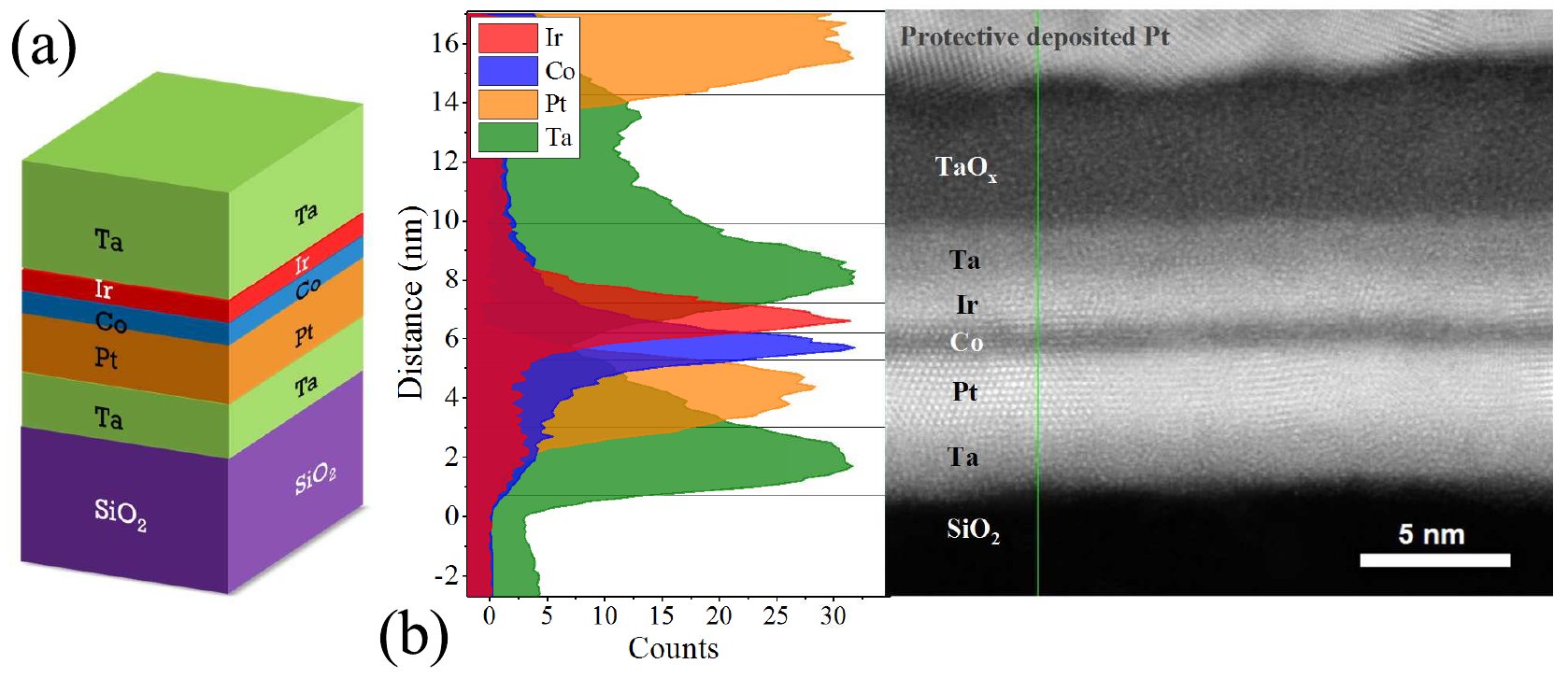}
  \caption{Magnetic multilayer structure. (a) Schematic of the multilayer stack, nominally Ta(2.0)/Pt(2.2)/Co(0.8)/Ir($t_\mathrm{Ir}$)/Ta(4.0) (layer thicknesses in nm). (b) EDX elemental line profiles across the multilayer structure with 1.0~nm Ir alongside a HAADF image of the same sample. The different interfaces and a smooth growth of the layers are clearly visible. \label{fig_stackTEM}}
\end{figure*}

Multilayers with a nominal structure of Ta(2.0)/Pt(2.2)/Co(0.8)/Ir($t_\mathrm{Ir}$)/Ta(4.0) (layer thicknesses in nm) were deposited onto thermally oxidized silicon substrates by dc magnetron sputtering (Fig.~\ref{fig_stackTEM}(a)). The Ir thickness $t_\mathrm{Ir}$ varied from 0 to 2.0~nm. The substrates were at room temperature and the base pressure was below $1.5\times10^{-7}$~Torr. The deposition Ar pressure was 3.0~mTorr.

High-resolution scanning transmission electron microscopy (STEM) and energy-dispersive X-ray (EDX) spectroscopy measurements were carried out to investigate the quality of the deposition and interface sharpness in the material stack. The measurements were performed at 300~kV employing a FEI Themis Titan equipped with ChemiSTEM technology. A probe semi-convergence angle of 24.6~mrad and an annular semi-detection range of the annular dark-field detector set to collect electrons scattered between 53 and 200~mrad were used. The cross-section lamellae for the STEM-EDX investigations were prepared with a FEI Helios Nanolab 450S focused ion-beam (FIB) instrument. To minimize possible damage to the stack structure, the sample was protected with a 200~nm-thick sputtered Pt layer before inserting it into the FIB. The high-angle annular dark-field (HAADF) STEM image in Fig.~\ref{fig_stackTEM}(b) shows easily distinguishable layers from the different materials, with the corresponding EDX elemental line profiles aligned with each layer. Partial oxidation of the Ta top layer due to the exposure to air can also be seen, proving that the capping layer was thick enough to prevent oxidation of the inner layers.

Perpendicular hysteresis loops of the samples were measured with the polar magneto-optical Kerr effect (P-MOKE). The anisotropy field, $H_\mathrm{K}$, and saturation magnetization, $M_\mathrm{s}$, were determined from hysteresis loops with an in-plane field measured using superconducting quantum interference device-vibrating sample magnetometry (SQUID-VSM). Symmetric bubble expansion, the growth of bubble domains in the presence only of an OoP driving field, was imaged using a wide-field Kerr microscope to study DW dynamics in the crossover from the creep to the viscous flow regime. To apply the OoP field, a small coil ($\sim 100$ turns and $\sim 1$~mm diameter) was carefully placed on top of the film surface. DW propagation using high driving fields could not be reached because of multiple domain nucleations and merging domains during pulse time. Using P-MOKE microscopy, one can measure the distance domain wall propagates during an OoP field pulse, and hence the velocity of the domain walls can be estimated. The results we show are average of 3-5 repeats for each applied field.

Asymmetric bubble expansion was studied by the same method as above, but with the addition of an extra electromagnet capable of supplying an InP field of up to 250~mT and a home-made OoP field coil that can apply field of up to 40~mT. For this part the symmetry-breaking InP field is kept constant and OoP driving field is pulsed to expand the bubble step-by-step. The DW velocity here is measured for the walls perpendicular to InP field, where the applied InP field enhances or cancels the DMI field, allowing the DMI strength $D$ to be measured.

The asymmetric frequency shift arising from the absortion or emission of spin-waves (SWs) in a ferromagnet with DMI was also used to measure $D$ using Brillouin light scattering (BLS) in the Damon–Eshbach geometry with a fixed wavevector of $k=16.7$~$\upmu\mathrm{m}^{-1}$.

\subsection{Magnetic Characterization}

\begin{figure}
	\includegraphics[width=8cm]{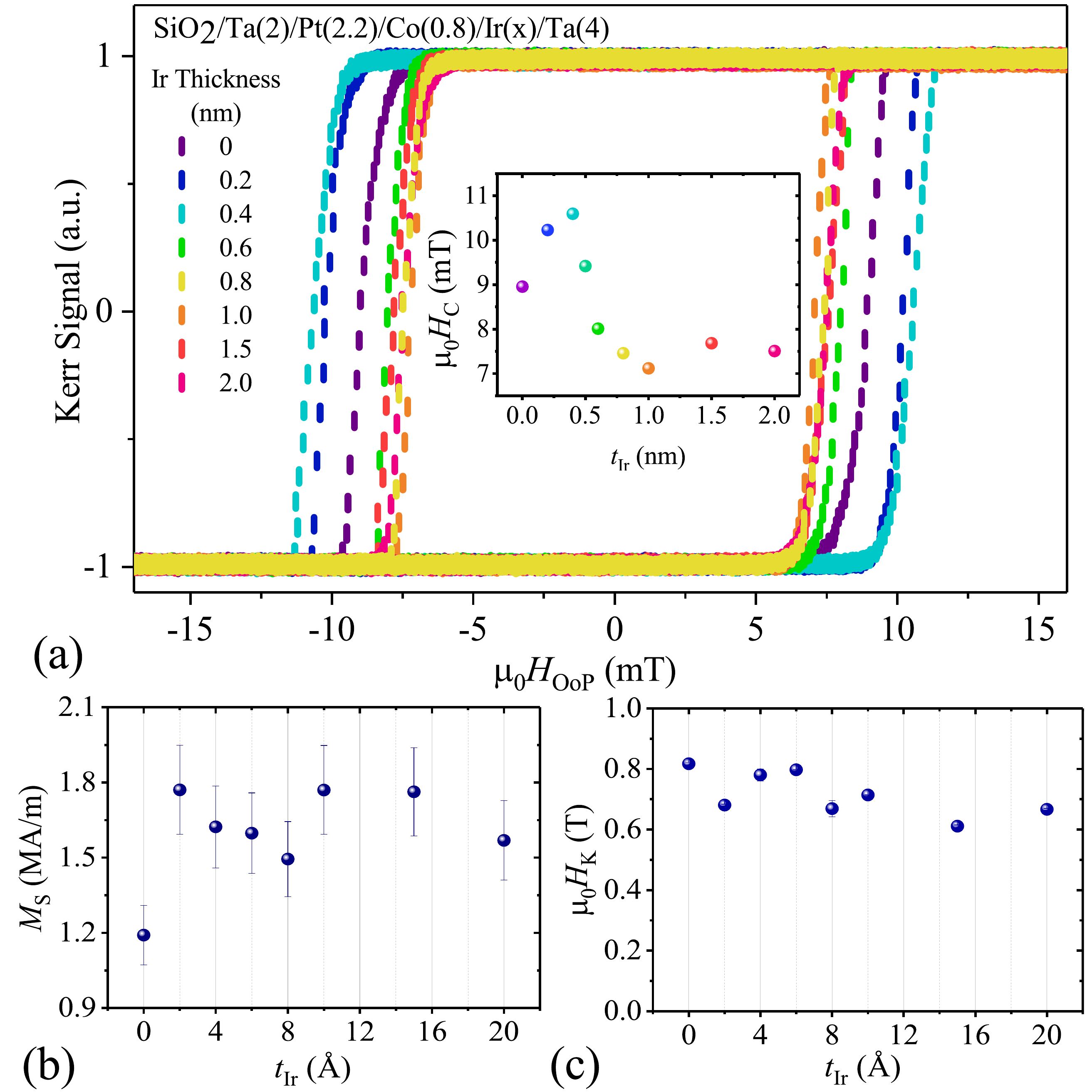}
	\caption{Magnetic characterization. (a) Polar Kerr measurements show clear square-shaped hysteresis loops confirming strong PMA. (b) Saturation magnetisation $M_\mathrm{S}$ and (c) anisotropy field $H_\mathrm{K}$ as a function of Ir thickness $t_\mathrm{Ir}$, both are determined from hysteresis loops acquired by SQUID-VSM with an in-plane field.
		\label{fig_magnetic}}
\end{figure}

All the Pt/Co(0.8)/Ir($t_\mathrm{Ir}$)/Ta multilayers showed square P-MOKE hysteresis loops for every value of $t_\mathrm{Ir}$, as presented in Fig.~\ref{fig_magnetic}(a). The OoP coercive field initially increases with $t_\mathrm{Ir}$, has a peak value for $t_\mathrm{Ir}=0.4$~nm but then decreases and reaches a constant level for $t_\mathrm{Ir}>0.6$~nm, around the point where Ir is expected to form a continuous layer.

Fig.~\ref{fig_magnetic}(b) and (c) also show the changes of saturation magnetization, $M_\mathrm{s}$, and anisotropy field, $H_\mathrm{K}$, with Ir thickness. $M_\mathrm{s}$ jumps up as soon as there is some Ir in the stack, and stays almost constant for higher $t_\mathrm{Ir}$. This suggests the presence of a dead layer between Co and top Ta layer, which is believed to be a result of intermixing between Ta and the ferromagnetic layer \cite{cheng11JAP,sinha13APL,Jang11JAP,bandiera11IEEE}, as well as a small degree of proximity magnetism in the Ir\cite{Ederer02PRB}. The dead layer causes a reduction of effective thickness of the magnetic material. The $M_\mathrm{s}$ of the samples with Ir has the average value of $M_\mathrm{s,avg}=1.7\pm0.1$~MA/m. $H_\mathrm{K}$ decreases slightly with $t_\mathrm{Ir}$, but the changes are not very significant.

A temperature dependent measurement of saturation magnetization was also fitted by the Bloch law to estimate the exchange stiffness constant, $A$. The measurements for multilayers with $t_\mathrm{Ir}=0.0$ and 0.4~nm resulted in an average value of $A = 17.0 \pm 0.2$~pJ/m, which is in good agreement with our previous measurements \cite{Shepley18PRB}. Example data are shown in Appendix~\ref{app_exch}.

\subsection{Domain Wall Velocity}

\begin{figure*}
  \includegraphics[width=12cm]{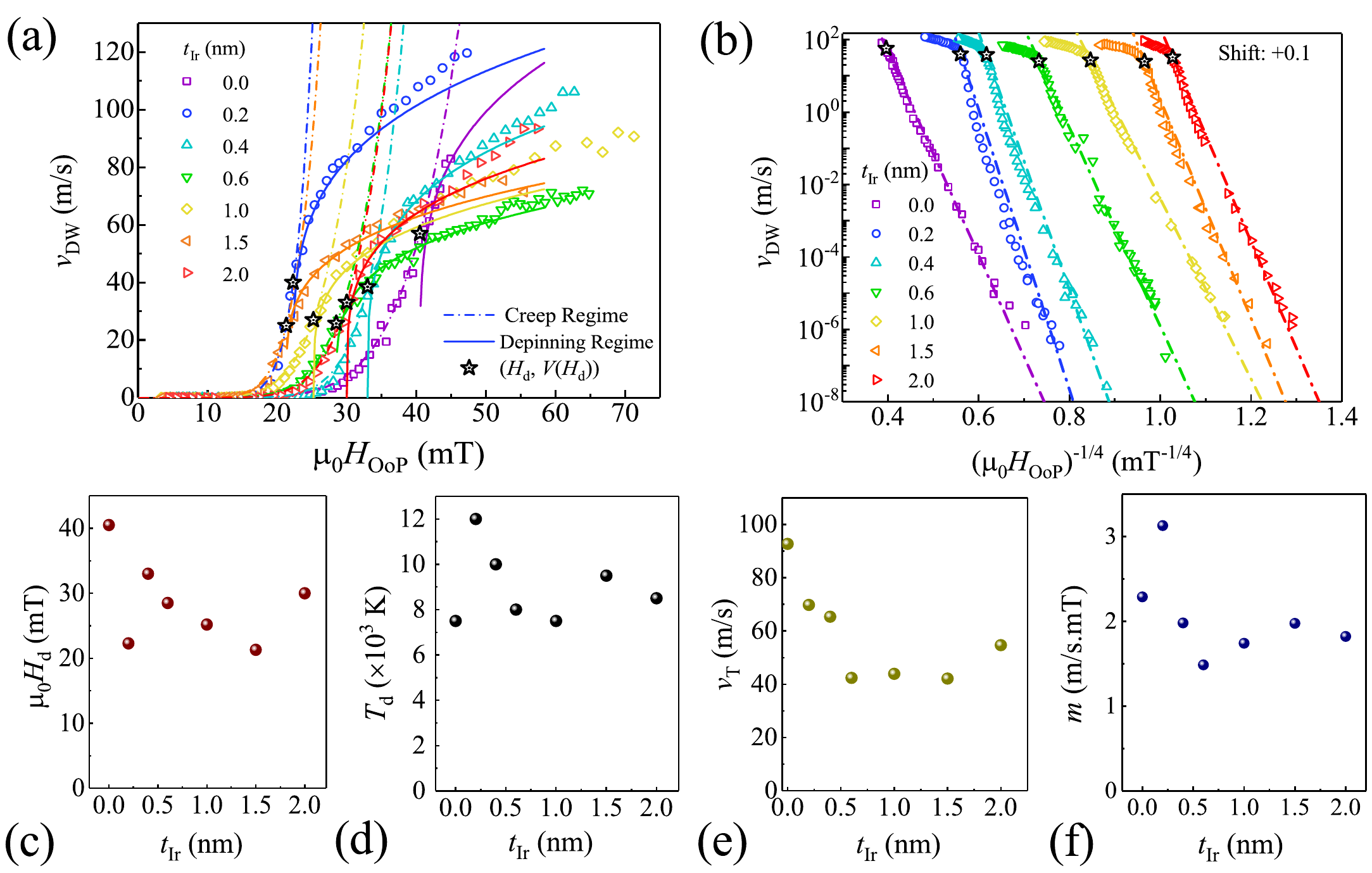}
  \caption{Field-induced domain wall motion. (a) DW velocity $v$ as a function of OoP applied field $\mu_0 H_\mathrm{OoP}$ for multilayers with increasing $t_\mathrm{Ir}$ along with the fits to creep and depinning universal functions (dashed-dot and solid lines, respectively). The stars show the inflection point which corresponds to the depinning field, $H_\mathrm{d}$. (b) $\log v$ as a function of scaled driving field to highlight the compatibility of experimental data to the universal creep law. Each set of data is horizontally shifted for a clearer presentation. The (c) depinning field $H_\mathrm{d}$, (d) depinning temperature $T_\mathrm{d}$, (e) disorder-free depinning velocity  $v_\mathrm{T}$, and (f) DW magnetic mobility $m_\mathrm{DW}$, each as a function of Ir thickness.
  \label{fig_DWM}}
\end{figure*}

As mentioned in the introduction, systematic studies of DW motion are needed to develop DW-based spintronic devices. Consequently, field induced domain wall motion (FIDWM) was studied for all of the films. Fig.~\ref{fig_DWM} shows the changes of DW velocity $v$ with increasing applied OoP field for stacks with different $t_\mathrm{Ir}$. DW dynamics is classified into different regimes of motion including the creep, depinning, and flow regimes \cite{metaxas07PRL}. In very low fields, the applied field is not enough to overcome the pinning barrier and move the DW. So, when $T\neq0$, thermal excitations can assist the field and cause a very slow motion of DWs known as creep. For fields higher than the so-called depinning field, $H_\mathrm{d}$, DW dynamics changes to a form known as the depinning regime. In both the creep and depinning regimes, thermal activation and the pinning potential dictate the DW movement. For a high enough drive field, the DW moves into a viscous flow regime that is independent of pinning force and temperature, and is only limited by dissipation.

As Fig.~\ref{fig_DWM}(b) shows, for fields lower than $H_\mathrm{d}$, $v$ rises by 9 orders of magnitude within a 10 mT field span, which is characteristic of creep regime behaviour. The linear changes of $\ln v $ vs. $H_\mathrm{OoP}^{-1/4}$ confirms the creep motion of the DWs in this field range \cite{lemerle98PRL}, confirming the validity of assuming DW as a 1D elastic interface progressing in a 2D medium with random-bond short-range pinning potential \cite{chauve00PRB,jeudy16PRL}. (In Appendix~\ref{app_creep} we show that this remains true even under the simultaneous application of an in-plane field.) For fields higher than $H_\mathrm{d}$, the measured velocities are in the depinning regime. All the data are fitted simultaneously with universal functions of creep and depinning motion of DW, to evaluate the three material-dependent pinning parameters: depinning temperature, $T_\mathrm{d}$, depinning field, $H_\mathrm{d}$, and disorder-free velocity at the depinning field, $v_\mathrm{T}$ \cite{jeudy16PRL,Jeudy18PRB}. These fits are shown in Fig.~\ref{fig_DWM}(a), and the parameters extracted from them are plotted in Fig.~\ref{fig_DWM}(c), (d), and (e). It appears that for larger Ir thicknesses, \textit{i.e.} $t_\mathrm{Ir}>0.5$~nm, each of these values stays roughly constant, while for $t_\mathrm{Ir}<0.5$~nm the parameters do not follow a monotonic change. This suggests that the Ir layer does not affect the DW dynamics significantly after it exceeds 0.5~nm in thickness. The estimated $T_\mathrm{d}$ and $H_\mathrm{d}$ are $\sim3$ and 4 times lower than the reported values for Au/Co/Au \cite{diazpardo17PRB} at room temperature, respectively, which indicates less average pinning in the films.

\begin{figure}
  \includegraphics[width=7cm]{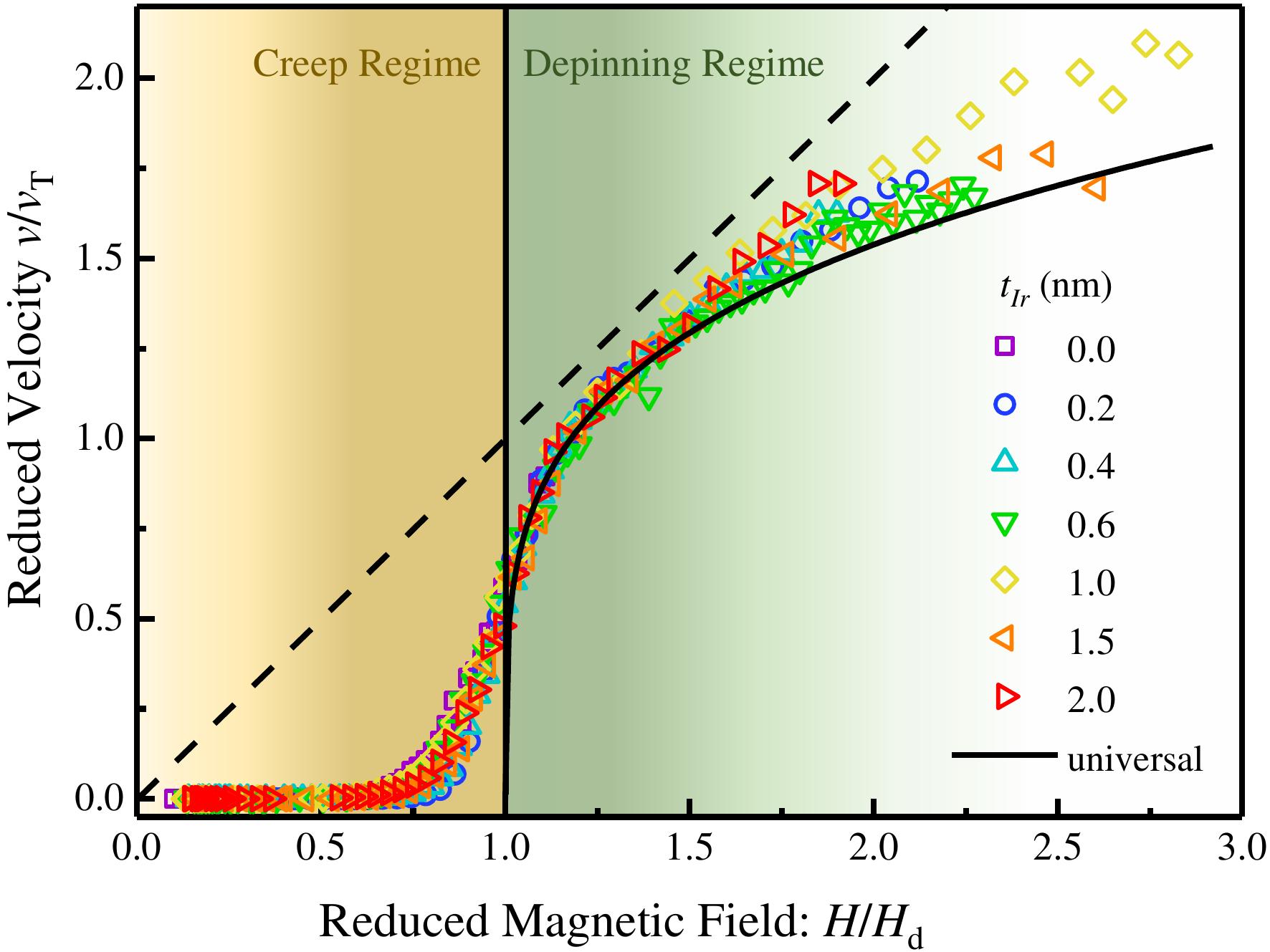}
  \caption{Reduced velocity vs. reduced field for different Ir thicknesses $t_\mathrm{Ir}$, showing a good universal collapse of the curves. The tilted dashed line represents the linear flow regime in which $v/v_{T}=H_\mathrm{OoP}/H_\mathrm{d}$. The vertical solid line separates the creep and depinning regime. The solid curve marks the universal depinning transition which is conforming to all experimental data in $1<H_\mathrm{OoP}/H_{\mathrm{d}}<1.5$ range. All the samples leave the universal depinning regime at some point after $H_\mathrm{OoP}/H_\mathrm{d} \approx 1.5$ and begin to approach the flow regime.
  \label{fig_universality}}
\end{figure}

By plotting the reduced velocity, $v/v_\mathrm{T}$, as a function of reduced driving field, $H_\mathrm{OoP}/H_\mathrm{d}$, shown in Fig~\ref{fig_universality}, one can see that all the data collapse onto one curve, emphasising the good agreement with universal depinning behaviour. This agreement is observed over a large range of $1 < H_\mathrm{OoP}/H_\mathrm{d} < 1.5$ that is comparable to the previous universality range (up to 1.3 at room temperature) reported for Pt/Co/Pt trilayers \cite{diazpardo17PRB}.

We were not able to fully enter the flow regime for any of the multilayers, due to multiple nucleation sites and the merging of bubble domains in high fields for the smallest available pulses. Nevertheless, we can estimate the DW dynamics in the flow regime using the material dependent parameters extracted from our fits. The depinning velocity, $v_\mathrm{T}$, is defined as the velocity of DW with $H_\mathrm{d}$ as driving force in the absence of pinning in the film. Knowing that, the mobility of DW can be calculated as $m_\mathrm{DW} = \mu_0^{-1} v_\mathrm{T}/H_\mathrm{d}$, according to Ref.~\onlinecite{Jeudy18PRB}. This is plotted in Fig.~\ref{fig_DWM}(f) for each of our samples. With the exception of the sample with $t_\mathrm{Ir}=0.2$~nm, this parameter is roughly constant, with an average of $m_\mathrm{DW,avg}=1.8\pm0.2$~ms$^{-1}$mT$^{-1}$, which is close to the DW mobility for Au/Co(0.8)/Au in Ref.~\onlinecite{kirilyuk97JMMM}. Using the DW mobility, the Gilbert damping can also be determined from $m_\mathrm{DW}=\gamma \Delta/\alpha$ for steady flow, or  $m_\mathrm{DW}=\gamma \Delta/{(\alpha+\alpha^{-1})}$ for precessional flow, $\Delta = \sqrt{A/K_\mathrm{eff}}$ is the DW thickness. As there was no solution for the precessional regime, the linear flow regime is proved to correspond to steady flow for all the samples, with the average damping value of $\alpha=0.48\pm0.01$, which, although high, is of a comparable order of magnitude to other results for Pt/Co multilayers \cite{metaxas07PRL}.

\subsection{Dzyaloshinskii-Moriya Interaction}

Multilayers of Co with PMA often have bubble domains. In the presence of DMI there will be an effective in-plane field $H_\mathrm{DMI}$ acting on the DW surrounding the bubble. The interaction of an applied in-plane field $H_\mathrm{InP}$ with $H_\mathrm{DMI}$ will affect the growth rate of opposite parts of the bubble domain, and so can be used for evaluation of the DMI strength and sign \cite{Je13PRB,Hrabec14PRB}. According to Je \textit{et al.} \cite{Je13PRB}, the energy density $\sigma_\mathrm{DW}$ of the DW can be written as
\begin{equation}
	\label{eq_BlochDW} \sigma_\mathrm{DW}(H_\mathrm{InP})=\sigma_0+\frac{\pi^2\Delta{M_\mathrm{S}}^2}{8K_\mathrm{D}}(H_\mathrm{InP}+H_\mathrm{DMI})^2,
\end{equation}
when $|H_\mathrm{InP}+H_\mathrm{DMI}|<({4K_\mathrm{D}}/{\pi M_\mathrm{S}})$, \textit{i.e.} where the sum of InP and DMI fields is not big enough to transform the wall configuration from Bloch to the N\'{e}el. On the other hand, in other conditions when the DW has the N\'{e}el structure, the DW energy density is
\begin{equation}
	\label{eq_NeelDW}
	\sigma_\mathrm{DW}(H_\mathrm{InP})=\sigma_0+2\Delta K_\mathrm{D}-\pi\Delta M_\mathrm{S}|H_\mathrm{InP}+H_\mathrm{DMI}|.
\end{equation}
In equations \ref{eq_BlochDW} and \ref{eq_NeelDW}, $\sigma_0$ is the Bloch DW energy, $M_\mathrm{S}$ is the saturation magnetization, $K_\mathrm{D}$ is the DW anisotropy energy density, and $\Delta$ is the DW width. Je \textit{et al.} argued that if the domain wall motion occurs in the creep regime then the DW velocity is \cite{Je13PRB}
\begin{equation}
	\label{eq_CreepVelocity}
	v=v_0\text{exp}(-\zeta H_{OoP}^{-\mu})
\end{equation}
where $v_0$ is the characteristic speed, $\mu$ is creep scaling exponent which is $1/4$ \cite{kim09Nat,lemerle98PRL}, and $\zeta$ is a scaling constant which exclusively is dependent on $H_\mathrm{InP}$ applied field via $\zeta=\zeta_0[\sigma(H_\mathrm{InP})/\sigma(0)]^{1/4}$. $\zeta_0$ is a scaling constant.

\begin{figure}
  \includegraphics[width=8cm]{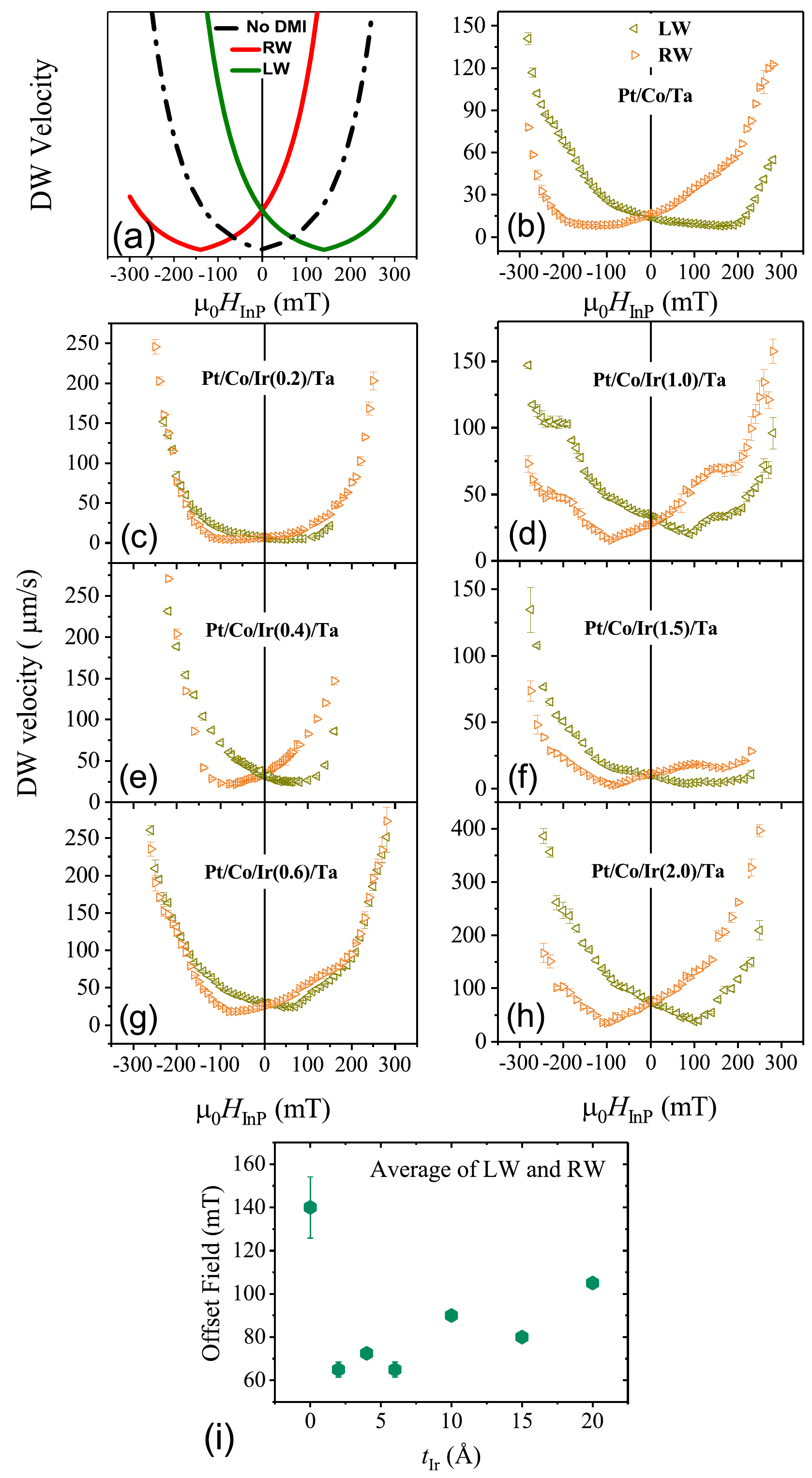}
  \caption{Variation of DW velocity $v$ with InP field $H_\mathrm{InP}$ for left- and right-moving DWs, LW and RW, respectively. (a) Simulated from Eq.~\ref{eq_CreepVelocity}. When there is no DMI, $v_\mathrm{DW}$ has a minimum at zero in-plane field and rises symmetrically for opposite field directions. When DMI is present the minimum point is shifted to $\pm H_\mathrm{offset}$ depending on the direction of the DMI vector inside the wall. (b)-(h) Experimental data for left wall (green triangles) and right wall (orange triangles) for multilayers with different $t_\mathrm{Ir}$. A lack of symmetry around the minimum points is obvious in most cases. (i) Average $H_\mathrm{offset}$ as a function of $t_\mathrm{Ir}$.
  \label{fig_BE}}
\end{figure}

\begin{figure*}
  \includegraphics[width=14cm]{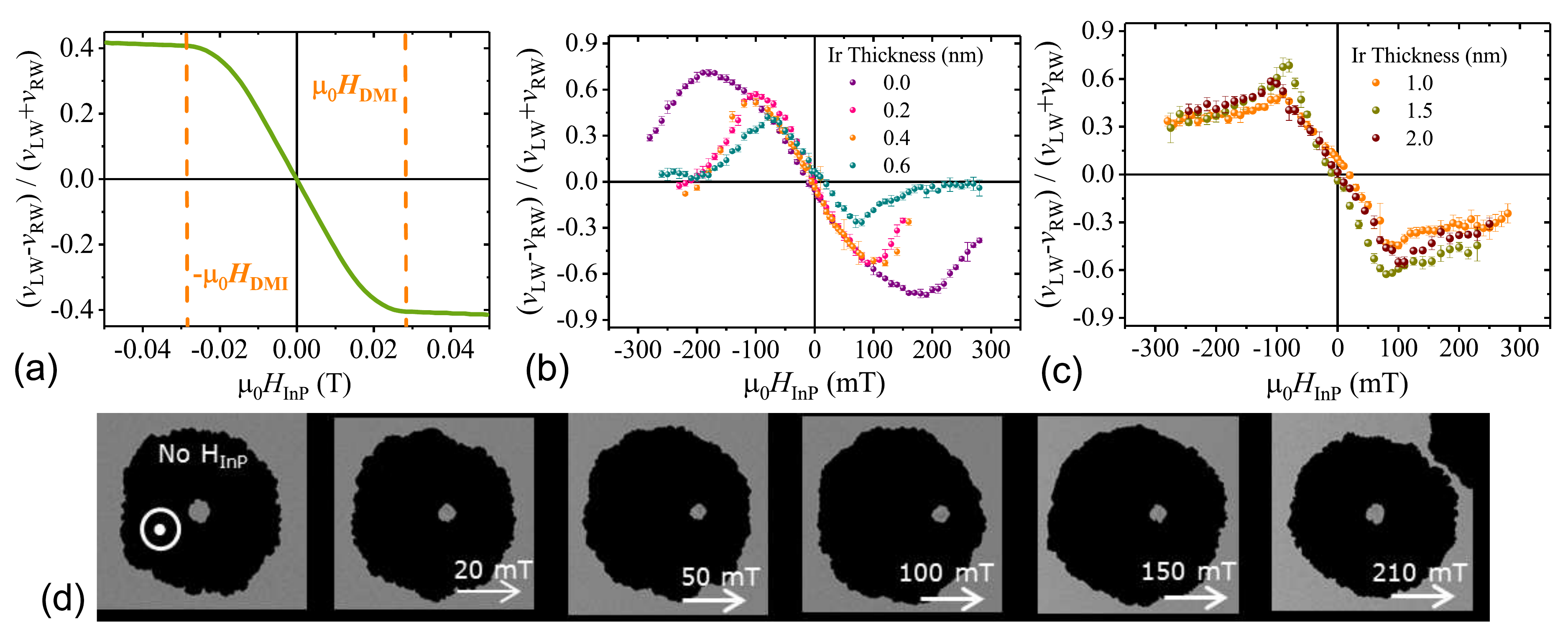}
  \caption{Difference between right and left DW velocities normalized to their sum, $(v_\mathrm{RW}-v_\mathrm{LW})/(v_\mathrm{RW}+v_\mathrm{LW})$, as a function of InP field, (a) extracted from Eq.~\ref{eq_CreepVelocity}, which predicts a rise in the velocity asymmetry with $H_\mathrm{InP}$ that reaches saturation at high enough fields. Experimental results for multilayers with (b) thin and (c) thick Ir layers (plotted separately for clarity), which show no such saturation. (d) Kerr microscope images from Pt/Co/Ir(0.6)/Ta  with increasing applied in-plane field. The propagated bubble domain changes from being symmetrical for zero InP field to asymmetrical growth for medium fields, then growth changes back to symmetrical for high enough magnetic fields.
  \label{fig_VelAsymm}}
\end{figure*}

In this way, when $H_\mathrm{InP}$ is equal and opposite to $H_\mathrm{DMI}$ the DW energy is maximum. Hence there will be a minimum in the velocity of the DW at that particular value of applied field, $H_\mathrm{offset}$. On the basis of this simple model, $H_\mathrm{offset}$ = $H_\mathrm{DMI}$, and the velocity of DW will increase symmetrically around this offset field. Consequently the radial symmetry of the DW creep is broken by $H_\mathrm{InP}$ and graphs of $v(H_\mathrm{InP})$ for DWs on opposite sides of the bubble will form mirrored offset pairs, as shown schematically in Fig~\ref{fig_BE}(a).

Results from asymmetric bubble expansion measurements on all the samples with different values of $t_\mathrm{Ir}$ are shown in Fig.~\ref{fig_BE}(b-h). Fig.~\ref{fig_BE}(i) shows the dependence of $H_\mathrm{offset}$, measured as an average of the values obtained from the curves for the left- and right-moving DWs, on $t_\mathrm{Ir}$. These data show the remarkable fact that $H_\mathrm{offset}$ drops significantly as soon as any Ir is introduced at the upper Co interface. Taking, for now, $H_\mathrm{offset}$ as an estimate for $H_\mathrm{DMI}$, this indicates a weaker overall DMI. This is at variance with \textit{ab initio} calculations \cite{Yang15PRL} and some experimental studies \cite{Chen13NatCom} that lead us to expect that Pt and Ir will induce DMI of opposite sign, leading to an overall additive effect when placed on either side of a ferromagnetic layer \cite{Moreau16NatNan}. Our result implies that the DMI induced by Ir is in fact more like that for Pt than that for Ta, which is expected to be small \cite{Torrejon14NatCom}. Germane to this, it is worth noting that recent BLS studies  also reported same sign of DMI for Co/Pt and Co/Ir interfaces \cite{Kim16APL}.

Unlike the curves expected from Eq.\ref{eq_BlochDW}-\ref{eq_CreepVelocity}--and other experimental results \cite{Hrabec14PRB, Khan16APL}--$v$ does not show a symmetrical change on either side of $H_\mathrm{offset}$ in any case. Furthermore, for some films there are step-like anomalies. Another notable feature is that the curves for the left- and right-moving DWs meet up at high enough in-plane fields, which is also not expected on the basis of Fig.~\ref{fig_BE}(a). Whilst none of these features can be explained with the theory of asymmetric bubble expansion in Ref.~\onlinecite{Je13PRB}, some have also been seen in other experiments on different structures \cite{Lavrijsen15PRB,Vanatka15JP,pellegren17PRL}. This suggests that the approach used in Eq.~\ref{eq_BlochDW} and \ref{eq_NeelDW} to define changes of DW energy with respect to $H_\mathrm{InP}$ is not universal and should be used with great care.

The in-plane field will eventually become strong enough to completely align the magnetization of the DW around the bubble in the direction of $H_\mathrm{InP}$. According to the simple creep model embodied in equations~\ref{eq_BlochDW}-\ref{eq_CreepVelocity}, beyond this point the two velocities for walls on either side of the bubble, $v_\mathrm{LW}$ and $v_\mathrm{RW}$, will have a constant ratio, meaning that their difference, normalized to their sum, will saturate, as shown in Fig.~\ref{fig_VelAsymm}(a). This behaviour is not observed in our multilayers. Fig.~\ref{fig_VelAsymm}(b) \& (c) show the variations of the velocity asymmetries for each sample with $t_\mathrm{Ir}$. Every curve has a peak (instead of a plateau) after which the asymmetry declines. In some cases we were able to apply a strong enough $H_\mathrm{InP}$ to bring the asymmetry back to zero, at which value it saturates. The details of DW velocity variation is also subtly different for samples with thinner and thicker Ir layers: for larger $t_\mathrm{Ir}$ the peak is more cusp-like. This return to symmetric expansion is also evident in Fig.~\ref{fig_VelAsymm}(d), which shows a series of Kerr images for increasing in-plane field. The growth asymmetry (\textit{i.e.} velocity asymmetry) initially increases, reaches a maximum value for $\mu_0 H_\mathrm{InP} \approx 100$~mT and then decreases until the propagation of DWs is symmetrical again at $\mu_0 H_\mathrm{InP} \approx 250$~mT. This behavior is not limited to one sample or one nucleation point, so it cannot be related to any spatial variation of the magnetic parameters.

\begin{figure}
  \includegraphics[width=8.5cm]{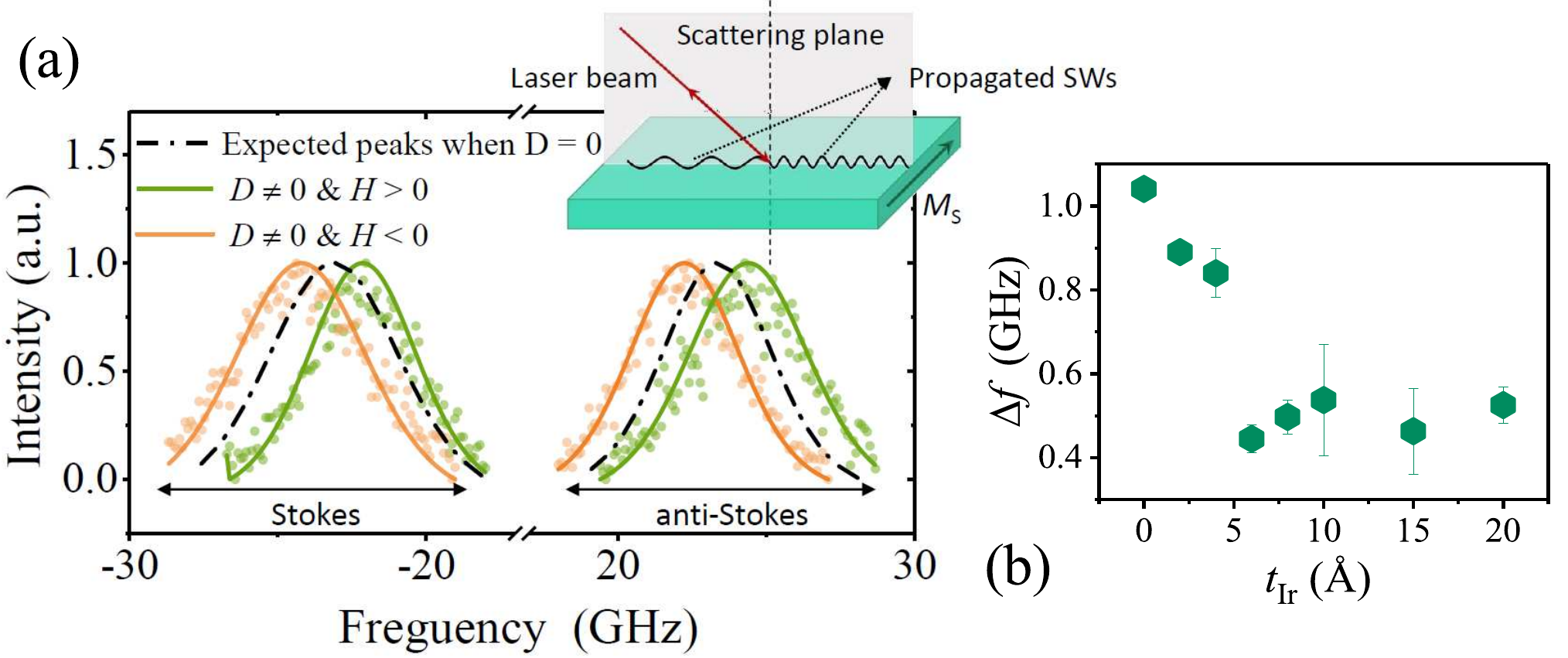}
  \caption{Brillouin light scattering. (a) Normalized BLS spectra measured for Pt/Co/Ta at two equal and opposite applied fields of $\sim1$ in orange and green. The black dashed line shows what is expected in case that there is no DMI in the sample. Symbols refer to experimental data and solid lines are Lorentzian fits. The panel at the top represents the Damon-Eschbach geometry which was used for measurements presented in this work. (b) Frequency shift $\Delta f$ against Ir thickness $t_\mathrm{Ir}$.
  \label{fig_bls}}
\end{figure}

In order to complement the asymmetric bubble expansion measurements, we also used BLS to evaluate the strength of the DMI in each of our multilayers. An example of the BLS spectra is shown in Fig.~\ref{fig_bls}(a), showing Stokes and anti-Stokes peaks. The nonreciprocal SW propagation in films with DMI leads to a frequency shift, $\Delta f$. This shift changes sign with magnetization direction. The black dashed-dotted line represents the expectation for the case when there is no DMI. Fig.~\ref{fig_bls}(b) shows $\Delta f$ averaged over the two frequency shifts applying opposite saturating fields and measured for each different value of $t_\mathrm{Ir}$. $\Delta f$ decreases slightly as the Ir is inserted between Co and Ta layers, but again remains almost constant for thicker Ir layers when $t_\mathrm{Ir}>0.5$ nm, reminiscent of Fig.~\ref{fig_BE}(i). Such a large nonreciprocity of the SWs cannot be because of surface anisotropy or dipolar effects. Surface anisotropy contributions come into play where $(k_\mathrm{SW}/t_\mathrm{FM}) \ll 1$ ($t_\mathrm{FM}$ is ferromagnetic thickness) \cite{stashkevich15PRB}, and thus are negligible here due to the ultrathin Co layer. On the other hand, frequency shifts resulting from dipolar effects do not change sign with respect to the magnetization direction of the sample in question \cite{moon13PRB}.

\section{Asymmetric Bubble Expansion Theory}

In this section, to go beyond the simple model expressed in Eq.\ref{eq_BlochDW}-\ref{eq_CreepVelocity}, we present a theoretical analysis of the dependence of the wall velocity as a function of in-plane applied fields, as shown in Fig.~\ref{fig_BE}. We consider the usual creep model,
\begin{equation}
	v = v_0 \exp\left( -\frac{\Delta E}{k_\mathrm{B} T} \right)
	\label{eq_creep}
\end{equation}
where the barrier energy has the universal form~\cite{jeudy16PRL}
\begin{equation}
	\Delta E = k_\mathrm{B} T_\mathrm{d} \left[ \left(\frac{H_\mathrm{OoP}}{H_\mathrm{d}} \right)^{-1/4} - 1 \right].
	\label{eq_barrier}
\end{equation}
Here, $v_0 = v(H=H_d)$, $H_\mathrm{d}$ is the depinning field, and $k_\mathrm{B} T_\mathrm{d}$ is the characteristic pinning energy scale. $H_\mathrm{OoP}$ is an OoP field driving the DW motion or bubble expansion. Our analysis is based on the assumption that the dominant contribution to the in-plane field dependence comes from the variation in the depinning field, $H_\mathrm{d}(H_\mathrm{InP})$. We compute this quantity numerically using micromagnetics simulations~\cite{Leliaert14JAP, Mumax14AIP} by following the method described in Ref.~\onlinecite{Kim17APL}. The simulation geometry comprises a 0.8 nm-thick ferromagnetic film with dimensions of $0.5~\upmu$m~$\times 1~\upmu$m that is discretized using $512 \times 1024 \times 1$ finite different cells. We used micromagnetic parameters consistent with the Pt/Co/Ta system, namely $M_\mathrm{s} = 1.19$~MA/m and $A = 20$~pJ/m. The magnetic disorder is modelled using a grain structure where the perpendicular anisotropy constant, $K_\mathrm{u} = K_\mathrm{eff} + \frac{1}{2} \mu_0 M_\mathrm{s}^2$, takes on a random value drawn from a uniform distribution centered about 1.38~MJ/m$^3$, with a 17.5\% spread in the values. The average grain size is taken to be 10 nm, consistent with our analysis of the TEM cross-sections shown in Fig.~\ref{fig_stackTEM}. The disorder parameters are chosen to give depinning fields at $H_\mathrm{InP}=0$ that match experimental values as closely as possible.

\begin{figure}
  \includegraphics[width=7.8cm]{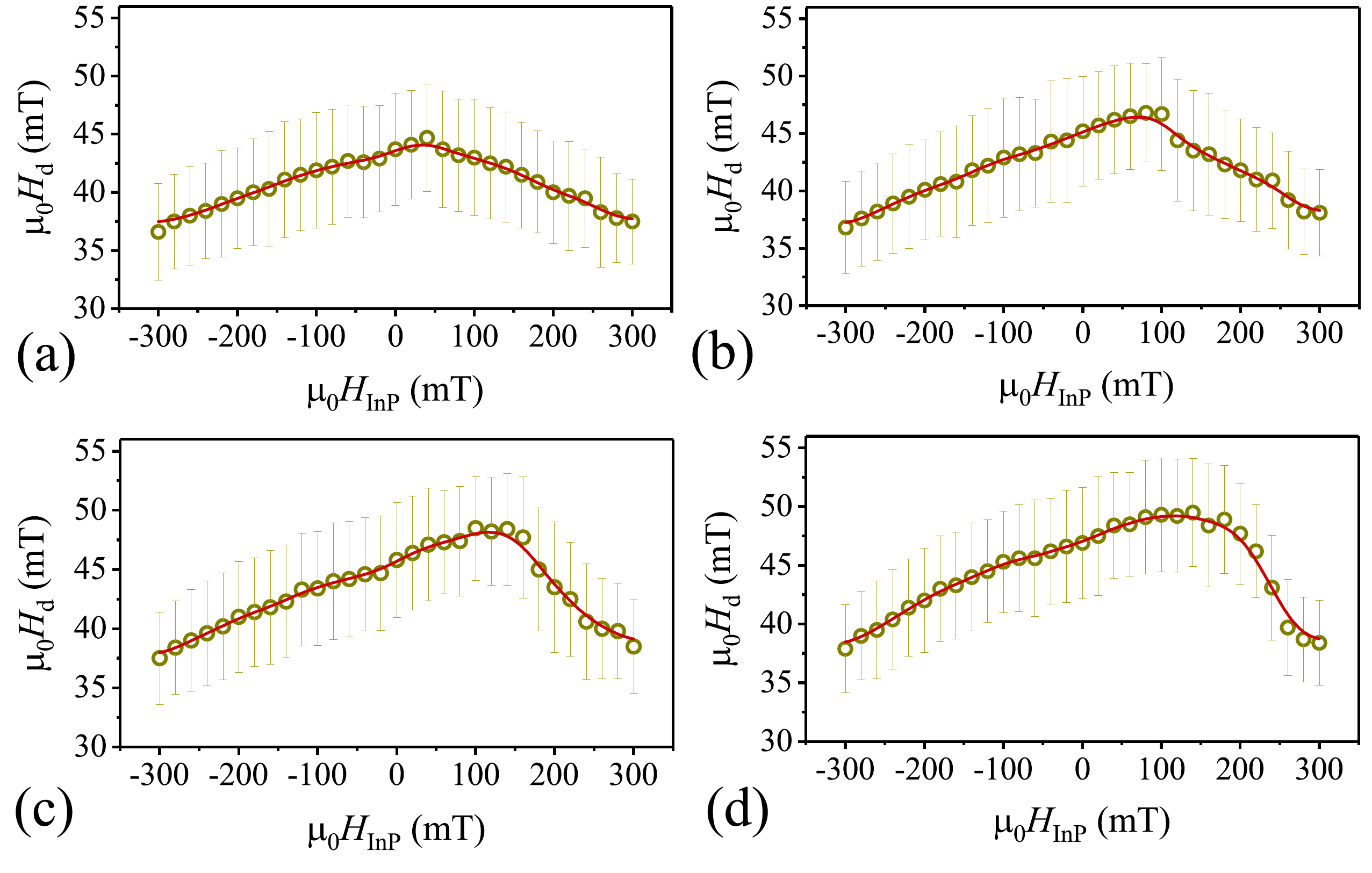}
  \caption{Depinning field as a function of in-plane applied field from micromagnetics simulations for different values of the DMI constant, $D$: (a) 0.5 mJ/m$^3$, (b) 1.0 mJ/m$^3$, (c) 1.5 mJ/m$^3$, and (d) 2.0 mJ/m$^3$. The circles represent the average $H_\mathrm{d}$ value and the error bars indicate one standard deviation. The solid (red) curve represents a smoothed function.
  \label{fig_HdvsHx_sim}}
\end{figure}

The depinning field is then estimated from the simulations as follows. For a given realization of the disorder, we relax an initially straight domain wall that runs along the width of the simulation grid in the $y$ direction. The wall is positioned close to the centre of the simulation grid ($x = 0$) and separates an up-domain to the left ($x < 0$) and a down-domain to the right ($x > 0$). To simulate an infinitely large system, periodic boundary conditions are applied along the $y$ direction, while the magnetization is assumed to be uniform outside the simulation grid in the $x$ direction. We include the dipolar fields from the magnetization outside the grid as an additional effective field. A uniform OoP external field is applied and is increased from zero in increments of 2 mT, where the magnetization is relaxed using an energy minimization procedure at each increment. During this procedure, the wall gradually roughens and the up-domain gradually expands toward the $x>0$ direction. The depinning field is assigned to be the highest field reached before the wall depins and sweeps through the system in the $x$ direction. This procedure is performed for 100 different realizations of the disorder for each value of the $H_\mathrm{InP}$ studied. The simulated variation in the depinning field, $H_\mathrm{d}(H_\mathrm{InP})$, for four different values of the DMI constant, $D$, is shown in Fig.~\ref{fig_HdvsHx_sim}. $H_\mathrm{d}$ has a maximum at a certain value of $H_\mathrm{InP}$ that increases as the DMI becomes stronger. The presence of the DMI leads to an asymmetric variation of $H_\mathrm{d}$ with respect to $H_\mathrm{InP}$, where the asymmetry becomes more pronounced as $D$ is increased. This variation can be in the tens of mT over the field range studied, which can lead to significant variations in the energy barrier given in Eq.~(\ref{eq_barrier}). We note that the functional form of $H_\mathrm{d}(H_\mathrm{InP})$ is reminiscent of the changes in the elastic energy of the domain wall~\cite{pellegren17PRL}.

\begin{figure}
  \includegraphics[width=8.5cm]{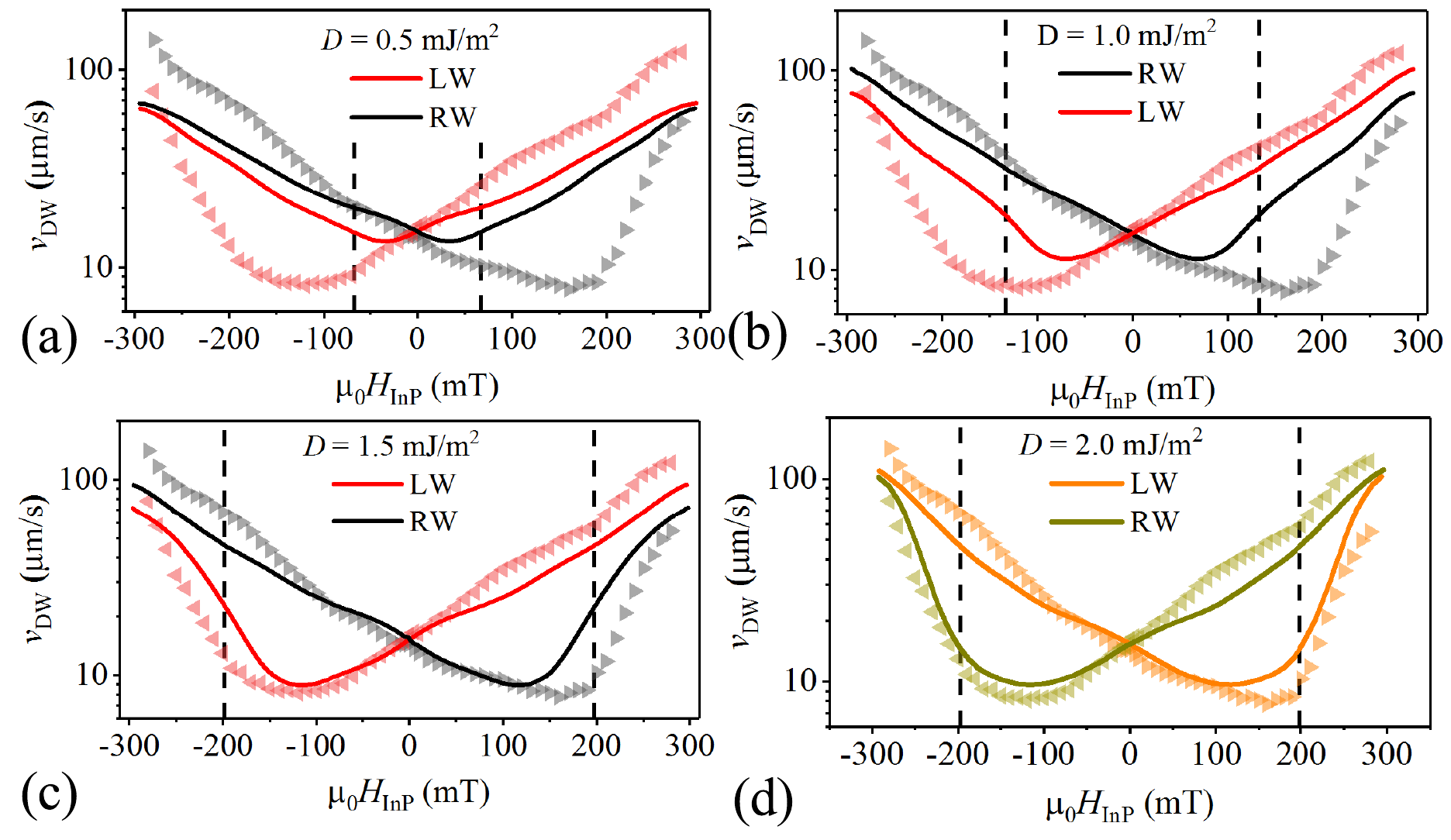}
  \caption{Domain wall velocity, $v$, as a function of in-plane applied field, $H_\mathrm{InP}$, for four different values of the DMI constant, $D$: (a) 0.5~mJ/m$^2$, (b) 1.0~mJ/m$^2$, (c) 1.5~mJ/m$^2$, and (d) 2.0~mJ/m$^2$. The triangles represent experimental data for the Pt/Co/Ta system (Fig.~\ref{fig_BE}). The dashed vertical lines represent the DMI field extracted from simulation.
  \label{fig_velvsHx_sim}}
\end{figure}

A prediction of how the wall velocity varies with $H_\mathrm{InP}$ can be made by using Eqs.~(\ref{eq_creep}) and (\ref{eq_barrier}) along with $H_\mathrm{d}(H_\mathrm{InP})$ from Fig.~\ref{fig_HdvsHx_sim}. The results are presented in Fig.~\ref{fig_velvsHx_sim} for different values of $D$. The velocity curves are computed for each $D$ value as follows. First, we perform a fit of Eq.~(\ref{eq_creep}) on the experimental $v(H_\mathrm{OoP})$ data with the value of $H_\mathrm{d} = H_{\mathrm{d,sim}}$ obtained from simulation, which allows us to determine $v_0 = v_0^*$ and $T_\mathrm{d} = T_\mathrm{d}^*$. Second, we use these parameters $(H_{\mathrm{d,sim}}, v_0^*, T_\mathrm{d}^*)$ to calculate the value of $H_\mathrm{OoP} = H_\mathrm{OoP}^*$ in Eq.~(\ref{eq_barrier}) such that the velocities match the experimental data at $H_\mathrm{InP} = 0$ in Fig.~\ref{fig_BE}b. The $v(H_\mathrm{InP})$ curves depicted in Fig.~\ref{fig_velvsHx_sim} were then obtained by using the smoothed function $H_\mathrm{d}(H_\mathrm{InP})$ (Fig.~\ref{fig_HdvsHx_sim}) in the expression for the energy barrier given in Eq.~\ref{eq_barrier} in which we insert the values $T_\mathrm{d} = T_\mathrm{d}^*$ and $ H = H_\mathrm{OoP}^*$. The only freely-adjustable parameter is $D$.

In the light of this, that the theoretical $v(H_\mathrm{InP})$ curve for $D = 2.0$ mJ/m$^2$ reproduces semiquantitatively the experimental data for Pt/Co/Ta system [Fig.~\ref{fig_velvsHx_sim}(d)] is remarkable. Besides capturing the overall shape of the asymmetry and the position of the velocity minimum, the curve also reproduces the fact that the velocities for the left and right wall converge toward one another as the magnitude of $H_\mathrm{InP}$ is increased. Such behaviour is absent in previous models for the DMI-induced changes in the wall velocity~\cite{Je13PRB}, where the DMI enters simply as an effective magnetic field (which results only in a shift of the velocity curve along the $H_\mathrm{InP}$ axis). It is also important to note that the position of the velocity minimum does not coincide with the effective DMI field, $\mu_0 H_\mathrm{DMI} = D/M_\mathrm{s} \Delta$, values for which are indicated in Fig.~\ref{fig_velvsHx_sim} by the vertical dashed lines (and computed from the micromagnetic parameters used in the simulations). This indicates that equating $H_\mathrm{DMI}$ with the field $H_\mathrm{InP}$ at which the velocity minimum occurs can lead to a significant underestimate of the DMI. These features are also present in the velocity curves for lower values of $D$, as shown in Figs.~\ref{fig_velvsHx_sim}(a-c).

\section{Discussion}

\begin{figure}
  \includegraphics[width=7cm]{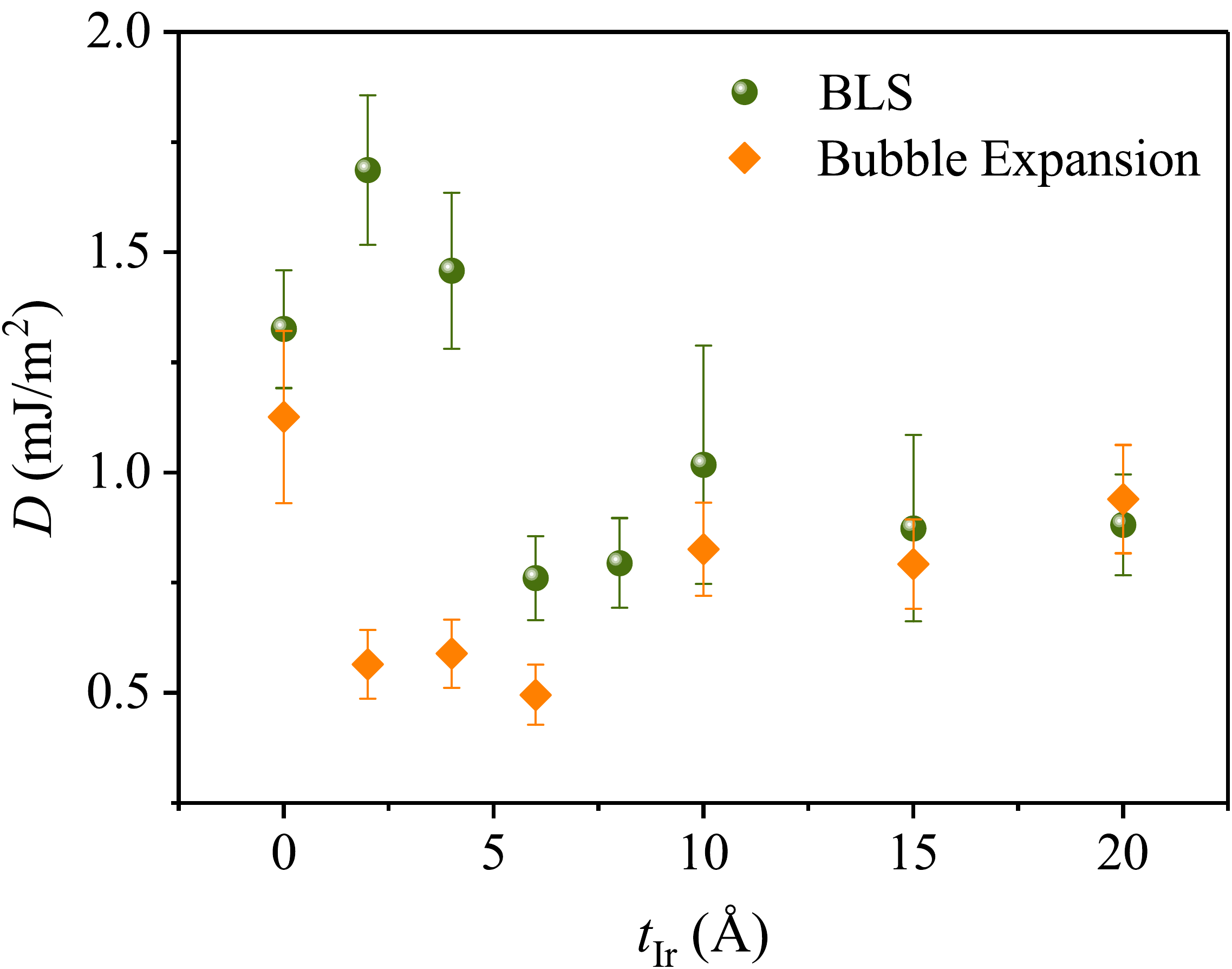}
  \caption{Comparison of DMI results for the series of samples. Orange dimonds circles are the results obtained from taking the field of minimum asymmetric bubble expansion velocity as $H_\mathrm{DMI}$. Green circles are the results from BLS. \label{fig_dmi}}
\end{figure}

Keeping all the above in mind, if we still assume that the $H_\mathrm{InP}=-H_\mathrm{DMI}$ for the velocity minimum, the DMI factor, $D$, can be calculated via $D=\mu_0 H_\mathrm{DMI}M_\mathrm{S}\Delta$ \cite{Hrabec14PRB,thiaville12EPL}. Here $\Delta$ is the DW width. The frequency shift of the spin waves can also be used to get $D$ as follows:
\begin{equation}
  \label{eq_BLS}
  \Delta f = \left| \frac{g^{\parallel}\mu_B}{h} \right| \mathrm{sgn}(M_\mathrm{OoP}) \frac{2D}{M_\mathrm{S}} k_\mathrm{InP}
\end{equation}
where $g^{\parallel}$ is the in-plane splitting factor, $\mu_\mathrm{B}$ the Bohr magneton and $h$ the Planck constant \cite{nembach15NatPhys}. The resulting DMI strengths assessed by the two methods are shown in Fig.~\ref{fig_dmi}.

In the past, results from BLS have shown stronger $D$ that results from asymmetric bubble expansion on the same sample \cite{Soucaille16PRB,zeissler17SciRep}. The same is true here for Ir thicknesses up to about 1~nm, although there is better agreement between the two methods for other values of $t_\mathrm{Ir}$. This might be due to the different ways that BLS and asymmetric bubble expansion probe the film. Asymmetric bubble expansion studies growth of a bubble domain, the nucleation and propagation of which is sensitive to spatial variation of the film's properties. Our simulation results also emphasise this sensitivity of DW creep to spatially fluctuating magnetic properties due to defects (\textit{i.e.} disorder distributions). There can be seen a large standard deviation of calculated $H_\mathrm{d}$ values in Fig.~\ref{fig_HdvsHx_sim} regardless of same input macroscopic experimental values. On the other hand, BLS measures difference in spin waves propagation in which local fluctuations of the properties are ineffective \cite{gross16PRB}. In this way, the resulting $D$ can be considered to be an average value for the film, not just at the strongest pinning sites that control the creep motion.

Taking a closer look at Fig.~\ref{fig_dmi} reveals that there are still two similarities between the BLS and asymmetric bubble expansion results. First, both show a reduction of the net $D$ value when comparing the samples having thick Ir --effectively a Pt/Co/Ir trilayer-- with the zero-Ir, \textit{i.e.} Pt/Co/Ta trilayer. As briefly dicussed above, this suggests that the DMI constant at an Ir/Co interface has the same sign as at a Pt/Co interface. The sign and strength changes of the DMI constant when one scans through $5d$ transition metals has been reported previously \cite{Ryu14NatCom,Torrejon14NatCom, Ma18PRL}. The DMI sign of a Pt/FM interface proved to be negative more often than not \cite{Yang15PRL,Khan16APL, Wells17PRB,Shepley18PRB}, which is equivalent to introducing left-handed chirality into the system. But the Ir/FM case is not as straightforward as Pt/FM interface. Initially, \textit{ab initio} calculations proposed that Ir introduces the opposite chirality to Pt \cite{kashid14PRB,Yang15PRL,Yamamoto17AIP}, which was supported by various experimental reports of additive effects\cite{Ryu14NatCom,Hrabec14PRB,Moreau16NatNan,Bacani16X,Ma17X}. Later on, the sign of DMI for Ir was debated when several experimental studies observed right-handed chirality in multilayers including Ir/Co \cite{Kim16APL, Ajejas17APL, Ma18PRL}. The curious case of Ir does not end here, as other \textit{ab initio} calculations showed that the chirality of Ir/FM interface differs when the adjacent ferromagnetic material changes from Fe (right-handed), to Co (left-handed), and Ni (right-handed) \cite{Belabbes16PRL}. Ma \textit{et al.} also measured opposite signs of $D$ for Ir/Co and Ir/CoFeB interfaces \cite{Ma18PRL}. Considering all these contradictory results about the sign of the DMI at Ir/FM interfaces, one will wonder about the possible physical reasons for it. In this type of system, the DMI is considered to be mostly an interfacial effect. Partly, this interfacial sensitivity comes from dependence on the HM $5d$ states filling. For example, the DMI has opposite sign for W and Ta, with less-than-half-filled $5d$ states, in comparison with Pt and Au with more-than-half-filled $5d$ states \cite{Ma18PRL}.  In addition, the same $5d$ states hybridise with $3d$ orbitals in the adjacent ferromagnet and the changes in hybridisation, as well as the alignment of Fermi levels across the FM/HM interface, will affect the DMI \cite{kashid14PRB,Belabbes16PRL}. As DMI is sensitive to slight changes in any hybridization or Fermi alignment, controlling the interface quality on the atomic level might be needed to fix the DMI strength and even sign. In polycrystalline thin films, such as the ones in this paper, this much control over deposition is almost impossible as there is always unavoidable interdiffusion, or interface roughness which is changing from sample to sample and system to system. The situation is more crucial for Ir, since it is placed in the middle of elements having opposite DMI signs in $5d$ heavy metals, having 7 electrons in its $5d$ orbitals.

Second, for $t_\mathrm{Ir} \gtrsim 0.5$~nm, not only are the measured $D$ values from both methods in close agreement (within error bars and despite all mentioned anomalies), but also only weakly, if at all, dependent on $t_\mathrm{Ir}$. This lack of dependence on $t_\mathrm{Ir}$ once it approaches this value was also observed for other material parameters such as those derived from field-induced DW motion fitted by universal creep and depinning regime functions (Fig~\ref{fig_DWM}(c-f)) or the coercive field (inset of Fig.~\ref{fig_magnetic}(a)). This seems reasonable in the light of the fact that the Ir layer is of extreme thinness when $t_\mathrm{Ir}$ is less than this value, and will not be completely continuous, such that the Co layer underneath remains in direct contact with Ta in some places. This will causes local variation in the value of interface dependent properties. Unfortunately, characterising such very thin Ir layers by means of TEM cross-sections is very challenging, and so we are not able to discuss the matter in a more quantitative way from an experimental point of view. Nevertheless, first principles calculations suggest that 80\% of the DMI strength is related to the first two monolayers of the HM layer \cite{Ma18PRL}. Two monolayers of Ir in the $(111)$ growth direction would be about 0.5~nm, the thickness after which the measured DMI constant does not change significantly. Changes of DMI with HM thickness and its saturation at high enough thicknesses was also reported for other multilayers \cite{Ma16PRB, Tacchi17PRL}. A case closer to ours is that reported by Rowan-Robinson \textit{et al.}, who observed this saturation for $t_\mathrm{Ir}>0.4$~nm for Pt/Co/Ir(0.0-2.5~nm)/Pt multilayers\cite{Robinson17SciRep}, which is in good agreement with our observations here.

It is good to keep in mind that for the case of $t_\mathrm{Ir}=0$, Co is in direct contact with Ta and reportedly this interface makes a magnetic deadlayer\cite{yu16SCiRep,bandiera11IEEE,zhang17APL,sato12APL,Jang10JAP,Jang11JAP,oguz08JAP,cheng11JAP,sinha13APL,wang06JAP,ingvarsson02JMMM}, which eventually leads to an underestimation of $M_\mathrm{S}$ due to the reduction of the effective thickness of the ferromagnetic material. This is reflected in the calculation of other parameters for this stack including the DMI strength, $D$. Nevertheless, if we take the average value of saturation magnetization as this sample's $M_\mathrm{S}$, the DMI strength will be $D_\mathrm{BLS}=1.8\pm0.1$~mJ/m$^2$ and $D_\mathrm{ABE}=1.6\pm0.3$~mJ/m$^2$ for BLS and asymmetric bubble expansion (na\"{\i}vely taking $H_\mathrm{DMI}$ to be the velocity minimum), respectively. The $D_\mathrm{BLS}$ value is very close to $D_\mathrm{sim}=2.0$~mJ/m$^2$ the value used in the simulations in Fig. \ref{fig_velvsHx_sim}(d).

\section{Conclusion}

The experimental data presented here gives a full picture of the DW dynamics and DMI of polycrystalline Pt/Co/Ir($t_\mathrm{Ir}$)/Ta multilayers. The chirality of the DWs proved to be left-handed using asymmetric bubble expansion, which is the usual behaviour reported for DWs in both theoretical \cite{kashid14PRB, Yang15PRL, Freimuth14JPhysCM} and experimental \cite{Emori13NatMat, Ryu14NatCom, Belmeguenai15PRB} studies of Co/Pt interfaces, also for Pt/Co/Ir multilayers \cite{Hrabec14PRB, zeissler17SciRep, Shepley18PRB}. The experimental $v(H_\mathrm{InP})$ curves for these films that were acquired using asymmetrical bubble expansion (Fig~\ref{fig_BE}(b-h)) do not have the form expected from the simple creep model \cite{Je13PRB} that is often used to analyse such data (Fig~\ref{fig_BE}(a)). In that model, the dependence of $v$ on $H_\mathrm{InP}$ appears exclusively in the DW energy. Meanwhile Soucaille \textit{et al.} \cite{Soucaille16PRB} reported variation of DW roughness with  $H_\mathrm{InP}$, and Pellegren \textit{et al.} considered the role of DW elastic energy \cite{pellegren17PRL}. Based on our observation of universal scaling (Fig.~\ref{fig_universality}), we introduced asymmetric variation of the depinning field, $H_\mathrm{d}$, with $H_\mathrm{InP}$. This model reproduces both lack of symmetry of the $v(H_\mathrm{InP})$ curve about their minima and the closing up of these curves for left- and right-moving DWs at high InP field. In the case of $t_\mathrm{Ir}=0$, where we have made a direct quantitative comparison, this approach to analysing the asymmetric bubble expansion data gives much better agreement with BLS results on the same sample than the na\"{\i}ve process of measuring the field at which a velocity minimum is observed.

\appendix

\section{Exchange Stiffness Measurement}
\label{app_exch}

\begin{figure}
	\includegraphics[width=6cm]{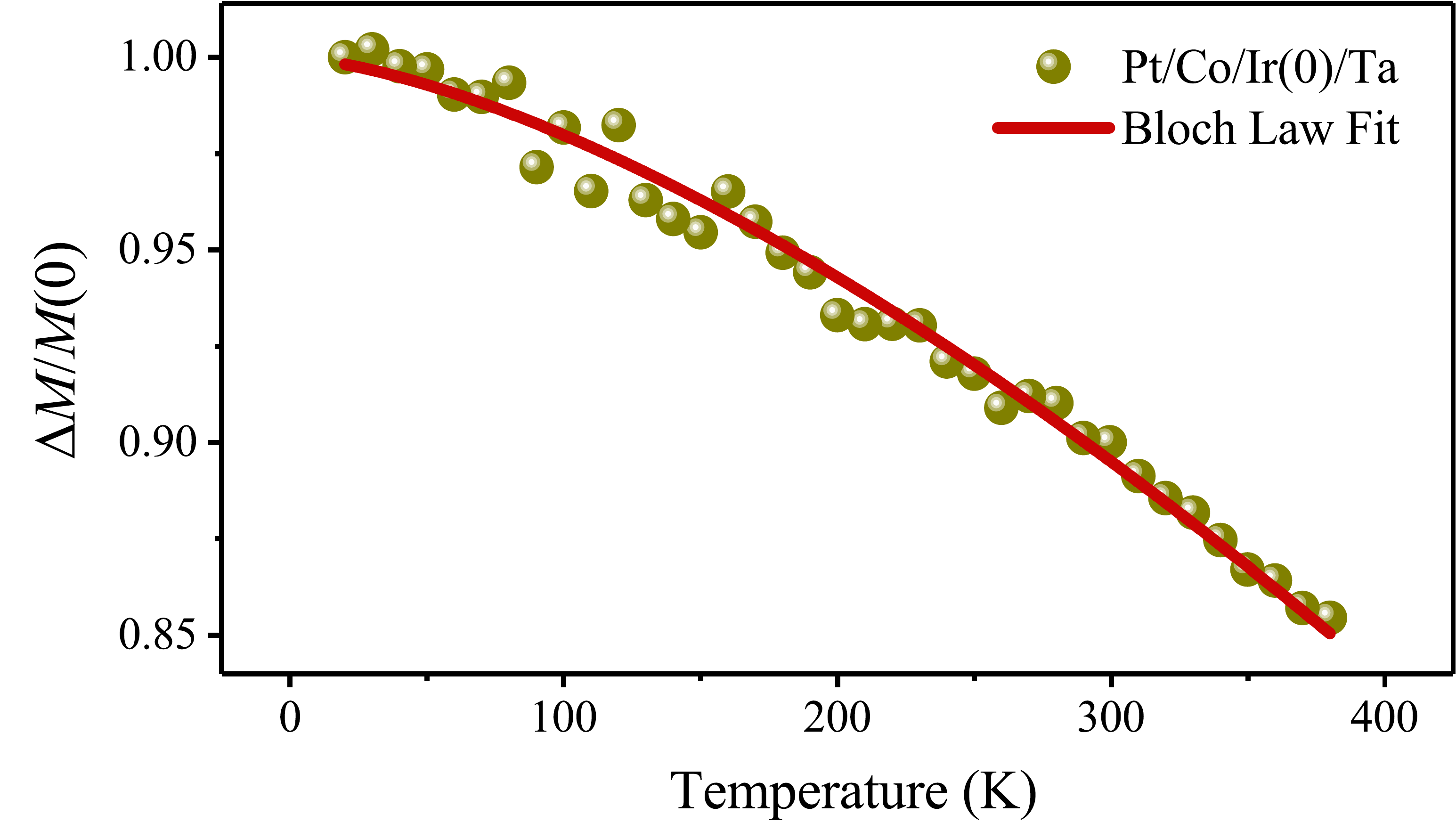}
	\caption{Temperature dependence of magnetization in Pt/Co/Ta. The red line shows the Bloch law fit. \label{fig_MvT}}
\end{figure}

We measured the temperature dependence of the saturation magnetisation $M_\textrm{s}(T)$ in order to determine the exchange stiffness $A$. This may be fitted to a Bloch law $M_\textrm{s}(T) = M_0 \left( 1 - bT^{3/2} \right)$, in which $M_0$ is the saturation magnetisation at zero~K, in order to extract the coefficient $b$. The exchange stiffness is then given by\cite{chikazumi1997}
\begin{equation}
  A = \frac{n k_{\mathrm{B}} S^2}{a}\left( \frac{C}{b} \right)^{2/3},
\end{equation}
where $n=4$ is the co-ordination number for an fcc lattice, $S$ is the spin per atom, $C$ is a constant that takes the value 0.0294 for an fcc lattice, and $a$ is the lattice constant. The data and fit for the Pt/Co/Ta sample (\textit{i.e.} Ir thickness $t_\mathrm{Ir} = 0$) are shown in Fig.~\ref{fig_MvT}. This fit yielded an average value of $A = 17.0 \pm 0.2$~pJ/m for multilayers with $t_\mathrm{Ir}=0.0$ and 0.4~nm.

\section{Confirming Creep Motion}
\label{app_creep}

\begin{figure}
 \includegraphics[width=9cm]{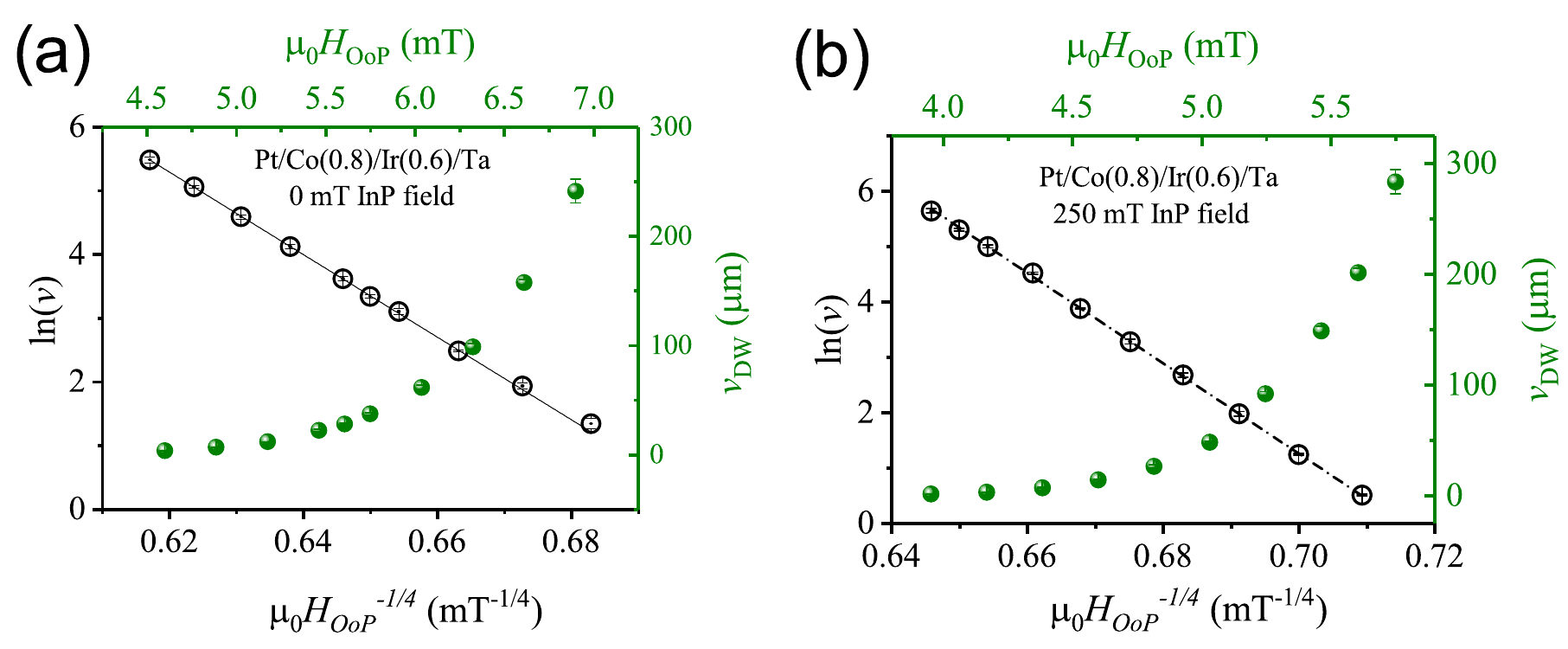}
 \caption{Investigation of creep propagation of the DWs in a bubble domain for $t_\mathrm{Ir} = 0.6$~nm. (a) Without any applied in-plane field and (b) in the presence of a high 250~mT in-plane field. In both cases the creep scaling law is well-obeyed. \label{inpcreep}}
\end{figure}

In this appendix we show additional data on DW creep motion. We checked that the samples were in the creep regime both in the absence and presence of (high) InP field in order to ensure that our asymmetric bubble expansion measurements were not affected by the changing InP field. A linear variation of $\ln v$ as a function of $H^{-1/4}$ verifies DW creep propagation \cite{lemerle98PRL}, as shown in Fig.~\ref{fig_DWM}(b) for zero in-plane field. We also performed measurements with an in-plane field of 250~mT, as shown here in Fig.~\ref{inpcreep} for the sample with $t_\mathrm{Ir} = 0.6$~nm. In both the zero-field and 250~mT field case, this scaling is well-obeyed by the sample.

\begin{acknowledgments}
The authors thank Simone Moretti and Rebeca D\'{\i}az-Pardo for their help. This work was supported by European Community under the Seventh Framework Programme – The People Programme, Multi-ITN “Wall” (Grant agreement no:608031).
\end{acknowledgments}


\bibliography{iridium_v4}

\begin{thebibliography}{77}%
\makeatletter
\providecommand \@ifxundefined [1]{%
 \@ifx{#1\undefined}
}%
\providecommand \@ifnum [1]{%
 \ifnum #1\expandafter \@firstoftwo
 \else \expandafter \@secondoftwo
 \fi
}%
\providecommand \@ifx [1]{%
 \ifx #1\expandafter \@firstoftwo
 \else \expandafter \@secondoftwo
 \fi
}%
\providecommand \natexlab [1]{#1}%
\providecommand \enquote  [1]{``#1''}%
\providecommand \bibnamefont  [1]{#1}%
\providecommand \bibfnamefont [1]{#1}%
\providecommand \citenamefont [1]{#1}%
\providecommand \href@noop [0]{\@secondoftwo}%
\providecommand \href [0]{\begingroup \@sanitize@url \@href}%
\providecommand \@href[1]{\@@startlink{#1}\@@href}%
\providecommand \@@href[1]{\endgroup#1\@@endlink}%
\providecommand \@sanitize@url [0]{\catcode `\\12\catcode `\$12\catcode
  `\&12\catcode `\#12\catcode `\^12\catcode `\_12\catcode `\%12\relax}%
\providecommand \@@startlink[1]{}%
\providecommand \@@endlink[0]{}%
\providecommand \url  [0]{\begingroup\@sanitize@url \@url }%
\providecommand \@url [1]{\endgroup\@href {#1}{\urlprefix }}%
\providecommand \urlprefix  [0]{URL }%
\providecommand \Eprint [0]{\href }%
\providecommand \doibase [0]{http://dx.doi.org/}%
\providecommand \selectlanguage [0]{\@gobble}%
\providecommand \bibinfo  [0]{\@secondoftwo}%
\providecommand \bibfield  [0]{\@secondoftwo}%
\providecommand \translation [1]{[#1]}%
\providecommand \BibitemOpen [0]{}%
\providecommand \bibitemStop [0]{}%
\providecommand \bibitemNoStop [0]{.\EOS\space}%
\providecommand \EOS [0]{\spacefactor3000\relax}%
\providecommand \BibitemShut  [1]{\csname bibitem#1\endcsname}%
\let\auto@bib@innerbib\@empty
\bibitem [{\citenamefont {Hilbert}\ and\ \citenamefont
  {Lopez}(2011)}]{Hilbert2011}%
  \BibitemOpen
  \bibfield  {author} {\bibinfo {author} {\bibfnamefont {M.}~\bibnamefont
  {Hilbert}}\ and\ \bibinfo {author} {\bibfnamefont {P.}~\bibnamefont
  {Lopez}},\ }\bibfield  {title} {\enquote {\bibinfo {title} {The world's
  technological capacity to store, communicate, and compute information},}\
  }\href {\doibase 10.1126/science.1200970} {\bibfield  {journal} {\bibinfo
  {journal} {Science}\ }\textbf {\bibinfo {volume} {332}},\ \bibinfo {pages}
  {60} (\bibinfo {year} {2011})}\BibitemShut {NoStop}%
\bibitem [{\citenamefont {Parkin}\ and\ \citenamefont
  {Yang}(2015)}]{Parkin2015}%
  \BibitemOpen
  \bibfield  {author} {\bibinfo {author} {\bibfnamefont {S.~S.~P.}\
  \bibnamefont {Parkin}}\ and\ \bibinfo {author} {\bibfnamefont {S.-H.}\
  \bibnamefont {Yang}},\ }\bibfield  {title} {\enquote {\bibinfo {title}
  {Memory on the racetrack},}\ }\href {\doibase 10.1038/nnano.2015.41}
  {\bibfield  {journal} {\bibinfo  {journal} {Nature Nanotechnol.}\ }\textbf
  {\bibinfo {volume} {10}},\ \bibinfo {pages} {195} (\bibinfo {year}
  {2015})}\BibitemShut {NoStop}%
\bibitem [{\citenamefont {Fert}\ \emph {et~al.}(2013)\citenamefont {Fert},
  \citenamefont {Cros},\ and\ \citenamefont {Sampaio}}]{Fert2013}%
  \BibitemOpen
  \bibfield  {author} {\bibinfo {author} {\bibfnamefont {A.}~\bibnamefont
  {Fert}}, \bibinfo {author} {\bibfnamefont {V.}~\bibnamefont {Cros}}, \ and\
  \bibinfo {author} {\bibfnamefont {J.}~\bibnamefont {Sampaio}},\ }\bibfield
  {title} {\enquote {\bibinfo {title} {Skyrmions on the track},}\ }\href@noop
  {} {\bibfield  {journal} {\bibinfo  {journal} {Nature Nanotech.}\ }\textbf
  {\bibinfo {volume} {8}},\ \bibinfo {pages} {152} (\bibinfo {year}
  {2013})}\BibitemShut {NoStop}%
\bibitem [{\citenamefont {Thiaville}\ \emph {et~al.}(2012)\citenamefont
  {Thiaville}, \citenamefont {Rohart}, \citenamefont {Ju{\'e}}, \citenamefont
  {Cros},\ and\ \citenamefont {Fert}}]{thiaville12EPL}%
  \BibitemOpen
  \bibfield  {author} {\bibinfo {author} {\bibfnamefont {A.}~\bibnamefont
  {Thiaville}}, \bibinfo {author} {\bibfnamefont {S.}~\bibnamefont {Rohart}},
  \bibinfo {author} {\bibfnamefont {{\'E}.}~\bibnamefont {Ju{\'e}}}, \bibinfo
  {author} {\bibfnamefont {V.}~\bibnamefont {Cros}}, \ and\ \bibinfo {author}
  {\bibfnamefont {A.}~\bibnamefont {Fert}},\ }\bibfield  {title} {\enquote
  {\bibinfo {title} {Dynamics of {Dzyaloshinskii} domain walls in ultrathin
  magnetic films},}\ }\href@noop {} {\bibfield  {journal} {\bibinfo  {journal}
  {EPL (Europhysics Letters)}\ }\textbf {\bibinfo {volume} {100}},\ \bibinfo
  {pages} {57002} (\bibinfo {year} {2012})}\BibitemShut {NoStop}%
\bibitem [{\citenamefont {Khvalkovskiy}\ \emph {et~al.}(2013)\citenamefont
  {Khvalkovskiy}, \citenamefont {Cros}, \citenamefont {Apalkov}, \citenamefont
  {Nikitin}, \citenamefont {Krounbi}, \citenamefont {Zvezdin}, \citenamefont
  {Anane}, \citenamefont {Grollier},\ and\ \citenamefont
  {Fert}}]{khvalkovskiy2013matching}%
  \BibitemOpen
  \bibfield  {author} {\bibinfo {author} {\bibfnamefont {A.~V.}\ \bibnamefont
  {Khvalkovskiy}}, \bibinfo {author} {\bibfnamefont {V.}~\bibnamefont {Cros}},
  \bibinfo {author} {\bibfnamefont {D.}~\bibnamefont {Apalkov}}, \bibinfo
  {author} {\bibfnamefont {V.}~\bibnamefont {Nikitin}}, \bibinfo {author}
  {\bibfnamefont {M.}~\bibnamefont {Krounbi}}, \bibinfo {author} {\bibfnamefont
  {K.~A.}\ \bibnamefont {Zvezdin}}, \bibinfo {author} {\bibfnamefont
  {A.}~\bibnamefont {Anane}}, \bibinfo {author} {\bibfnamefont
  {J.}~\bibnamefont {Grollier}}, \ and\ \bibinfo {author} {\bibfnamefont
  {A.}~\bibnamefont {Fert}},\ }\bibfield  {title} {\enquote {\bibinfo {title}
  {Matching domain-wall configuration and spin-orbit torques for efficient
  domain-wall motion},}\ }\href@noop {} {\bibfield  {journal} {\bibinfo
  {journal} {Phys. Rev. B}\ }\textbf {\bibinfo {volume} {87}},\ \bibinfo
  {pages} {020402} (\bibinfo {year} {2013})}\BibitemShut {NoStop}%
\bibitem [{\citenamefont {Ryu}\ \emph {et~al.}(2013)\citenamefont {Ryu},
  \citenamefont {Thomas}, \citenamefont {Yang},\ and\ \citenamefont
  {Parkin}}]{ryu2013chiral}%
  \BibitemOpen
  \bibfield  {author} {\bibinfo {author} {\bibfnamefont {Kwang-Su}\
  \bibnamefont {Ryu}}, \bibinfo {author} {\bibfnamefont {Luc}\ \bibnamefont
  {Thomas}}, \bibinfo {author} {\bibfnamefont {See-Hun}\ \bibnamefont {Yang}},
  \ and\ \bibinfo {author} {\bibfnamefont {Stuart S.~P.}\ \bibnamefont
  {Parkin}},\ }\bibfield  {title} {\enquote {\bibinfo {title} {Chiral spin
  torque at magnetic domain walls},}\ }\href@noop {} {\bibfield  {journal}
  {\bibinfo  {journal} {Nature Nano.}\ }\textbf {\bibinfo {volume} {8}},\
  \bibinfo {pages} {527--533} (\bibinfo {year} {2013})}\BibitemShut {NoStop}%
\bibitem [{\citenamefont {Emori}\ \emph {et~al.}(2013)\citenamefont {Emori},
  \citenamefont {Bauer}, \citenamefont {Ahn}, \citenamefont {Martinez},\ and\
  \citenamefont {Beach}}]{Emori13NatMat}%
  \BibitemOpen
  \bibfield  {author} {\bibinfo {author} {\bibfnamefont {S.}~\bibnamefont
  {Emori}}, \bibinfo {author} {\bibfnamefont {U.}~\bibnamefont {Bauer}},
  \bibinfo {author} {\bibfnamefont {S.-M.}\ \bibnamefont {Ahn}}, \bibinfo
  {author} {\bibfnamefont {E.}~\bibnamefont {Martinez}}, \ and\ \bibinfo
  {author} {\bibfnamefont {G.~S.~D.}\ \bibnamefont {Beach}},\ }\bibfield
  {title} {\enquote {\bibinfo {title} {Current-driven dynamics of chiral
  ferromagnetic domain walls},}\ }\href {\doibase 10.1038/nmat3675} {\bibfield
  {journal} {\bibinfo  {journal} {Nature Mater.}\ }\textbf {\bibinfo {volume}
  {12}},\ \bibinfo {pages} {611} (\bibinfo {year} {2013})}\BibitemShut
  {NoStop}%
\bibitem [{\citenamefont {Martinez}\ \emph {et~al.}(2013)\citenamefont
  {Martinez}, \citenamefont {Emori},\ and\ \citenamefont
  {Beach}}]{Martinez13APL}%
  \BibitemOpen
  \bibfield  {author} {\bibinfo {author} {\bibfnamefont {E.}~\bibnamefont
  {Martinez}}, \bibinfo {author} {\bibfnamefont {S.}~\bibnamefont {Emori}}, \
  and\ \bibinfo {author} {\bibfnamefont {G.~S.~D.}\ \bibnamefont {Beach}},\
  }\bibfield  {title} {\enquote {\bibinfo {title} {Current-driven domain wall
  motion along high perpendicular anisotropy multilayers: {T}he role of the
  {R}ashba field, the spin {H}all effect, and the {D}zyaloshinskii-{M}oriya
  interaction},}\ }\href {\doibase 10.1063/1.4818723} {\bibfield  {journal}
  {\bibinfo  {journal} {Appl. Phys. Lett.}\ }\textbf {\bibinfo {volume}
  {103}},\ \bibinfo {pages} {072406} (\bibinfo {year} {2013})}\BibitemShut
  {NoStop}%
\bibitem [{\citenamefont {Torrejon}\ \emph {et~al.}(2014)\citenamefont
  {Torrejon}, \citenamefont {Kim}, \citenamefont {Sinha}, \citenamefont
  {Mitani}, \citenamefont {Hayashi}, \citenamefont {Yamanouchi},\ and\
  \citenamefont {Ohno}}]{Torrejon14NatCom}%
  \BibitemOpen
  \bibfield  {author} {\bibinfo {author} {\bibfnamefont {J.}~\bibnamefont
  {Torrejon}}, \bibinfo {author} {\bibfnamefont {J.}~\bibnamefont {Kim}},
  \bibinfo {author} {\bibfnamefont {J.}~\bibnamefont {Sinha}}, \bibinfo
  {author} {\bibfnamefont {S.}~\bibnamefont {Mitani}}, \bibinfo {author}
  {\bibfnamefont {M.}~\bibnamefont {Hayashi}}, \bibinfo {author} {\bibfnamefont
  {M.}~\bibnamefont {Yamanouchi}}, \ and\ \bibinfo {author} {\bibfnamefont
  {H.}~\bibnamefont {Ohno}},\ }\bibfield  {title} {\enquote {\bibinfo {title}
  {Interface control of the magnetic chirality in {CoFeB/MgO} heterostructures
  with heavy-metal underlayers},}\ }\href {\doibase 10.1038/ncomms5655}
  {\bibfield  {journal} {\bibinfo  {journal} {Nature Commun.}\ }\textbf
  {\bibinfo {volume} {5}},\ \bibinfo {pages} {4655} (\bibinfo {year}
  {2014})}\BibitemShut {NoStop}%
\bibitem [{\citenamefont {Garello}\ \emph {et~al.}(2013)\citenamefont
  {Garello}, \citenamefont {Miron}, \citenamefont {Avci}, \citenamefont
  {Freimuth}, \citenamefont {Mokrousov}, \citenamefont {Bl\"{u}gel},
  \citenamefont {Auffret}, \citenamefont {Boulle}, \citenamefont {Gaudin},\
  and\ \citenamefont {Gambardella}}]{Garello2013}%
  \BibitemOpen
  \bibfield  {author} {\bibinfo {author} {\bibfnamefont {K.}~\bibnamefont
  {Garello}}, \bibinfo {author} {\bibfnamefont {I.~M.}\ \bibnamefont {Miron}},
  \bibinfo {author} {\bibfnamefont {C.~O.}\ \bibnamefont {Avci}}, \bibinfo
  {author} {\bibfnamefont {F.}~\bibnamefont {Freimuth}}, \bibinfo {author}
  {\bibfnamefont {Y.}~\bibnamefont {Mokrousov}}, \bibinfo {author}
  {\bibfnamefont {S.}~\bibnamefont {Bl\"{u}gel}}, \bibinfo {author}
  {\bibfnamefont {S.}~\bibnamefont {Auffret}}, \bibinfo {author} {\bibfnamefont
  {O.}~\bibnamefont {Boulle}}, \bibinfo {author} {\bibfnamefont
  {G.}~\bibnamefont {Gaudin}}, \ and\ \bibinfo {author} {\bibfnamefont
  {P.}~\bibnamefont {Gambardella}},\ }\bibfield  {title} {\enquote {\bibinfo
  {title} {Symmetry and magnitude of spin{\textendash}orbit torques in
  ferromagnetic heterostructures},}\ }\href {\doibase 10.1038/nnano.2013.145}
  {\bibfield  {journal} {\bibinfo  {journal} {Nature Nanotech}\ }\textbf
  {\bibinfo {volume} {8}},\ \bibinfo {pages} {587} (\bibinfo {year}
  {2013})}\BibitemShut {NoStop}%
\bibitem [{\citenamefont {Meckler}\ \emph {et~al.}(2009)\citenamefont
  {Meckler}, \citenamefont {Mikuszeit}, \citenamefont {Pre{\ss}ler},
  \citenamefont {Vedmedenko}, \citenamefont {Pietzsch},\ and\ \citenamefont
  {Wiesendanger}}]{Meckler09PRL}%
  \BibitemOpen
  \bibfield  {author} {\bibinfo {author} {\bibfnamefont {S.}~\bibnamefont
  {Meckler}}, \bibinfo {author} {\bibfnamefont {N.}~\bibnamefont {Mikuszeit}},
  \bibinfo {author} {\bibfnamefont {A.}~\bibnamefont {Pre{\ss}ler}}, \bibinfo
  {author} {\bibfnamefont {E.~Y.}\ \bibnamefont {Vedmedenko}}, \bibinfo
  {author} {\bibfnamefont {O.}~\bibnamefont {Pietzsch}}, \ and\ \bibinfo
  {author} {\bibfnamefont {R.}~\bibnamefont {Wiesendanger}},\ }\bibfield
  {title} {\enquote {\bibinfo {title} {Real-space observation of a
  right-rotating inhomogeneous cycloidal spin spiral by spin-polarized scanning
  tunneling microscopy in a triple axes vector magnet},}\ }\href@noop {}
  {\bibfield  {journal} {\bibinfo  {journal} {Phys. Rev. Lett.}\ }\textbf
  {\bibinfo {volume} {103}} (\bibinfo {year} {2009})}\BibitemShut {NoStop}%
\bibitem [{\citenamefont {Chen}\ \emph {et~al.}(2013)\citenamefont {Chen},
  \citenamefont {Ma}, \citenamefont {N'Diaye}, \citenamefont {Kwon},
  \citenamefont {Won}, \citenamefont {Wu},\ and\ \citenamefont
  {Schmid}}]{Chen13NatCom}%
  \BibitemOpen
  \bibfield  {author} {\bibinfo {author} {\bibfnamefont {G.}~\bibnamefont
  {Chen}}, \bibinfo {author} {\bibfnamefont {T.}~\bibnamefont {Ma}}, \bibinfo
  {author} {\bibfnamefont {A.~T.}\ \bibnamefont {N'Diaye}}, \bibinfo {author}
  {\bibfnamefont {H.}~\bibnamefont {Kwon}}, \bibinfo {author} {\bibfnamefont
  {C.}~\bibnamefont {Won}}, \bibinfo {author} {\bibfnamefont {Y.}~\bibnamefont
  {Wu}}, \ and\ \bibinfo {author} {\bibfnamefont {A.~K.}\ \bibnamefont
  {Schmid}},\ }\bibfield  {title} {\enquote {\bibinfo {title} {Tailoring the
  chirality of magnetic domain walls by interface engineering},}\ }\href@noop
  {} {\bibfield  {journal} {\bibinfo  {journal} {Nature Commun.}\ }\textbf
  {\bibinfo {volume} {4}},\ \bibinfo {pages} {2671} (\bibinfo {year}
  {2013})}\BibitemShut {NoStop}%
\bibitem [{\citenamefont {Boulle}\ \emph {et~al.}(2016)\citenamefont {Boulle},
  \citenamefont {Vogel}, \citenamefont {Yang}, \citenamefont {Pizzini},
  \citenamefont {de~Souza~Chaves}, \citenamefont {Locatelli}, \citenamefont
  {Mente{\c{s}}}, \citenamefont {Sala}, \citenamefont {Buda-Prejbeanu},
  \citenamefont {Klein}, \citenamefont {Belmeguenai}, \citenamefont
  {Roussign{\'{e}}}, \citenamefont {Stashkevich}, \citenamefont {Ch{\'{e}}rif},
  \citenamefont {Aballe}, \citenamefont {Foerster}, \citenamefont {Chshiev},
  \citenamefont {Auffret}, \citenamefont {Miron},\ and\ \citenamefont
  {Gaudin}}]{Boulle16NatNano}%
  \BibitemOpen
  \bibfield  {author} {\bibinfo {author} {\bibfnamefont {O.}~\bibnamefont
  {Boulle}}, \bibinfo {author} {\bibfnamefont {J.}~\bibnamefont {Vogel}},
  \bibinfo {author} {\bibfnamefont {H.}~\bibnamefont {Yang}}, \bibinfo {author}
  {\bibfnamefont {S.}~\bibnamefont {Pizzini}}, \bibinfo {author} {\bibfnamefont
  {D.}~\bibnamefont {de~Souza~Chaves}}, \bibinfo {author} {\bibfnamefont
  {A.}~\bibnamefont {Locatelli}}, \bibinfo {author} {\bibfnamefont {T.~O.}\
  \bibnamefont {Mente{\c{s}}}}, \bibinfo {author} {\bibfnamefont
  {A.}~\bibnamefont {Sala}}, \bibinfo {author} {\bibfnamefont {L.~D.}\
  \bibnamefont {Buda-Prejbeanu}}, \bibinfo {author} {\bibfnamefont
  {O.}~\bibnamefont {Klein}}, \bibinfo {author} {\bibfnamefont
  {M.}~\bibnamefont {Belmeguenai}}, \bibinfo {author} {\bibfnamefont
  {Y.}~\bibnamefont {Roussign{\'{e}}}}, \bibinfo {author} {\bibfnamefont
  {A.}~\bibnamefont {Stashkevich}}, \bibinfo {author} {\bibfnamefont {S.~M.}\
  \bibnamefont {Ch{\'{e}}rif}}, \bibinfo {author} {\bibfnamefont
  {L.}~\bibnamefont {Aballe}}, \bibinfo {author} {\bibfnamefont
  {M.}~\bibnamefont {Foerster}}, \bibinfo {author} {\bibfnamefont
  {M.}~\bibnamefont {Chshiev}}, \bibinfo {author} {\bibfnamefont
  {S.}~\bibnamefont {Auffret}}, \bibinfo {author} {\bibfnamefont {I.~M.}\
  \bibnamefont {Miron}}, \ and\ \bibinfo {author} {\bibfnamefont
  {G.}~\bibnamefont {Gaudin}},\ }\bibfield  {title} {\enquote {\bibinfo {title}
  {Room-temperature chiral magnetic skyrmions in ultrathin magnetic
  nanostructures},}\ }\href {\doibase 10.1038/nnano.2015.315} {\bibfield
  {journal} {\bibinfo  {journal} {Nature Nanotech.}\ }\textbf {\bibinfo
  {volume} {11}},\ \bibinfo {pages} {449} (\bibinfo {year} {2016})}\BibitemShut
  {NoStop}%
\bibitem [{\citenamefont {Nembach}\ \emph {et~al.}(2015)\citenamefont
  {Nembach}, \citenamefont {Shaw}, \citenamefont {Weiler}, \citenamefont
  {Ju{\'e}},\ and\ \citenamefont {Silva}}]{nembach15NatPhys}%
  \BibitemOpen
  \bibfield  {author} {\bibinfo {author} {\bibfnamefont {H.~T.}\ \bibnamefont
  {Nembach}}, \bibinfo {author} {\bibfnamefont {J.~M.}\ \bibnamefont {Shaw}},
  \bibinfo {author} {\bibfnamefont {M.}~\bibnamefont {Weiler}}, \bibinfo
  {author} {\bibfnamefont {E.}~\bibnamefont {Ju{\'e}}}, \ and\ \bibinfo
  {author} {\bibfnamefont {T.~J.}\ \bibnamefont {Silva}},\ }\bibfield  {title}
  {\enquote {\bibinfo {title} {Linear relation between {Heisenberg} exchange
  and interfacial {Dzyaloshinskii-Moriya} interaction in metal films},}\ }\href
  {\doibase 10.1038/nphys3418} {\bibfield  {journal} {\bibinfo  {journal}
  {Nature Phys.}\ }\textbf {\bibinfo {volume} {11}},\ \bibinfo {pages} {825}
  (\bibinfo {year} {2015})}\BibitemShut {NoStop}%
\bibitem [{\citenamefont {Belmeguenai}\ \emph {et~al.}(2015)\citenamefont
  {Belmeguenai}, \citenamefont {Adam}, \citenamefont {Roussign\'e},
  \citenamefont {Eimer}, \citenamefont {Devolder}, \citenamefont {Kim},
  \citenamefont {Cherif}, \citenamefont {Stashkevich},\ and\ \citenamefont
  {Thiaville}}]{Belmeguenai15PRB}%
  \BibitemOpen
  \bibfield  {author} {\bibinfo {author} {\bibfnamefont {M.}~\bibnamefont
  {Belmeguenai}}, \bibinfo {author} {\bibfnamefont {J.-P.}\ \bibnamefont
  {Adam}}, \bibinfo {author} {\bibfnamefont {Y.}~\bibnamefont {Roussign\'e}},
  \bibinfo {author} {\bibfnamefont {S.}~\bibnamefont {Eimer}}, \bibinfo
  {author} {\bibfnamefont {T.}~\bibnamefont {Devolder}}, \bibinfo {author}
  {\bibfnamefont {J.-Von.}\ \bibnamefont {Kim}}, \bibinfo {author}
  {\bibfnamefont {S.~M.}\ \bibnamefont {Cherif}}, \bibinfo {author}
  {\bibfnamefont {A.}~\bibnamefont {Stashkevich}}, \ and\ \bibinfo {author}
  {\bibfnamefont {A.}~\bibnamefont {Thiaville}},\ }\bibfield  {title} {\enquote
  {\bibinfo {title} {Interfacial {Dzyaloshinskii-Moriya} interaction in
  perpendicularly magnetized {Pt/Co/AlO}$_{x}$ ultrathin films measured by
  {Brillouin} light spectroscopy},}\ }\href {\doibase
  10.1103/PhysRevB.91.180405} {\bibfield  {journal} {\bibinfo  {journal} {Phys.
  Rev. B}\ }\textbf {\bibinfo {volume} {91}},\ \bibinfo {pages} {180405}
  (\bibinfo {year} {2015})}\BibitemShut {NoStop}%
\bibitem [{\citenamefont {Je}\ \emph {et~al.}(2013)\citenamefont {Je},
  \citenamefont {Kim}, \citenamefont {Yoo}, \citenamefont {Min}, \citenamefont
  {Lee},\ and\ \citenamefont {Choe}}]{Je13PRB}%
  \BibitemOpen
  \bibfield  {author} {\bibinfo {author} {\bibfnamefont {S.-G.}\ \bibnamefont
  {Je}}, \bibinfo {author} {\bibfnamefont {D.-Ho.}\ \bibnamefont {Kim}},
  \bibinfo {author} {\bibfnamefont {S.-C.}\ \bibnamefont {Yoo}}, \bibinfo
  {author} {\bibfnamefont {B.-Chul.}\ \bibnamefont {Min}}, \bibinfo {author}
  {\bibfnamefont {K.-J.}\ \bibnamefont {Lee}}, \ and\ \bibinfo {author}
  {\bibfnamefont {S.-B.}\ \bibnamefont {Choe}},\ }\bibfield  {title} {\enquote
  {\bibinfo {title} {Asymmetric magnetic domain-wall motion by the
  {Dzyaloshinskii-Moriya} interaction},}\ }\href {\doibase
  https://doi.org/10.1103/PhysRevB.88.214401} {\bibfield  {journal} {\bibinfo
  {journal} {Phys. Rev. B}\ }\textbf {\bibinfo {volume} {88}},\ \bibinfo
  {pages} {214401} (\bibinfo {year} {2013})}\BibitemShut {NoStop}%
\bibitem [{\citenamefont {Hrabec}\ \emph {et~al.}(2014)\citenamefont {Hrabec},
  \citenamefont {Porter}, \citenamefont {Wells}, \citenamefont {Benitez},
  \citenamefont {Burnell}, \citenamefont {McVitie}, \citenamefont {McGrouther},
  \citenamefont {Moore},\ and\ \citenamefont {Marrows}}]{Hrabec14PRB}%
  \BibitemOpen
  \bibfield  {author} {\bibinfo {author} {\bibfnamefont {A.}~\bibnamefont
  {Hrabec}}, \bibinfo {author} {\bibfnamefont {N.~A.}\ \bibnamefont {Porter}},
  \bibinfo {author} {\bibfnamefont {A.~J.}\ \bibnamefont {Wells}}, \bibinfo
  {author} {\bibfnamefont {M.-J.}\ \bibnamefont {Benitez}}, \bibinfo {author}
  {\bibfnamefont {G.}~\bibnamefont {Burnell}}, \bibinfo {author} {\bibfnamefont
  {S.}~\bibnamefont {McVitie}}, \bibinfo {author} {\bibfnamefont
  {D.}~\bibnamefont {McGrouther}}, \bibinfo {author} {\bibfnamefont {T.~A.}\
  \bibnamefont {Moore}}, \ and\ \bibinfo {author} {\bibfnamefont {C.~H.}\
  \bibnamefont {Marrows}},\ }\bibfield  {title} {\enquote {\bibinfo {title}
  {Measuring and tailoring the {Dzyaloshinskii-Moriya} interaction in
  perpendicularly magnetized thin films},}\ }\href {\doibase
  https://doi.org/10.1103/PhysRevB.90.020402} {\bibfield  {journal} {\bibinfo
  {journal} {Phys. Rev. B}\ }\textbf {\bibinfo {volume} {90}},\ \bibinfo
  {pages} {020402} (\bibinfo {year} {2014})}\BibitemShut {NoStop}%
\bibitem [{\citenamefont {Gallagher}\ \emph {et~al.}(1979)\citenamefont
  {Gallagher}, \citenamefont {Ju},\ and\ \citenamefont
  {Humphrey}}]{Gallagher79JAP}%
  \BibitemOpen
  \bibfield  {author} {\bibinfo {author} {\bibfnamefont {T.~J.}\ \bibnamefont
  {Gallagher}}, \bibinfo {author} {\bibfnamefont {K.}~\bibnamefont {Ju}}, \
  and\ \bibinfo {author} {\bibfnamefont {F.~B.}\ \bibnamefont {Humphrey}},\
  }\bibfield  {title} {\enquote {\bibinfo {title} {State identification and
  stability of magnetic bubbles with unit winding number},}\ }\href {\doibase
  https://doi.org/10.1063/1.325991} {\bibfield  {journal} {\bibinfo  {journal}
  {J. Appl. Phys.}\ }\textbf {\bibinfo {volume} {50}},\ \bibinfo {pages} {997}
  (\bibinfo {year} {1979})}\BibitemShut {NoStop}%
\bibitem [{\citenamefont {Leeuw}\ \emph {et~al.}(1980)\citenamefont {Leeuw},
  \citenamefont {Doel},\ and\ \citenamefont {Enz}}]{deleeuw80RPP}%
  \BibitemOpen
  \bibfield  {author} {\bibinfo {author} {\bibfnamefont {F.~H.~De}\
  \bibnamefont {Leeuw}}, \bibinfo {author} {\bibfnamefont {R.~Van~Den}\
  \bibnamefont {Doel}}, \ and\ \bibinfo {author} {\bibfnamefont
  {U.}~\bibnamefont {Enz}},\ }\bibfield  {title} {\enquote {\bibinfo {title}
  {Dynamic properties of magnetic domain walls and magnetic bubbles},}\ }\href
  {\doibase https://doi.org/10.1088/0034-4885/43/6/001} {\bibfield  {journal}
  {\bibinfo  {journal} {Rep. Prog. Phys.}\ }\textbf {\bibinfo {volume} {43}},\
  \bibinfo {pages} {689} (\bibinfo {year} {1980})}\BibitemShut {NoStop}%
\bibitem [{\citenamefont {Kabanov}\ \emph {et~al.}(2010)\citenamefont
  {Kabanov}, \citenamefont {Iunin}, \citenamefont {Nikitenko}, \citenamefont
  {Shapiro}, \citenamefont {Shull}, \citenamefont {Zhu},\ and\ \citenamefont
  {Chien}}]{Kabanov10IEEE}%
  \BibitemOpen
  \bibfield  {author} {\bibinfo {author} {\bibfnamefont {Y.~P.}\ \bibnamefont
  {Kabanov}}, \bibinfo {author} {\bibfnamefont {Y.~L.}\ \bibnamefont {Iunin}},
  \bibinfo {author} {\bibfnamefont {V.~I.}\ \bibnamefont {Nikitenko}}, \bibinfo
  {author} {\bibfnamefont {A.~J.}\ \bibnamefont {Shapiro}}, \bibinfo {author}
  {\bibfnamefont {R.~D.}\ \bibnamefont {Shull}}, \bibinfo {author}
  {\bibfnamefont {L.~Y.}\ \bibnamefont {Zhu}}, \ and\ \bibinfo {author}
  {\bibfnamefont {C.~L.}\ \bibnamefont {Chien}},\ }\bibfield  {title} {\enquote
  {\bibinfo {title} {In-plane field effects on the dynamics of domain walls in
  ultrathin {Co} films with perpendicular anisotropy},}\ }\href {\doibase
  10.1109/TMAG.2010.2045740} {\bibfield  {journal} {\bibinfo  {journal} {IEEE
  Trans. Magn.}\ }\textbf {\bibinfo {volume} {46}},\ \bibinfo {pages} {2220}
  (\bibinfo {year} {2010})}\BibitemShut {NoStop}%
\bibitem [{\citenamefont {Petit}\ \emph {et~al.}(2015)\citenamefont {Petit},
  \citenamefont {Seem}, \citenamefont {Tillette}, \citenamefont {Mansell},\
  and\ \citenamefont {Cowburn}}]{Petit15APL}%
  \BibitemOpen
  \bibfield  {author} {\bibinfo {author} {\bibfnamefont {D.}~\bibnamefont
  {Petit}}, \bibinfo {author} {\bibfnamefont {P.~R.}\ \bibnamefont {Seem}},
  \bibinfo {author} {\bibfnamefont {M.}~\bibnamefont {Tillette}}, \bibinfo
  {author} {\bibfnamefont {R.}~\bibnamefont {Mansell}}, \ and\ \bibinfo
  {author} {\bibfnamefont {R.~P.}\ \bibnamefont {Cowburn}},\ }\bibfield
  {title} {\enquote {\bibinfo {title} {Two-dimensional control of field-driven
  magnetic bubble movement using {Dzyaloshinskii-Moriya} interactions},}\
  }\href@noop {} {\bibfield  {journal} {\bibinfo  {journal} {Appl. Phys.
  Lett.}\ }\textbf {\bibinfo {volume} {106}},\ \bibinfo {pages} {022402}
  (\bibinfo {year} {2015})}\BibitemShut {NoStop}%
\bibitem [{\citenamefont {Khan}\ \emph {et~al.}(2016)\citenamefont {Khan},
  \citenamefont {Shepley}, \citenamefont {Hrabec}, \citenamefont {Wells},
  \citenamefont {Ocker}, \citenamefont {Marrows},\ and\ \citenamefont
  {Moore}}]{Khan16APL}%
  \BibitemOpen
  \bibfield  {author} {\bibinfo {author} {\bibfnamefont {R.~A.}\ \bibnamefont
  {Khan}}, \bibinfo {author} {\bibfnamefont {P.~M.}\ \bibnamefont {Shepley}},
  \bibinfo {author} {\bibfnamefont {A.}~\bibnamefont {Hrabec}}, \bibinfo
  {author} {\bibfnamefont {A.~W.~J.}\ \bibnamefont {Wells}}, \bibinfo {author}
  {\bibfnamefont {B.}~\bibnamefont {Ocker}}, \bibinfo {author} {\bibfnamefont
  {C.~H.}\ \bibnamefont {Marrows}}, \ and\ \bibinfo {author} {\bibfnamefont
  {T.~A.}\ \bibnamefont {Moore}},\ }\bibfield  {title} {\enquote {\bibinfo
  {title} {Effect of annealing on the interfacial {Dzyaloshinskii-Moriya}
  interaction in {Ta/CoFeB/MgO} trilayers},}\ }\href@noop {} {\bibfield
  {journal} {\bibinfo  {journal} {Appl. Phys. Lett.}\ }\textbf {\bibinfo
  {volume} {109}},\ \bibinfo {pages} {132404} (\bibinfo {year}
  {2016})}\BibitemShut {NoStop}%
\bibitem [{\citenamefont {Pham}\ \emph {et~al.}(2016)\citenamefont {Pham},
  \citenamefont {Vogel}, \citenamefont {Sampaio}, \citenamefont {Va\v{n}atka},
  \citenamefont {Rojas-S\'{a}nchez}, \citenamefont {Bonfim}, \citenamefont
  {Chaves}, \citenamefont {Choueikani}, \citenamefont {Ohresser}, \citenamefont
  {Otero}, \citenamefont {Thiaville},\ and\ \citenamefont
  {Pizzini}}]{Pham16EPL}%
  \BibitemOpen
  \bibfield  {author} {\bibinfo {author} {\bibfnamefont {T.~H.}\ \bibnamefont
  {Pham}}, \bibinfo {author} {\bibfnamefont {J.}~\bibnamefont {Vogel}},
  \bibinfo {author} {\bibfnamefont {J.}~\bibnamefont {Sampaio}}, \bibinfo
  {author} {\bibfnamefont {M.}~\bibnamefont {Va\v{n}atka}}, \bibinfo {author}
  {\bibfnamefont {J.-C.}\ \bibnamefont {Rojas-S\'{a}nchez}}, \bibinfo {author}
  {\bibfnamefont {M.}~\bibnamefont {Bonfim}}, \bibinfo {author} {\bibfnamefont
  {D.~S.}\ \bibnamefont {Chaves}}, \bibinfo {author} {\bibfnamefont
  {F.}~\bibnamefont {Choueikani}}, \bibinfo {author} {\bibfnamefont
  {P.}~\bibnamefont {Ohresser}}, \bibinfo {author} {\bibfnamefont
  {E.}~\bibnamefont {Otero}}, \bibinfo {author} {\bibfnamefont
  {A.}~\bibnamefont {Thiaville}}, \ and\ \bibinfo {author} {\bibfnamefont
  {S.}~\bibnamefont {Pizzini}},\ }\bibfield  {title} {\enquote {\bibinfo
  {title} {Very large domain wall velocities in {Pt/Co/GdO}$x$ and {Pt/Co/Gd}
  trilayers with {Dzyaloshinskii-Moriya} interaction},}\ }\href@noop {}
  {\bibfield  {journal} {\bibinfo  {journal} {EPL (Europhys. Lett.)}\ }\textbf
  {\bibinfo {volume} {113}},\ \bibinfo {pages} {67001} (\bibinfo {year}
  {2016})}\BibitemShut {NoStop}%
\bibitem [{\citenamefont {Kim}\ \emph {et~al.}(2015)\citenamefont {Kim},
  \citenamefont {Kim}, \citenamefont {Moon},\ and\ \citenamefont
  {Choe}}]{kim15APL}%
  \BibitemOpen
  \bibfield  {author} {\bibinfo {author} {\bibfnamefont {D.-Y.}\ \bibnamefont
  {Kim}}, \bibinfo {author} {\bibfnamefont {D.-H.}\ \bibnamefont {Kim}},
  \bibinfo {author} {\bibfnamefont {J.}~\bibnamefont {Moon}}, \ and\ \bibinfo
  {author} {\bibfnamefont {S.-B.}\ \bibnamefont {Choe}},\ }\bibfield  {title}
  {\enquote {\bibinfo {title} {Determination of magnetic domain-wall types
  using {Dzyaloshinskii-Moriya}-interaction-induced domain patterns},}\
  }\href@noop {} {\bibfield  {journal} {\bibinfo  {journal} {App. Phys. Lett.}\
  }\textbf {\bibinfo {volume} {106}},\ \bibinfo {pages} {262403} (\bibinfo
  {year} {2015})}\BibitemShut {NoStop}%
\bibitem [{\citenamefont {Soucaille}\ \emph {et~al.}(2016)\citenamefont
  {Soucaille}, \citenamefont {Belmeguenai}, \citenamefont {Torrejon},
  \citenamefont {Kim}, \citenamefont {Devolder}, \citenamefont {Roussign\'e},
  \citenamefont {Ch\'erif}, \citenamefont {Stashkevich}, \citenamefont
  {Hayashi},\ and\ \citenamefont {Adam}}]{Soucaille16PRB}%
  \BibitemOpen
  \bibfield  {author} {\bibinfo {author} {\bibfnamefont {R.}~\bibnamefont
  {Soucaille}}, \bibinfo {author} {\bibfnamefont {M.}~\bibnamefont
  {Belmeguenai}}, \bibinfo {author} {\bibfnamefont {J.}~\bibnamefont
  {Torrejon}}, \bibinfo {author} {\bibfnamefont {J.-V.}\ \bibnamefont {Kim}},
  \bibinfo {author} {\bibfnamefont {T.}~\bibnamefont {Devolder}}, \bibinfo
  {author} {\bibfnamefont {Y.}~\bibnamefont {Roussign\'e}}, \bibinfo {author}
  {\bibfnamefont {S.-M.}\ \bibnamefont {Ch\'erif}}, \bibinfo {author}
  {\bibfnamefont {A.~A.}\ \bibnamefont {Stashkevich}}, \bibinfo {author}
  {\bibfnamefont {M.}~\bibnamefont {Hayashi}}, \ and\ \bibinfo {author}
  {\bibfnamefont {J.-P.}\ \bibnamefont {Adam}},\ }\bibfield  {title} {\enquote
  {\bibinfo {title} {Probing the {Dzyaloshinskii-Moriya} interaction in {CoFeB}
  ultrathin films using domain wall creep and {Brillouin} light
  spectroscopy},}\ }\href {\doibase 10.1103/PhysRevB.94.104431} {\bibfield
  {journal} {\bibinfo  {journal} {Phys. Rev. B}\ }\textbf {\bibinfo {volume}
  {94}},\ \bibinfo {pages} {104431} (\bibinfo {year} {2016})}\BibitemShut
  {NoStop}%
\bibitem [{\citenamefont {Lavrijsen}\ \emph {et~al.}(2015)\citenamefont
  {Lavrijsen}, \citenamefont {Hartmann}, \citenamefont {van~den Brink},
  \citenamefont {Yin}, \citenamefont {Barcones}, \citenamefont {Duine},
  \citenamefont {Verheijen}, \citenamefont {Swagten},\ and\ \citenamefont
  {Koopmans}}]{Lavrijsen15PRB}%
  \BibitemOpen
  \bibfield  {author} {\bibinfo {author} {\bibfnamefont {R.}~\bibnamefont
  {Lavrijsen}}, \bibinfo {author} {\bibfnamefont {D.~M.~F.}\ \bibnamefont
  {Hartmann}}, \bibinfo {author} {\bibfnamefont {A.}~\bibnamefont {van~den
  Brink}}, \bibinfo {author} {\bibfnamefont {Y.}~\bibnamefont {Yin}}, \bibinfo
  {author} {\bibfnamefont {B.}~\bibnamefont {Barcones}}, \bibinfo {author}
  {\bibfnamefont {R.~A.}\ \bibnamefont {Duine}}, \bibinfo {author}
  {\bibfnamefont {M.~A.}\ \bibnamefont {Verheijen}}, \bibinfo {author}
  {\bibfnamefont {H.~J.~M.}\ \bibnamefont {Swagten}}, \ and\ \bibinfo {author}
  {\bibfnamefont {B.}~\bibnamefont {Koopmans}},\ }\bibfield  {title} {\enquote
  {\bibinfo {title} {Asymmetric magnetic bubble expansion under in-plane field
  in {Pt/Co/Pt}: Effect of interface engineering},}\ }\href {\doibase
  10.1103/PhysRevB.91.104414} {\bibfield  {journal} {\bibinfo  {journal} {Phys.
  Rev. B}\ }\textbf {\bibinfo {volume} {91}},\ \bibinfo {pages} {104414}
  (\bibinfo {year} {2015})}\BibitemShut {NoStop}%
\bibitem [{\citenamefont {Va{\v{n}}atka}\ \emph {et~al.}(2015)\citenamefont
  {Va{\v{n}}atka}, \citenamefont {Rojas-S{\'a}nchez}, \citenamefont {Vogel},
  \citenamefont {Bonfim}, \citenamefont {Belmeguenai}, \citenamefont
  {Roussign{\'e}}, \citenamefont {Stashkevich}, \citenamefont {Thiaville},\
  and\ \citenamefont {Pizzini}}]{Vanatka15JP}%
  \BibitemOpen
  \bibfield  {author} {\bibinfo {author} {\bibfnamefont {M.}~\bibnamefont
  {Va{\v{n}}atka}}, \bibinfo {author} {\bibfnamefont {J.-C.}\ \bibnamefont
  {Rojas-S{\'a}nchez}}, \bibinfo {author} {\bibfnamefont {J.}~\bibnamefont
  {Vogel}}, \bibinfo {author} {\bibfnamefont {M.}~\bibnamefont {Bonfim}},
  \bibinfo {author} {\bibfnamefont {M.}~\bibnamefont {Belmeguenai}}, \bibinfo
  {author} {\bibfnamefont {Y.}~\bibnamefont {Roussign{\'e}}}, \bibinfo {author}
  {\bibfnamefont {A.}~\bibnamefont {Stashkevich}}, \bibinfo {author}
  {\bibfnamefont {A.}~\bibnamefont {Thiaville}}, \ and\ \bibinfo {author}
  {\bibfnamefont {S.}~\bibnamefont {Pizzini}},\ }\bibfield  {title} {\enquote
  {\bibinfo {title} {Velocity asymmetry of {Dzyaloshinskii} domain walls in the
  creep and flow regimes},}\ }\href {\doibase
  https://doi.org/10.1088/0953-8984/27/32/326002} {\bibfield  {journal}
  {\bibinfo  {journal} {J. Phys.: Condens. Matt.}\ }\textbf {\bibinfo {volume}
  {27}},\ \bibinfo {pages} {326002} (\bibinfo {year} {2015})}\BibitemShut
  {NoStop}%
\bibitem [{\citenamefont {Lau}\ \emph {et~al.}(2016)\citenamefont {Lau},
  \citenamefont {Sundar}, \citenamefont {Zhu},\ and\ \citenamefont
  {Sokalski}}]{lau16PRB}%
  \BibitemOpen
  \bibfield  {author} {\bibinfo {author} {\bibfnamefont {Derek}\ \bibnamefont
  {Lau}}, \bibinfo {author} {\bibfnamefont {Vignesh}\ \bibnamefont {Sundar}},
  \bibinfo {author} {\bibfnamefont {Jian-Gang}\ \bibnamefont {Zhu}}, \ and\
  \bibinfo {author} {\bibfnamefont {Vincent}\ \bibnamefont {Sokalski}},\
  }\bibfield  {title} {\enquote {\bibinfo {title} {Energetic molding of chiral
  magnetic bubbles},}\ }\href {\doibase
  https://doi.org/10.1103/PhysRevB.94.060401} {\bibfield  {journal} {\bibinfo
  {journal} {Phys. Rev. {B}}\ }\textbf {\bibinfo {volume} {94}},\ \bibinfo
  {pages} {060401} (\bibinfo {year} {2016})}\BibitemShut {NoStop}%
\bibitem [{\citenamefont {Ajejas}\ \emph {et~al.}(2017)\citenamefont {Ajejas},
  \citenamefont {K\v{r}i\v{z}\'{a}kov\'{a}}, \citenamefont {de~Souza~Chaves},
  \citenamefont {Vogel}, \citenamefont {Perna}, \citenamefont {Guerrero},
  \citenamefont {Gudin}, \citenamefont {Camarero},\ and\ \citenamefont
  {Pizzini}}]{Ajejas17APL}%
  \BibitemOpen
  \bibfield  {author} {\bibinfo {author} {\bibfnamefont {F.}~\bibnamefont
  {Ajejas}}, \bibinfo {author} {\bibfnamefont {V.}~\bibnamefont
  {K\v{r}i\v{z}\'{a}kov\'{a}}}, \bibinfo {author} {\bibfnamefont
  {D.}~\bibnamefont {de~Souza~Chaves}}, \bibinfo {author} {\bibfnamefont
  {J.}~\bibnamefont {Vogel}}, \bibinfo {author} {\bibfnamefont
  {P.}~\bibnamefont {Perna}}, \bibinfo {author} {\bibfnamefont
  {R.}~\bibnamefont {Guerrero}}, \bibinfo {author} {\bibfnamefont
  {A.}~\bibnamefont {Gudin}}, \bibinfo {author} {\bibfnamefont
  {J.}~\bibnamefont {Camarero}}, \ and\ \bibinfo {author} {\bibfnamefont
  {S.}~\bibnamefont {Pizzini}},\ }\bibfield  {title} {\enquote {\bibinfo
  {title} {Tuning domain wall velocity with {Dzyaloshinskii-Moriya}
  interaction},}\ }\href@noop {} {\bibfield  {journal} {\bibinfo  {journal}
  {Appl. Phys. Lett.}\ }\textbf {\bibinfo {volume} {111}},\ \bibinfo {pages}
  {202402} (\bibinfo {year} {2017})}\BibitemShut {NoStop}%
\bibitem [{\citenamefont {Shepley}\ \emph {et~al.}(2018)\citenamefont
  {Shepley}, \citenamefont {Tunnicliffe}, \citenamefont {Shahbazi},
  \citenamefont {Burnell},\ and\ \citenamefont {Moore}}]{Shepley18PRB}%
  \BibitemOpen
  \bibfield  {author} {\bibinfo {author} {\bibfnamefont {P.~M.}\ \bibnamefont
  {Shepley}}, \bibinfo {author} {\bibfnamefont {H.}~\bibnamefont
  {Tunnicliffe}}, \bibinfo {author} {\bibfnamefont {K.}~\bibnamefont
  {Shahbazi}}, \bibinfo {author} {\bibfnamefont {G.}~\bibnamefont {Burnell}}, \
  and\ \bibinfo {author} {\bibfnamefont {T.~A.}\ \bibnamefont {Moore}},\
  }\bibfield  {title} {\enquote {\bibinfo {title} {Magnetic properties,
  domain-wall creep motion, and the {Dzyaloshinskii-Moriya} interaction in
  {Pt/Co/Ir} thin films},}\ }\href {\doibase 10.1103/PhysRevB.97.134417}
  {\bibfield  {journal} {\bibinfo  {journal} {Phys. Rev. B}\ }\textbf {\bibinfo
  {volume} {97}},\ \bibinfo {pages} {134417} (\bibinfo {year}
  {2018})}\BibitemShut {NoStop}%
\bibitem [{\citenamefont {Pellegren}\ \emph {et~al.}(2017)\citenamefont
  {Pellegren}, \citenamefont {Lau},\ and\ \citenamefont
  {Sokalski}}]{pellegren17PRL}%
  \BibitemOpen
  \bibfield  {author} {\bibinfo {author} {\bibfnamefont {J.~P.}\ \bibnamefont
  {Pellegren}}, \bibinfo {author} {\bibfnamefont {D.}~\bibnamefont {Lau}}, \
  and\ \bibinfo {author} {\bibfnamefont {V.}~\bibnamefont {Sokalski}},\
  }\bibfield  {title} {\enquote {\bibinfo {title} {Dispersive stiffness of
  {Dzyaloshinskii} domain walls},}\ }\href {\doibase
  10.1103/PhysRevLett.119.027203} {\bibfield  {journal} {\bibinfo  {journal}
  {Phys. Rev. Lett.}\ }\textbf {\bibinfo {volume} {119}},\ \bibinfo {pages}
  {027203} (\bibinfo {year} {2017})}\BibitemShut {NoStop}%
\bibitem [{\citenamefont {Woo}\ \emph {et~al.}(2014)\citenamefont {Woo},
  \citenamefont {Mann}, \citenamefont {Tan}, \citenamefont {Caretta},\ and\
  \citenamefont {Beach}}]{woo14eAPL}%
  \BibitemOpen
  \bibfield  {author} {\bibinfo {author} {\bibfnamefont {S.}~\bibnamefont
  {Woo}}, \bibinfo {author} {\bibfnamefont {M.}~\bibnamefont {Mann}}, \bibinfo
  {author} {\bibfnamefont {A.~J.}\ \bibnamefont {Tan}}, \bibinfo {author}
  {\bibfnamefont {L.}~\bibnamefont {Caretta}}, \ and\ \bibinfo {author}
  {\bibfnamefont {G.~S.~D.}\ \bibnamefont {Beach}},\ }\bibfield  {title}
  {\enquote {\bibinfo {title} {Enhanced spin-orbit torques in {Pt/Co/Ta}
  heterostructures},}\ }\href {\doibase https://doi.org/10.1063/1.4902529}
  {\bibfield  {journal} {\bibinfo  {journal} {Appl. Phys. Lett.}\ }\textbf
  {\bibinfo {volume} {105}},\ \bibinfo {pages} {212404} (\bibinfo {year}
  {2014})}\BibitemShut {NoStop}%
\bibitem [{\citenamefont {Woo}\ \emph {et~al.}(2016)\citenamefont {Woo},
  \citenamefont {Litzius}, \citenamefont {Kr{\"u}ger}, \citenamefont {Im},
  \citenamefont {Caretta}, \citenamefont {Richter}, \citenamefont {Mann},
  \citenamefont {Krone}, \citenamefont {Reeve}, \citenamefont {Weigand},
  \citenamefont {Agrawal}, \citenamefont {Lemesh}, \citenamefont {Mawass},
  \citenamefont {Fischer}, \citenamefont {Kl\"{a}ui},\ and\ \citenamefont
  {Beach}}]{woo16NatMat}%
  \BibitemOpen
  \bibfield  {author} {\bibinfo {author} {\bibfnamefont {S.}~\bibnamefont
  {Woo}}, \bibinfo {author} {\bibfnamefont {K.}~\bibnamefont {Litzius}},
  \bibinfo {author} {\bibfnamefont {B.}~\bibnamefont {Kr{\"u}ger}}, \bibinfo
  {author} {\bibfnamefont {M.-Y.}\ \bibnamefont {Im}}, \bibinfo {author}
  {\bibfnamefont {L.}~\bibnamefont {Caretta}}, \bibinfo {author} {\bibfnamefont
  {K.}~\bibnamefont {Richter}}, \bibinfo {author} {\bibfnamefont
  {M.}~\bibnamefont {Mann}}, \bibinfo {author} {\bibfnamefont {A.}~\bibnamefont
  {Krone}}, \bibinfo {author} {\bibfnamefont {R.~M.}\ \bibnamefont {Reeve}},
  \bibinfo {author} {\bibfnamefont {M.}~\bibnamefont {Weigand}}, \bibinfo
  {author} {\bibfnamefont {P.}~\bibnamefont {Agrawal}}, \bibinfo {author}
  {\bibfnamefont {I.}~\bibnamefont {Lemesh}}, \bibinfo {author} {\bibfnamefont
  {M.-A.}\ \bibnamefont {Mawass}}, \bibinfo {author} {\bibfnamefont
  {P.}~\bibnamefont {Fischer}}, \bibinfo {author} {\bibfnamefont
  {M.}~\bibnamefont {Kl\"{a}ui}}, \ and\ \bibinfo {author} {\bibfnamefont
  {G.~S.~D.}\ \bibnamefont {Beach}},\ }\bibfield  {title} {\enquote {\bibinfo
  {title} {Observation of room-temperature magnetic skyrmions and their
  current-driven dynamics in ultrathin metallic ferromagnets},}\ }\href
  {\doibase 10.1038/nmat4593} {\bibfield  {journal} {\bibinfo  {journal}
  {Nature Mater.}\ }\textbf {\bibinfo {volume} {15}},\ \bibinfo {pages} {501}
  (\bibinfo {year} {2016})}\BibitemShut {NoStop}%
\bibitem [{\citenamefont {Kashid}\ \emph {et~al.}(2014)\citenamefont {Kashid},
  \citenamefont {Schena}, \citenamefont {Zimmermann}, \citenamefont
  {Mokrousov}, \citenamefont {Bl{\"u}gel}, \citenamefont {Shah},\ and\
  \citenamefont {Salunke}}]{kashid14PRB}%
  \BibitemOpen
  \bibfield  {author} {\bibinfo {author} {\bibfnamefont {V.}~\bibnamefont
  {Kashid}}, \bibinfo {author} {\bibfnamefont {T.}~\bibnamefont {Schena}},
  \bibinfo {author} {\bibfnamefont {B.}~\bibnamefont {Zimmermann}}, \bibinfo
  {author} {\bibfnamefont {Y.}~\bibnamefont {Mokrousov}}, \bibinfo {author}
  {\bibfnamefont {S.}~\bibnamefont {Bl{\"u}gel}}, \bibinfo {author}
  {\bibfnamefont {V.}~\bibnamefont {Shah}}, \ and\ \bibinfo {author}
  {\bibfnamefont {H.~G.}\ \bibnamefont {Salunke}},\ }\bibfield  {title}
  {\enquote {\bibinfo {title} {{Dzyaloshinskii-Moriya} interaction and chiral
  magnetism in $3d-5d$ zigzag chains: Tight-binding model and ab initio
  calculations},}\ }\href@noop {} {\bibfield  {journal} {\bibinfo  {journal}
  {Phys. Rev. B}\ }\textbf {\bibinfo {volume} {90}},\ \bibinfo {pages} {054412}
  (\bibinfo {year} {2014})}\BibitemShut {NoStop}%
\bibitem [{\citenamefont {Yang}\ \emph {et~al.}(2015)\citenamefont {Yang},
  \citenamefont {Thiaville}, \citenamefont {Rohart}, \citenamefont {Fert},\
  and\ \citenamefont {Chshiev}}]{Yang15PRL}%
  \BibitemOpen
  \bibfield  {author} {\bibinfo {author} {\bibfnamefont {Hongxin}\ \bibnamefont
  {Yang}}, \bibinfo {author} {\bibfnamefont {Andr\'e}\ \bibnamefont
  {Thiaville}}, \bibinfo {author} {\bibfnamefont {Stanislas}\ \bibnamefont
  {Rohart}}, \bibinfo {author} {\bibfnamefont {Albert}\ \bibnamefont {Fert}}, \
  and\ \bibinfo {author} {\bibfnamefont {Mairbek}\ \bibnamefont {Chshiev}},\
  }\bibfield  {title} {\enquote {\bibinfo {title} {Anatomy of
  {Dzyaloshinskii-Moriya} interaction at $\mathrm{Co}/\mathrm{Pt}$
  interfaces},}\ }\href {\doibase 10.1103/PhysRevLett.115.267210} {\bibfield
  {journal} {\bibinfo  {journal} {Phys. Rev. Lett.}\ }\textbf {\bibinfo
  {volume} {115}},\ \bibinfo {pages} {267210} (\bibinfo {year}
  {2015})}\BibitemShut {NoStop}%
\bibitem [{\citenamefont {Moreau-Luchaire}\ \emph {et~al.}(2016)\citenamefont
  {Moreau-Luchaire}, \citenamefont {Moutafis}, \citenamefont {Reyren},
  \citenamefont {Sampaio}, \citenamefont {Vaz}, \citenamefont {Van~Horne},
  \citenamefont {Bouzehouane}, \citenamefont {Garcia}, \citenamefont
  {Deranlot}, \citenamefont {Warnicke}, \citenamefont {Wohlh\"{u}ter},
  \citenamefont {George}, \citenamefont {Weigand}, \citenamefont {Raabe},
  \citenamefont {Cros},\ and\ \citenamefont {Fert}}]{Moreau16NatNan}%
  \BibitemOpen
  \bibfield  {author} {\bibinfo {author} {\bibfnamefont {C.}~\bibnamefont
  {Moreau-Luchaire}}, \bibinfo {author} {\bibfnamefont {C.}~\bibnamefont
  {Moutafis}}, \bibinfo {author} {\bibfnamefont {N.}~\bibnamefont {Reyren}},
  \bibinfo {author} {\bibfnamefont {J.}~\bibnamefont {Sampaio}}, \bibinfo
  {author} {\bibfnamefont {C.~A.~F.}\ \bibnamefont {Vaz}}, \bibinfo {author}
  {\bibfnamefont {N.}~\bibnamefont {Van~Horne}}, \bibinfo {author}
  {\bibfnamefont {K.}~\bibnamefont {Bouzehouane}}, \bibinfo {author}
  {\bibfnamefont {K.}~\bibnamefont {Garcia}}, \bibinfo {author} {\bibfnamefont
  {C.}~\bibnamefont {Deranlot}}, \bibinfo {author} {\bibfnamefont
  {P.}~\bibnamefont {Warnicke}}, \bibinfo {author} {\bibfnamefont
  {P.}~\bibnamefont {Wohlh\"{u}ter}}, \bibinfo {author} {\bibfnamefont {J.-M.}\
  \bibnamefont {George}}, \bibinfo {author} {\bibfnamefont {M.}~\bibnamefont
  {Weigand}}, \bibinfo {author} {\bibfnamefont {J.}~\bibnamefont {Raabe}},
  \bibinfo {author} {\bibfnamefont {V.}~\bibnamefont {Cros}}, \ and\ \bibinfo
  {author} {\bibfnamefont {A.}~\bibnamefont {Fert}},\ }\bibfield  {title}
  {\enquote {\bibinfo {title} {Additive interfacial chiral interaction in
  multilayers for stabilization of small individual skyrmions at room
  temperature},}\ }\href@noop {} {\bibfield  {journal} {\bibinfo  {journal}
  {Nature Nanotech.}\ }\textbf {\bibinfo {volume} {11}},\ \bibinfo {pages}
  {444} (\bibinfo {year} {2016})}\BibitemShut {NoStop}%
\bibitem [{\citenamefont {Soumyanarayanan}\ \emph {et~al.}(2017)\citenamefont
  {Soumyanarayanan}, \citenamefont {Raju}, \citenamefont {Oyarce},
  \citenamefont {Tan}, \citenamefont {Im}, \citenamefont {Petrovi{\'{c}}},
  \citenamefont {Ho}, \citenamefont {Khoo}, \citenamefont {Tran}, \citenamefont
  {Gan}, \citenamefont {Ernult},\ and\ \citenamefont
  {Panagopoulos}}]{Soumyanarayanan2017}%
  \BibitemOpen
  \bibfield  {author} {\bibinfo {author} {\bibfnamefont {A.}~\bibnamefont
  {Soumyanarayanan}}, \bibinfo {author} {\bibfnamefont {M.}~\bibnamefont
  {Raju}}, \bibinfo {author} {\bibfnamefont {A.~L.~Gonzalez}\ \bibnamefont
  {Oyarce}}, \bibinfo {author} {\bibfnamefont {A.~K.~C.}\ \bibnamefont {Tan}},
  \bibinfo {author} {\bibfnamefont {M.-Y.}\ \bibnamefont {Im}}, \bibinfo
  {author} {\bibfnamefont {A.~P.}\ \bibnamefont {Petrovi{\'{c}}}}, \bibinfo
  {author} {\bibfnamefont {P.}~\bibnamefont {Ho}}, \bibinfo {author}
  {\bibfnamefont {K.~H.}\ \bibnamefont {Khoo}}, \bibinfo {author}
  {\bibfnamefont {M.}~\bibnamefont {Tran}}, \bibinfo {author} {\bibfnamefont
  {C.~K.}\ \bibnamefont {Gan}}, \bibinfo {author} {\bibfnamefont
  {F.}~\bibnamefont {Ernult}}, \ and\ \bibinfo {author} {\bibfnamefont
  {C.}~\bibnamefont {Panagopoulos}},\ }\bibfield  {title} {\enquote {\bibinfo
  {title} {Tunable room-temperature magnetic skyrmions in {Ir/Fe/Co/Pt}
  multilayers},}\ }\href {\doibase 10.1038/nmat4934} {\bibfield  {journal}
  {\bibinfo  {journal} {Nature Mater.}\ }\textbf {\bibinfo {volume} {16}},\
  \bibinfo {pages} {898} (\bibinfo {year} {2017})}\BibitemShut {NoStop}%
\bibitem [{\citenamefont {Kim}\ \emph {et~al.}(2016)\citenamefont {Kim},
  \citenamefont {Jung}, \citenamefont {Cho}, \citenamefont {Han}, \citenamefont
  {Yin}, \citenamefont {Kim}, \citenamefont {Swagten},\ and\ \citenamefont
  {You}}]{Kim16APL}%
  \BibitemOpen
  \bibfield  {author} {\bibinfo {author} {\bibfnamefont {N.-H.}\ \bibnamefont
  {Kim}}, \bibinfo {author} {\bibfnamefont {J.}~\bibnamefont {Jung}}, \bibinfo
  {author} {\bibfnamefont {J.}~\bibnamefont {Cho}}, \bibinfo {author}
  {\bibfnamefont {D.-S.}\ \bibnamefont {Han}}, \bibinfo {author} {\bibfnamefont
  {Y.}~\bibnamefont {Yin}}, \bibinfo {author} {\bibfnamefont {J.-S.}\
  \bibnamefont {Kim}}, \bibinfo {author} {\bibfnamefont {H.~J.~M.}\
  \bibnamefont {Swagten}}, \ and\ \bibinfo {author} {\bibfnamefont {C.-Y.}\
  \bibnamefont {You}},\ }\bibfield  {title} {\enquote {\bibinfo {title}
  {Interfacial {Dzyaloshinskii-Moriya} interaction, surface anisotropy energy,
  and spin pumping at spin orbit coupled {Ir/Co} interface},}\ }\href@noop {}
  {\bibfield  {journal} {\bibinfo  {journal} {Appl. Phys. Lett.}\ }\textbf
  {\bibinfo {volume} {108}},\ \bibinfo {pages} {142406} (\bibinfo {year}
  {2016})}\BibitemShut {NoStop}%
\bibitem [{\citenamefont {Cheng}\ \emph {et~al.}(2011)\citenamefont {Cheng},
  \citenamefont {Feng}, \citenamefont {Chern}, \citenamefont {Lee},\ and\
  \citenamefont {Wu}}]{cheng11JAP}%
  \BibitemOpen
  \bibfield  {author} {\bibinfo {author} {\bibfnamefont {C.-W.}\ \bibnamefont
  {Cheng}}, \bibinfo {author} {\bibfnamefont {W.}~\bibnamefont {Feng}},
  \bibinfo {author} {\bibfnamefont {G.}~\bibnamefont {Chern}}, \bibinfo
  {author} {\bibfnamefont {C.~M.}\ \bibnamefont {Lee}}, \ and\ \bibinfo
  {author} {\bibfnamefont {T.-H.}\ \bibnamefont {Wu}},\ }\bibfield  {title}
  {\enquote {\bibinfo {title} {Effect of cap layer thickness on the
  perpendicular magnetic anisotropy in top {MgO/CoFeB/Ta} structures},}\
  }\href@noop {} {\bibfield  {journal} {\bibinfo  {journal} {J. Appl. Phys.}\
  }\textbf {\bibinfo {volume} {110}},\ \bibinfo {pages} {033916} (\bibinfo
  {year} {2011})}\BibitemShut {NoStop}%
\bibitem [{\citenamefont {Sinha}\ \emph {et~al.}(2013)\citenamefont {Sinha},
  \citenamefont {Hayashi}, \citenamefont {Kellock}, \citenamefont {Fukami},
  \citenamefont {Yamanouchi}, \citenamefont {Sato}, \citenamefont {Ikeda},
  \citenamefont {Mitani}, \citenamefont {Yang}, \citenamefont {Parkin},\ and\
  \citenamefont {Ohno}}]{sinha13APL}%
  \BibitemOpen
  \bibfield  {author} {\bibinfo {author} {\bibfnamefont {J.}~\bibnamefont
  {Sinha}}, \bibinfo {author} {\bibfnamefont {M.}~\bibnamefont {Hayashi}},
  \bibinfo {author} {\bibfnamefont {A.~J.}\ \bibnamefont {Kellock}}, \bibinfo
  {author} {\bibfnamefont {S.}~\bibnamefont {Fukami}}, \bibinfo {author}
  {\bibfnamefont {M.}~\bibnamefont {Yamanouchi}}, \bibinfo {author}
  {\bibfnamefont {H.}~\bibnamefont {Sato}}, \bibinfo {author} {\bibfnamefont
  {S.}~\bibnamefont {Ikeda}}, \bibinfo {author} {\bibfnamefont
  {S.}~\bibnamefont {Mitani}}, \bibinfo {author} {\bibfnamefont {S.-H.}\
  \bibnamefont {Yang}}, \bibinfo {author} {\bibfnamefont {S.~S.~P.}\
  \bibnamefont {Parkin}}, \ and\ \bibinfo {author} {\bibfnamefont
  {H.}~\bibnamefont {Ohno}},\ }\bibfield  {title} {\enquote {\bibinfo {title}
  {Enhanced interface perpendicular magnetic anisotropy in {Ta/CoFeB/MgO} using
  nitrogen doped {Ta} underlayers},}\ }\href@noop {} {\bibfield  {journal}
  {\bibinfo  {journal} {Appl. Phys. Lett.}\ }\textbf {\bibinfo {volume}
  {102}},\ \bibinfo {pages} {242405} (\bibinfo {year} {2013})}\BibitemShut
  {NoStop}%
\bibitem [{\citenamefont {Jang}\ \emph {et~al.}(2011)\citenamefont {Jang},
  \citenamefont {You}, \citenamefont {Lim},\ and\ \citenamefont
  {Lee}}]{Jang11JAP}%
  \BibitemOpen
  \bibfield  {author} {\bibinfo {author} {\bibfnamefont {S.~Y.}\ \bibnamefont
  {Jang}}, \bibinfo {author} {\bibfnamefont {C.-Y.}\ \bibnamefont {You}},
  \bibinfo {author} {\bibfnamefont {S.~H.}\ \bibnamefont {Lim}}, \ and\
  \bibinfo {author} {\bibfnamefont {S.~R.}\ \bibnamefont {Lee}},\ }\bibfield
  {title} {\enquote {\bibinfo {title} {Annealing effects on the magnetic dead
  layer and saturation magnetization in unit structures relevant to a synthetic
  ferrimagnetic free structure},}\ }\href@noop {} {\bibfield  {journal}
  {\bibinfo  {journal} {J. Appl. Phys.}\ }\textbf {\bibinfo {volume} {109}},\
  \bibinfo {pages} {013901} (\bibinfo {year} {2011})}\BibitemShut {NoStop}%
\bibitem [{\citenamefont {Bandiera}\ \emph {et~al.}(2011)\citenamefont
  {Bandiera}, \citenamefont {Sousa}, \citenamefont {Rodmacq},\ and\
  \citenamefont {Dieny}}]{bandiera11IEEE}%
  \BibitemOpen
  \bibfield  {author} {\bibinfo {author} {\bibfnamefont {S.}~\bibnamefont
  {Bandiera}}, \bibinfo {author} {\bibfnamefont {R.~C.}\ \bibnamefont {Sousa}},
  \bibinfo {author} {\bibfnamefont {B.}~\bibnamefont {Rodmacq}}, \ and\
  \bibinfo {author} {\bibfnamefont {B.}~\bibnamefont {Dieny}},\ }\bibfield
  {title} {\enquote {\bibinfo {title} {Asymmetric interfacial perpendicular
  magnetic anisotropy in {Pt/Co/Pt} trilayers},}\ }\href {\doibase
  10.1109/LMAG.2011.2174032} {\bibfield  {journal} {\bibinfo  {journal} {IEEE
  Magn. Lett.}\ }\textbf {\bibinfo {volume} {2}},\ \bibinfo {pages} {3000504}
  (\bibinfo {year} {2011})}\BibitemShut {NoStop}%
\bibitem [{\citenamefont {Ederer}\ \emph {et~al.}(2002)\citenamefont {Ederer},
  \citenamefont {Komelj}, \citenamefont {Fähnle},\ and\ \citenamefont
  {Schütz}}]{Ederer02PRB}%
  \BibitemOpen
  \bibfield  {author} {\bibinfo {author} {\bibfnamefont {Claude}\ \bibnamefont
  {Ederer}}, \bibinfo {author} {\bibfnamefont {Matej}\ \bibnamefont {Komelj}},
  \bibinfo {author} {\bibfnamefont {Manfred}\ \bibnamefont {Fähnle}}, \ and\
  \bibinfo {author} {\bibfnamefont {Gisela}\ \bibnamefont {Schütz}},\
  }\bibfield  {title} {\enquote {\bibinfo {title} {Theory of induced magnetic
  moments and x-ray magnetic circular dichroism in {Co-Pt} multilayers},}\
  }\href {\doibase 10.1103/physrevb.66.094413} {\bibfield  {journal} {\bibinfo
  {journal} {Physical Review B}\ }\textbf {\bibinfo {volume} {66}},\ \bibinfo
  {pages} {094413} (\bibinfo {year} {2002})}\BibitemShut {NoStop}%
\bibitem [{\citenamefont {Metaxas}\ \emph {et~al.}(2007)\citenamefont
  {Metaxas}, \citenamefont {Jamet}, \citenamefont {Mougin}, \citenamefont
  {Cormier}, \citenamefont {Ferr\'e}, \citenamefont {Baltz}, \citenamefont
  {Rodmacq}, \citenamefont {Dieny},\ and\ \citenamefont
  {Stamps}}]{metaxas07PRL}%
  \BibitemOpen
  \bibfield  {author} {\bibinfo {author} {\bibfnamefont {P.~J.}\ \bibnamefont
  {Metaxas}}, \bibinfo {author} {\bibfnamefont {J.~P.}\ \bibnamefont {Jamet}},
  \bibinfo {author} {\bibfnamefont {A.}~\bibnamefont {Mougin}}, \bibinfo
  {author} {\bibfnamefont {M.}~\bibnamefont {Cormier}}, \bibinfo {author}
  {\bibfnamefont {J.}~\bibnamefont {Ferr\'e}}, \bibinfo {author} {\bibfnamefont
  {V.}~\bibnamefont {Baltz}}, \bibinfo {author} {\bibfnamefont
  {B.}~\bibnamefont {Rodmacq}}, \bibinfo {author} {\bibfnamefont
  {B.}~\bibnamefont {Dieny}}, \ and\ \bibinfo {author} {\bibfnamefont {R.~L.}\
  \bibnamefont {Stamps}},\ }\bibfield  {title} {\enquote {\bibinfo {title}
  {Creep and flow regimes of magnetic domain-wall motion in ultrathin
  $\mathrm{Pt}/\mathrm{Co}/\mathrm{Pt}$ films with perpendicular anisotropy},}\
  }\href {\doibase 10.1103/PhysRevLett.99.217208} {\bibfield  {journal}
  {\bibinfo  {journal} {Phys. Rev. Lett.}\ }\textbf {\bibinfo {volume} {99}},\
  \bibinfo {pages} {217208} (\bibinfo {year} {2007})}\BibitemShut {NoStop}%
\bibitem [{\citenamefont {Lemerle}\ \emph {et~al.}(1998)\citenamefont
  {Lemerle}, \citenamefont {Ferr\'{e}}, \citenamefont {Chappert}, \citenamefont
  {Mathet}, \citenamefont {Giamarchi},\ and\ \citenamefont
  {Le~Doussal}}]{lemerle98PRL}%
  \BibitemOpen
  \bibfield  {author} {\bibinfo {author} {\bibfnamefont {S.}~\bibnamefont
  {Lemerle}}, \bibinfo {author} {\bibfnamefont {J.}~\bibnamefont {Ferr\'{e}}},
  \bibinfo {author} {\bibfnamefont {C.}~\bibnamefont {Chappert}}, \bibinfo
  {author} {\bibfnamefont {V.}~\bibnamefont {Mathet}}, \bibinfo {author}
  {\bibfnamefont {T.}~\bibnamefont {Giamarchi}}, \ and\ \bibinfo {author}
  {\bibfnamefont {P.}~\bibnamefont {Le~Doussal}},\ }\bibfield  {title}
  {\enquote {\bibinfo {title} {Domain wall creep in an {Ising} ultrathin
  magnetic film},}\ }\href {\doibase
  https://doi.org/10.1103/PhysRevLett.80.849} {\bibfield  {journal} {\bibinfo
  {journal} {Phys. Rev. Lett.}\ }\textbf {\bibinfo {volume} {80}},\ \bibinfo
  {pages} {849} (\bibinfo {year} {1998})}\BibitemShut {NoStop}%
\bibitem [{\citenamefont {Chauve}\ \emph {et~al.}(2000)\citenamefont {Chauve},
  \citenamefont {Giamarchi},\ and\ \citenamefont {Le~Doussal}}]{chauve00PRB}%
  \BibitemOpen
  \bibfield  {author} {\bibinfo {author} {\bibfnamefont {P.}~\bibnamefont
  {Chauve}}, \bibinfo {author} {\bibfnamefont {T.}~\bibnamefont {Giamarchi}}, \
  and\ \bibinfo {author} {\bibfnamefont {P.}~\bibnamefont {Le~Doussal}},\
  }\bibfield  {title} {\enquote {\bibinfo {title} {Creep and depinning in
  disordered media},}\ }\href@noop {} {\bibfield  {journal} {\bibinfo
  {journal} {Phys. Rev. B}\ }\textbf {\bibinfo {volume} {62}},\ \bibinfo
  {pages} {6241} (\bibinfo {year} {2000})}\BibitemShut {NoStop}%
\bibitem [{\citenamefont {Jeudy}\ \emph {et~al.}(2016)\citenamefont {Jeudy},
  \citenamefont {Mougin}, \citenamefont {Bustingorry}, \citenamefont
  {Savero~Torres}, \citenamefont {Gorchon}, \citenamefont {Kolton},
  \citenamefont {Lema\^{\i}tre},\ and\ \citenamefont {Jamet}}]{jeudy16PRL}%
  \BibitemOpen
  \bibfield  {author} {\bibinfo {author} {\bibfnamefont {V.}~\bibnamefont
  {Jeudy}}, \bibinfo {author} {\bibfnamefont {A.}~\bibnamefont {Mougin}},
  \bibinfo {author} {\bibfnamefont {S.}~\bibnamefont {Bustingorry}}, \bibinfo
  {author} {\bibfnamefont {W.}~\bibnamefont {Savero~Torres}}, \bibinfo {author}
  {\bibfnamefont {J.}~\bibnamefont {Gorchon}}, \bibinfo {author} {\bibfnamefont
  {A.~B.}\ \bibnamefont {Kolton}}, \bibinfo {author} {\bibfnamefont
  {A.}~\bibnamefont {Lema\^{\i}tre}}, \ and\ \bibinfo {author} {\bibfnamefont
  {J.-P.}\ \bibnamefont {Jamet}},\ }\bibfield  {title} {\enquote {\bibinfo
  {title} {Universal pinning energy barrier for driven domain walls in thin
  ferromagnetic films},}\ }\href {\doibase 10.1103/PhysRevLett.117.057201}
  {\bibfield  {journal} {\bibinfo  {journal} {Phys. Rev. Lett.}\ }\textbf
  {\bibinfo {volume} {117}},\ \bibinfo {pages} {057201} (\bibinfo {year}
  {2016})}\BibitemShut {NoStop}%
\bibitem [{\citenamefont {Jeudy}\ \emph {et~al.}(2018)\citenamefont {Jeudy},
  \citenamefont {Pardo}, \citenamefont {Torres}, \citenamefont {Bustingorry},\
  and\ \citenamefont {Kolton}}]{Jeudy18PRB}%
  \BibitemOpen
  \bibfield  {author} {\bibinfo {author} {\bibfnamefont {V.}~\bibnamefont
  {Jeudy}}, \bibinfo {author} {\bibfnamefont {R.~D{\'{\i}}az}\ \bibnamefont
  {Pardo}}, \bibinfo {author} {\bibfnamefont {W.~Savero}\ \bibnamefont
  {Torres}}, \bibinfo {author} {\bibfnamefont {S.}~\bibnamefont {Bustingorry}},
  \ and\ \bibinfo {author} {\bibfnamefont {A.~B.}\ \bibnamefont {Kolton}},\
  }\bibfield  {title} {\enquote {\bibinfo {title} {Pinning of domain walls in
  thin ferromagnetic films},}\ }\href {\doibase 10.1103/physrevb.98.054406}
  {\bibfield  {journal} {\bibinfo  {journal} {Physical Review B}\ }\textbf
  {\bibinfo {volume} {98}},\ \bibinfo {pages} {054406} (\bibinfo {year}
  {2018})}\BibitemShut {NoStop}%
\bibitem [{\citenamefont {Diaz~Pardo}\ \emph {et~al.}(2017)\citenamefont
  {Diaz~Pardo}, \citenamefont {Savero~Torres}, \citenamefont {Kolton},
  \citenamefont {Bustingorry},\ and\ \citenamefont {Jeudy}}]{diazpardo17PRB}%
  \BibitemOpen
  \bibfield  {author} {\bibinfo {author} {\bibfnamefont {R.}~\bibnamefont
  {Diaz~Pardo}}, \bibinfo {author} {\bibfnamefont {W.}~\bibnamefont
  {Savero~Torres}}, \bibinfo {author} {\bibfnamefont {A.~B.}\ \bibnamefont
  {Kolton}}, \bibinfo {author} {\bibfnamefont {S.}~\bibnamefont {Bustingorry}},
  \ and\ \bibinfo {author} {\bibfnamefont {V.}~\bibnamefont {Jeudy}},\
  }\bibfield  {title} {\enquote {\bibinfo {title} {Universal depinning
  transition of domain walls in ultrathin ferromagnets},}\ }\href {\doibase
  10.1103/PhysRevB.95.184434} {\bibfield  {journal} {\bibinfo  {journal} {Phys.
  Rev. B}\ }\textbf {\bibinfo {volume} {95}},\ \bibinfo {pages} {184434}
  (\bibinfo {year} {2017})}\BibitemShut {NoStop}%
\bibitem [{\citenamefont {Kirilyuk}\ \emph {et~al.}(1997)\citenamefont
  {Kirilyuk}, \citenamefont {Ferr{\'e}}, \citenamefont {Grolier}, \citenamefont
  {Jamet},\ and\ \citenamefont {Renard}}]{kirilyuk97JMMM}%
  \BibitemOpen
  \bibfield  {author} {\bibinfo {author} {\bibfnamefont {A.}~\bibnamefont
  {Kirilyuk}}, \bibinfo {author} {\bibfnamefont {J.}~\bibnamefont {Ferr{\'e}}},
  \bibinfo {author} {\bibfnamefont {V.}~\bibnamefont {Grolier}}, \bibinfo
  {author} {\bibfnamefont {J.P.}\ \bibnamefont {Jamet}}, \ and\ \bibinfo
  {author} {\bibfnamefont {D.}~\bibnamefont {Renard}},\ }\bibfield  {title}
  {\enquote {\bibinfo {title} {Magnetization reversal in ultrathin
  ferromagnetic films with perpendicular anisotropy},}\ }\href {\doibase
  https://doi.org/10.1016/S0304-8853(96)00744-5} {\bibfield  {journal}
  {\bibinfo  {journal} {J. Magn. Magn. Mater.}\ }\textbf {\bibinfo {volume}
  {171}},\ \bibinfo {pages} {45} (\bibinfo {year} {1997})}\BibitemShut
  {NoStop}%
\bibitem [{\citenamefont {Kim}\ \emph {et~al.}(2009)\citenamefont {Kim},
  \citenamefont {Lee}, \citenamefont {Ahn}, \citenamefont {Lee}, \citenamefont
  {Lee}, \citenamefont {Cho}, \citenamefont {Seo}, \citenamefont {Shin},
  \citenamefont {Choe},\ and\ \citenamefont {Lee}}]{kim09Nat}%
  \BibitemOpen
  \bibfield  {author} {\bibinfo {author} {\bibfnamefont {K.-J.}\ \bibnamefont
  {Kim}}, \bibinfo {author} {\bibfnamefont {J.-C.}\ \bibnamefont {Lee}},
  \bibinfo {author} {\bibfnamefont {S.-M.}\ \bibnamefont {Ahn}}, \bibinfo
  {author} {\bibfnamefont {K.-S.}\ \bibnamefont {Lee}}, \bibinfo {author}
  {\bibfnamefont {C.-W.}\ \bibnamefont {Lee}}, \bibinfo {author} {\bibfnamefont
  {Y.~J.}\ \bibnamefont {Cho}}, \bibinfo {author} {\bibfnamefont
  {S.}~\bibnamefont {Seo}}, \bibinfo {author} {\bibfnamefont {K.-H.}\
  \bibnamefont {Shin}}, \bibinfo {author} {\bibfnamefont {S.-B.}\ \bibnamefont
  {Choe}}, \ and\ \bibinfo {author} {\bibfnamefont {H.-W.}\ \bibnamefont
  {Lee}},\ }\bibfield  {title} {\enquote {\bibinfo {title} {Interdimensional
  universality of dynamic interfaces},}\ }\href {\doibase 10.1038/nature07874}
  {\bibfield  {journal} {\bibinfo  {journal} {Nature}\ }\textbf {\bibinfo
  {volume} {458}},\ \bibinfo {pages} {740} (\bibinfo {year}
  {2009})}\BibitemShut {NoStop}%
\bibitem [{\citenamefont {Stashkevich}\ \emph {et~al.}(2015)\citenamefont
  {Stashkevich}, \citenamefont {Belmeguenai}, \citenamefont {Roussign\'e},
  \citenamefont {Cherif}, \citenamefont {Kostylev}, \citenamefont {Gabor},
  \citenamefont {Lacour}, \citenamefont {Tiusan},\ and\ \citenamefont
  {Hehn}}]{stashkevich15PRB}%
  \BibitemOpen
  \bibfield  {author} {\bibinfo {author} {\bibfnamefont {A.~A.}\ \bibnamefont
  {Stashkevich}}, \bibinfo {author} {\bibfnamefont {M.}~\bibnamefont
  {Belmeguenai}}, \bibinfo {author} {\bibfnamefont {Y.}~\bibnamefont
  {Roussign\'e}}, \bibinfo {author} {\bibfnamefont {S.~M.}\ \bibnamefont
  {Cherif}}, \bibinfo {author} {\bibfnamefont {M.}~\bibnamefont {Kostylev}},
  \bibinfo {author} {\bibfnamefont {M.}~\bibnamefont {Gabor}}, \bibinfo
  {author} {\bibfnamefont {D.}~\bibnamefont {Lacour}}, \bibinfo {author}
  {\bibfnamefont {C.}~\bibnamefont {Tiusan}}, \ and\ \bibinfo {author}
  {\bibfnamefont {M.}~\bibnamefont {Hehn}},\ }\bibfield  {title} {\enquote
  {\bibinfo {title} {Experimental study of spin-wave dispersion in {Py/Pt} film
  structures in the presence of an interface {Dzyaloshinskii-Moriya}
  interaction},}\ }\href {\doibase 10.1103/PhysRevB.91.214409} {\bibfield
  {journal} {\bibinfo  {journal} {Phys. Rev. B}\ }\textbf {\bibinfo {volume}
  {91}},\ \bibinfo {pages} {214409} (\bibinfo {year} {2015})}\BibitemShut
  {NoStop}%
\bibitem [{\citenamefont {Moon}\ \emph {et~al.}(2013)\citenamefont {Moon},
  \citenamefont {Seo}, \citenamefont {Lee}, \citenamefont {Kim}, \citenamefont
  {Ryu}, \citenamefont {Lee}, \citenamefont {McMichael},\ and\ \citenamefont
  {Stiles}}]{moon13PRB}%
  \BibitemOpen
  \bibfield  {author} {\bibinfo {author} {\bibfnamefont {J.-H.}\ \bibnamefont
  {Moon}}, \bibinfo {author} {\bibfnamefont {S.-M.}\ \bibnamefont {Seo}},
  \bibinfo {author} {\bibfnamefont {K.-J.}\ \bibnamefont {Lee}}, \bibinfo
  {author} {\bibfnamefont {K.-W.}\ \bibnamefont {Kim}}, \bibinfo {author}
  {\bibfnamefont {J.}~\bibnamefont {Ryu}}, \bibinfo {author} {\bibfnamefont
  {H.-W.}\ \bibnamefont {Lee}}, \bibinfo {author} {\bibfnamefont {R.~D}\
  \bibnamefont {McMichael}}, \ and\ \bibinfo {author} {\bibfnamefont {M.~D.}\
  \bibnamefont {Stiles}},\ }\bibfield  {title} {\enquote {\bibinfo {title}
  {Spin-wave propagation in the presence of interfacial {Dzyaloshinskii-Moriya}
  interaction},}\ }\href {\doibase https://doi.org/10.1103/PhysRevB.88.184404}
  {\bibfield  {journal} {\bibinfo  {journal} {Phys. Rev. B}\ }\textbf {\bibinfo
  {volume} {88}},\ \bibinfo {pages} {184404} (\bibinfo {year}
  {2013})}\BibitemShut {NoStop}%
\bibitem [{\citenamefont {Leliaert}\ \emph {et~al.}(2014)\citenamefont
  {Leliaert}, \citenamefont {Van~de Wiele}, \citenamefont {Vansteenkiste},
  \citenamefont {Laurson}, \citenamefont {Durin}, \citenamefont {Dupré},\ and\
  \citenamefont {Van~Waeyenberge}}]{Leliaert14JAP}%
  \BibitemOpen
  \bibfield  {author} {\bibinfo {author} {\bibfnamefont {J.}~\bibnamefont
  {Leliaert}}, \bibinfo {author} {\bibfnamefont {B.}~\bibnamefont {Van~de
  Wiele}}, \bibinfo {author} {\bibfnamefont {A.}~\bibnamefont {Vansteenkiste}},
  \bibinfo {author} {\bibfnamefont {L.}~\bibnamefont {Laurson}}, \bibinfo
  {author} {\bibfnamefont {G.}~\bibnamefont {Durin}}, \bibinfo {author}
  {\bibfnamefont {L.}~\bibnamefont {Dupré}}, \ and\ \bibinfo {author}
  {\bibfnamefont {B.}~\bibnamefont {Van~Waeyenberge}},\ }\bibfield  {title}
  {\enquote {\bibinfo {title} {Current-driven domain wall mobility in
  polycrystalline permalloy nanowires: A numerical study},}\ }\href@noop {}
  {\bibfield  {journal} {\bibinfo  {journal} {J. Appl. Phys.}\ }\textbf
  {\bibinfo {volume} {115}},\ \bibinfo {pages} {233903} (\bibinfo {year}
  {2014})}\BibitemShut {NoStop}%
\bibitem [{\citenamefont {Vansteenkiste}\ \emph {et~al.}(2014)\citenamefont
  {Vansteenkiste}, \citenamefont {Leliaert}, \citenamefont {Dvornik},
  \citenamefont {Helsen}, \citenamefont {Garcia-Sanchez},\ and\ \citenamefont
  {Waeyenberge}}]{Mumax14AIP}%
  \BibitemOpen
  \bibfield  {author} {\bibinfo {author} {\bibfnamefont {A.}~\bibnamefont
  {Vansteenkiste}}, \bibinfo {author} {\bibfnamefont {J.}~\bibnamefont
  {Leliaert}}, \bibinfo {author} {\bibfnamefont {M.}~\bibnamefont {Dvornik}},
  \bibinfo {author} {\bibfnamefont {M.}~\bibnamefont {Helsen}}, \bibinfo
  {author} {\bibfnamefont {F.}~\bibnamefont {Garcia-Sanchez}}, \ and\ \bibinfo
  {author} {\bibfnamefont {B.~Van}\ \bibnamefont {Waeyenberge}},\ }\bibfield
  {title} {\enquote {\bibinfo {title} {The design and verification of
  {MuMax3}},}\ }\href@noop {} {\bibfield  {journal} {\bibinfo  {journal} {{AIP}
  Advances}\ }\textbf {\bibinfo {volume} {4}},\ \bibinfo {pages} {107133}
  (\bibinfo {year} {2014})}\BibitemShut {NoStop}%
\bibitem [{\citenamefont {Kim}\ and\ \citenamefont {Yoo}(2017)}]{Kim17APL}%
  \BibitemOpen
  \bibfield  {author} {\bibinfo {author} {\bibfnamefont {J.-V.}\ \bibnamefont
  {Kim}}\ and\ \bibinfo {author} {\bibfnamefont {M.-W.}\ \bibnamefont {Yoo}},\
  }\bibfield  {title} {\enquote {\bibinfo {title} {Current-driven skyrmion
  dynamics in disordered films},}\ }\href@noop {} {\bibfield  {journal}
  {\bibinfo  {journal} {Appl. Phys. Lett.}\ }\textbf {\bibinfo {volume}
  {110}},\ \bibinfo {pages} {132404} (\bibinfo {year} {2017})}\BibitemShut
  {NoStop}%
\bibitem [{\citenamefont {Zeissler}\ \emph {et~al.}(2017)\citenamefont
  {Zeissler}, \citenamefont {Mruczkiewicz}, \citenamefont {Finizio},
  \citenamefont {Raabe}, \citenamefont {Shepley}, \citenamefont {Sadovnikov},
  \citenamefont {Nikitov}, \citenamefont {Fallon}, \citenamefont {McFadzean},
  \citenamefont {McVitie}, \citenamefont {Moore}, \citenamefont {Burnell},\
  and\ \citenamefont {Marrows}}]{zeissler17SciRep}%
  \BibitemOpen
  \bibfield  {author} {\bibinfo {author} {\bibfnamefont {K.}~\bibnamefont
  {Zeissler}}, \bibinfo {author} {\bibfnamefont {M.}~\bibnamefont
  {Mruczkiewicz}}, \bibinfo {author} {\bibfnamefont {S.}~\bibnamefont
  {Finizio}}, \bibinfo {author} {\bibfnamefont {J.}~\bibnamefont {Raabe}},
  \bibinfo {author} {\bibfnamefont {P.~M.}\ \bibnamefont {Shepley}}, \bibinfo
  {author} {\bibfnamefont {A.~V.}\ \bibnamefont {Sadovnikov}}, \bibinfo
  {author} {\bibfnamefont {S.~A.}\ \bibnamefont {Nikitov}}, \bibinfo {author}
  {\bibfnamefont {K.}~\bibnamefont {Fallon}}, \bibinfo {author} {\bibfnamefont
  {S.}~\bibnamefont {McFadzean}}, \bibinfo {author} {\bibfnamefont
  {S.}~\bibnamefont {McVitie}}, \bibinfo {author} {\bibfnamefont {T.~A.}\
  \bibnamefont {Moore}}, \bibinfo {author} {\bibfnamefont {G.}~\bibnamefont
  {Burnell}}, \ and\ \bibinfo {author} {\bibfnamefont {C.~H.}\ \bibnamefont
  {Marrows}},\ }\bibfield  {title} {\enquote {\bibinfo {title} {Pinning and
  hysteresis in the field dependent diameter evolution of skyrmions in
  {Pt/Co/Ir} superlattice stacks},}\ }\href@noop {} {\bibfield  {journal}
  {\bibinfo  {journal} {Sci, Rep.}\ }\textbf {\bibinfo {volume} {7}},\ \bibinfo
  {pages} {15125} (\bibinfo {year} {2017})}\BibitemShut {NoStop}%
\bibitem [{\citenamefont {Gross}\ \emph {et~al.}(2016)\citenamefont {Gross},
  \citenamefont {Mart\'{\i}nez}, \citenamefont {Tetienne}, \citenamefont
  {Hingant}, \citenamefont {Roch}, \citenamefont {Garcia}, \citenamefont
  {Soucaille}, \citenamefont {Adam}, \citenamefont {Kim}, \citenamefont
  {Rohart}, \citenamefont {Thiaville}, \citenamefont {Torrejon}, \citenamefont
  {Hayashi},\ and\ \citenamefont {Jacques}}]{gross16PRB}%
  \BibitemOpen
  \bibfield  {author} {\bibinfo {author} {\bibfnamefont {I.}~\bibnamefont
  {Gross}}, \bibinfo {author} {\bibfnamefont {L.~J.}\ \bibnamefont
  {Mart\'{\i}nez}}, \bibinfo {author} {\bibfnamefont {J.-P.}\ \bibnamefont
  {Tetienne}}, \bibinfo {author} {\bibfnamefont {T.}~\bibnamefont {Hingant}},
  \bibinfo {author} {\bibfnamefont {J.-F.}\ \bibnamefont {Roch}}, \bibinfo
  {author} {\bibfnamefont {K.}~\bibnamefont {Garcia}}, \bibinfo {author}
  {\bibfnamefont {R.}~\bibnamefont {Soucaille}}, \bibinfo {author}
  {\bibfnamefont {J.~P.}\ \bibnamefont {Adam}}, \bibinfo {author}
  {\bibfnamefont {J.-V.}\ \bibnamefont {Kim}}, \bibinfo {author} {\bibfnamefont
  {S.}~\bibnamefont {Rohart}}, \bibinfo {author} {\bibfnamefont
  {A.}~\bibnamefont {Thiaville}}, \bibinfo {author} {\bibfnamefont
  {J.}~\bibnamefont {Torrejon}}, \bibinfo {author} {\bibfnamefont
  {M.}~\bibnamefont {Hayashi}}, \ and\ \bibinfo {author} {\bibfnamefont
  {V.}~\bibnamefont {Jacques}},\ }\bibfield  {title} {\enquote {\bibinfo
  {title} {Direct measurement of interfacial {Dzyaloshinskii-Moriya}
  interaction in $x|\text{CoFeB}|\text{MgO}$ heterostructures with a scanning
  nv magnetometer $(x=\text{Ta},\text{TaN}, \text{and W})$},}\ }\href {\doibase
  10.1103/PhysRevB.94.064413} {\bibfield  {journal} {\bibinfo  {journal} {Phys.
  Rev. B}\ }\textbf {\bibinfo {volume} {94}},\ \bibinfo {pages} {064413}
  (\bibinfo {year} {2016})}\BibitemShut {NoStop}%
\bibitem [{\citenamefont {Ryu}\ \emph {et~al.}(2014)\citenamefont {Ryu},
  \citenamefont {Yang}, \citenamefont {Thomas},\ and\ \citenamefont
  {Parkin}}]{Ryu14NatCom}%
  \BibitemOpen
  \bibfield  {author} {\bibinfo {author} {\bibfnamefont {K.-S.}\ \bibnamefont
  {Ryu}}, \bibinfo {author} {\bibfnamefont {S.-H.}\ \bibnamefont {Yang}},
  \bibinfo {author} {\bibfnamefont {L.}~\bibnamefont {Thomas}}, \ and\ \bibinfo
  {author} {\bibfnamefont {S.~S.~P.}\ \bibnamefont {Parkin}},\ }\bibfield
  {title} {\enquote {\bibinfo {title} {Chiral spin torque arising from
  proximity-induced magnetization},}\ }\href@noop {} {\bibfield  {journal}
  {\bibinfo  {journal} {Nature Commun.}\ }\textbf {\bibinfo {volume} {5}},\
  \bibinfo {pages} {3910} (\bibinfo {year} {2014})}\BibitemShut {NoStop}%
\bibitem [{\citenamefont {Ma}\ \emph {et~al.}(2018)\citenamefont {Ma},
  \citenamefont {Yu}, \citenamefont {Tang}, \citenamefont {Li}, \citenamefont
  {He}, \citenamefont {Shi}, \citenamefont {Wang},\ and\ \citenamefont
  {Li}}]{Ma18PRL}%
  \BibitemOpen
  \bibfield  {author} {\bibinfo {author} {\bibfnamefont {X.}~\bibnamefont
  {Ma}}, \bibinfo {author} {\bibfnamefont {G.}~\bibnamefont {Yu}}, \bibinfo
  {author} {\bibfnamefont {C.}~\bibnamefont {Tang}}, \bibinfo {author}
  {\bibfnamefont {X.}~\bibnamefont {Li}}, \bibinfo {author} {\bibfnamefont
  {C.}~\bibnamefont {He}}, \bibinfo {author} {\bibfnamefont {J.}~\bibnamefont
  {Shi}}, \bibinfo {author} {\bibfnamefont {K.~L.}\ \bibnamefont {Wang}}, \
  and\ \bibinfo {author} {\bibfnamefont {X.}~\bibnamefont {Li}},\ }\bibfield
  {title} {\enquote {\bibinfo {title} {Interfacial {Dzyaloshinskii-Moriya}
  interaction: {E}ffect of $5d$ band filling and correlation with spin mixing
  conductance},}\ }\href@noop {} {\bibfield  {journal} {\bibinfo  {journal}
  {Phys. Rev. Lett.}\ }\textbf {\bibinfo {volume} {120}},\ \bibinfo {pages}
  {157204} (\bibinfo {year} {2018})}\BibitemShut {NoStop}%
\bibitem [{\citenamefont {Wells}\ \emph {et~al.}(2017)\citenamefont {Wells},
  \citenamefont {Shepley}, \citenamefont {Marrows},\ and\ \citenamefont
  {Moore}}]{Wells17PRB}%
  \BibitemOpen
  \bibfield  {author} {\bibinfo {author} {\bibfnamefont {A.~W.~J.}\
  \bibnamefont {Wells}}, \bibinfo {author} {\bibfnamefont {P.~M.}\ \bibnamefont
  {Shepley}}, \bibinfo {author} {\bibfnamefont {C.~H.}\ \bibnamefont
  {Marrows}}, \ and\ \bibinfo {author} {\bibfnamefont {T.~A.}\ \bibnamefont
  {Moore}},\ }\bibfield  {title} {\enquote {\bibinfo {title} {Effect of
  interfacial intermixing on the {Dzyaloshinskii-Moriya} interaction in
  {Pt/Co/Pt}},}\ }\href@noop {} {\bibfield  {journal} {\bibinfo  {journal}
  {Phys. Rev. B}\ }\textbf {\bibinfo {volume} {95}},\ \bibinfo {pages} {054428}
  (\bibinfo {year} {2017})}\BibitemShut {NoStop}%
\bibitem [{\citenamefont {Yamamoto}\ \emph {et~al.}(2017)\citenamefont
  {Yamamoto}, \citenamefont {Pradipto}, \citenamefont {Nawa}, \citenamefont
  {Akiyama}, \citenamefont {Ito}, \citenamefont {Ono},\ and\ \citenamefont
  {Nakamura}}]{Yamamoto17AIP}%
  \BibitemOpen
  \bibfield  {author} {\bibinfo {author} {\bibfnamefont {K.}~\bibnamefont
  {Yamamoto}}, \bibinfo {author} {\bibfnamefont {A.-M.}\ \bibnamefont
  {Pradipto}}, \bibinfo {author} {\bibfnamefont {K.}~\bibnamefont {Nawa}},
  \bibinfo {author} {\bibfnamefont {T.}~\bibnamefont {Akiyama}}, \bibinfo
  {author} {\bibfnamefont {T.}~\bibnamefont {Ito}}, \bibinfo {author}
  {\bibfnamefont {T.}~\bibnamefont {Ono}}, \ and\ \bibinfo {author}
  {\bibfnamefont {K.}~\bibnamefont {Nakamura}},\ }\bibfield  {title} {\enquote
  {\bibinfo {title} {Interfacial dzyaloshinskii-moriya interaction and orbital
  magnetic moments of metallic multilayer films},}\ }\href@noop {} {\bibfield
  {journal} {\bibinfo  {journal} {AIP Advances}\ }\textbf {\bibinfo {volume}
  {7}},\ \bibinfo {pages} {056302} (\bibinfo {year} {2017})}\BibitemShut
  {NoStop}%
\bibitem [{\citenamefont {Ba{\'c}ani}\ \emph {et~al.}(2016)\citenamefont
  {Ba{\'c}ani}, \citenamefont {Marioni}, \citenamefont {Schwenk},\ and\
  \citenamefont {Hug}}]{Bacani16X}%
  \BibitemOpen
  \bibfield  {author} {\bibinfo {author} {\bibfnamefont {M.}~\bibnamefont
  {Ba{\'c}ani}}, \bibinfo {author} {\bibfnamefont {M.~A.}\ \bibnamefont
  {Marioni}}, \bibinfo {author} {\bibfnamefont {J.}~\bibnamefont {Schwenk}}, \
  and\ \bibinfo {author} {\bibfnamefont {H.~J.}\ \bibnamefont {Hug}},\
  }\href@noop {} {\enquote {\bibinfo {title} {How to measure the local
  {Dzyaloshinskii Moriya} interaction in skyrmion thin film multilayers},}\ }
  (\bibinfo {year} {2016}),\ \Eprint {http://arxiv.org/abs/1609.01615}
  {arXiv:1609.01615 [cond-mat.mtrl-sci]} \BibitemShut {NoStop}%
\bibitem [{\citenamefont {Ma}\ \emph {et~al.}(2017)\citenamefont {Ma},
  \citenamefont {Yu}, \citenamefont {Tang}, \citenamefont {Li}, \citenamefont
  {He}, \citenamefont {Shi}, \citenamefont {Wang},\ and\ \citenamefont
  {Li}}]{Ma17X}%
  \BibitemOpen
  \bibfield  {author} {\bibinfo {author} {\bibfnamefont {X.}~\bibnamefont
  {Ma}}, \bibinfo {author} {\bibfnamefont {G.}~\bibnamefont {Yu}}, \bibinfo
  {author} {\bibfnamefont {C.}~\bibnamefont {Tang}}, \bibinfo {author}
  {\bibfnamefont {X.}~\bibnamefont {Li}}, \bibinfo {author} {\bibfnamefont
  {C.}~\bibnamefont {He}}, \bibinfo {author} {\bibfnamefont {J.}~\bibnamefont
  {Shi}}, \bibinfo {author} {\bibfnamefont {K.~L}\ \bibnamefont {Wang}}, \ and\
  \bibinfo {author} {\bibfnamefont {X.}~\bibnamefont {Li}},\ }\href@noop {}
  {\enquote {\bibinfo {title} {Scaling of {Dzyaloshinskii Moriya} interaction
  at heavy metal and ferromagnetic metal interfaces},}\ } (\bibinfo {year}
  {2017}),\ \Eprint {http://arxiv.org/abs/1709.03961} {arXiv:1709.03961
  [cond-mat.mtrl-sci]} \BibitemShut {NoStop}%
\bibitem [{\citenamefont {Belabbes}\ \emph {et~al.}(2016)\citenamefont
  {Belabbes}, \citenamefont {Bihlmayer}, \citenamefont {Bechstedt},
  \citenamefont {Bl\"ugel},\ and\ \citenamefont {Manchon}}]{Belabbes16PRL}%
  \BibitemOpen
  \bibfield  {author} {\bibinfo {author} {\bibfnamefont {A.}~\bibnamefont
  {Belabbes}}, \bibinfo {author} {\bibfnamefont {G.}~\bibnamefont {Bihlmayer}},
  \bibinfo {author} {\bibfnamefont {F.}~\bibnamefont {Bechstedt}}, \bibinfo
  {author} {\bibfnamefont {S.}~\bibnamefont {Bl\"ugel}}, \ and\ \bibinfo
  {author} {\bibfnamefont {A.}~\bibnamefont {Manchon}},\ }\bibfield  {title}
  {\enquote {\bibinfo {title} {Hund's rule-driven {Dzyaloshinskii-Moriya}
  interaction at $3d\text{\ensuremath{-}}5d$ interfaces},}\ }\href {\doibase
  10.1103/PhysRevLett.117.247202} {\bibfield  {journal} {\bibinfo  {journal}
  {Phys. Rev. Lett.}\ }\textbf {\bibinfo {volume} {117}},\ \bibinfo {pages}
  {247202} (\bibinfo {year} {2016})}\BibitemShut {NoStop}%
\bibitem [{\citenamefont {Ma}\ \emph {et~al.}(2016)\citenamefont {Ma},
  \citenamefont {Yu}, \citenamefont {Li}, \citenamefont {Wang}, \citenamefont
  {Wu}, \citenamefont {Olsson}, \citenamefont {Chu}, \citenamefont {An},
  \citenamefont {Xiao}, \citenamefont {Wang},\ and\ \citenamefont
  {Li}}]{Ma16PRB}%
  \BibitemOpen
  \bibfield  {author} {\bibinfo {author} {\bibfnamefont {X.}~\bibnamefont
  {Ma}}, \bibinfo {author} {\bibfnamefont {G.}~\bibnamefont {Yu}}, \bibinfo
  {author} {\bibfnamefont {X.}~\bibnamefont {Li}}, \bibinfo {author}
  {\bibfnamefont {T.}~\bibnamefont {Wang}}, \bibinfo {author} {\bibfnamefont
  {D.}~\bibnamefont {Wu}}, \bibinfo {author} {\bibfnamefont {K.~S.}\
  \bibnamefont {Olsson}}, \bibinfo {author} {\bibfnamefont {Z.}~\bibnamefont
  {Chu}}, \bibinfo {author} {\bibfnamefont {K.}~\bibnamefont {An}}, \bibinfo
  {author} {\bibfnamefont {J.~Q.}\ \bibnamefont {Xiao}}, \bibinfo {author}
  {\bibfnamefont {K.~L.}\ \bibnamefont {Wang}}, \ and\ \bibinfo {author}
  {\bibfnamefont {X.}~\bibnamefont {Li}},\ }\bibfield  {title} {\enquote
  {\bibinfo {title} {Interfacial control of {Dzyaloshinskii-Moriya} interaction
  in heavy metal/ferromagnetic metal thin film heterostructures},}\ }\href@noop
  {} {\bibfield  {journal} {\bibinfo  {journal} {Phys. Rev. B}\ }\textbf
  {\bibinfo {volume} {94}},\ \bibinfo {pages} {180408} (\bibinfo {year}
  {2016})}\BibitemShut {NoStop}%
\bibitem [{\citenamefont {Tacchi}\ \emph {et~al.}(2017)\citenamefont {Tacchi},
  \citenamefont {Troncoso}, \citenamefont {Ahlberg}, \citenamefont {Gubbiotti},
  \citenamefont {Madami}, \citenamefont {\AA{}kerman},\ and\ \citenamefont
  {Landeros}}]{Tacchi17PRL}%
  \BibitemOpen
  \bibfield  {author} {\bibinfo {author} {\bibfnamefont {S.}~\bibnamefont
  {Tacchi}}, \bibinfo {author} {\bibfnamefont {R.~E.}\ \bibnamefont
  {Troncoso}}, \bibinfo {author} {\bibfnamefont {M.}~\bibnamefont {Ahlberg}},
  \bibinfo {author} {\bibfnamefont {G.}~\bibnamefont {Gubbiotti}}, \bibinfo
  {author} {\bibfnamefont {M.}~\bibnamefont {Madami}}, \bibinfo {author}
  {\bibfnamefont {J.}~\bibnamefont {\AA{}kerman}}, \ and\ \bibinfo {author}
  {\bibfnamefont {P.}~\bibnamefont {Landeros}},\ }\bibfield  {title} {\enquote
  {\bibinfo {title} {Interfacial {Dzyaloshinskii-Moriya} interaction in
  $\mathrm{Pt}/\mathrm{CoFeB}$ films: Effect of the heavy-metal thickness},}\
  }\href {\doibase 10.1103/PhysRevLett.118.147201} {\bibfield  {journal}
  {\bibinfo  {journal} {Phys. Rev. Lett.}\ }\textbf {\bibinfo {volume} {118}},\
  \bibinfo {pages} {147201} (\bibinfo {year} {2017})}\BibitemShut {NoStop}%
\bibitem [{\citenamefont {Rowan-Robinson}\ \emph {et~al.}(2017)\citenamefont
  {Rowan-Robinson}, \citenamefont {Stashkevich}, \citenamefont
  {Roussign{\'{e}}}, \citenamefont {Belmeguenai}, \citenamefont {Ch{\'{e}}rif},
  \citenamefont {Thiaville}, \citenamefont {Hase}, \citenamefont {Hindmarch},\
  and\ \citenamefont {Atkinson}}]{Robinson17SciRep}%
  \BibitemOpen
  \bibfield  {author} {\bibinfo {author} {\bibfnamefont {R.~M.}\ \bibnamefont
  {Rowan-Robinson}}, \bibinfo {author} {\bibfnamefont {A.~A.}\ \bibnamefont
  {Stashkevich}}, \bibinfo {author} {\bibfnamefont {Y.}~\bibnamefont
  {Roussign{\'{e}}}}, \bibinfo {author} {\bibfnamefont {M.}~\bibnamefont
  {Belmeguenai}}, \bibinfo {author} {\bibfnamefont {S.-M.}\ \bibnamefont
  {Ch{\'{e}}rif}}, \bibinfo {author} {\bibfnamefont {A.}~\bibnamefont
  {Thiaville}}, \bibinfo {author} {\bibfnamefont {T.~P.~A.}\ \bibnamefont
  {Hase}}, \bibinfo {author} {\bibfnamefont {A.~T.}\ \bibnamefont {Hindmarch}},
  \ and\ \bibinfo {author} {\bibfnamefont {D.}~\bibnamefont {Atkinson}},\
  }\bibfield  {title} {\enquote {\bibinfo {title} {The interfacial nature of
  proximity-induced magnetism and the {Dzyaloshinskii-Moriya} interaction at
  the {Pt/Co} interface},}\ }\href@noop {} {\bibfield  {journal} {\bibinfo
  {journal} {Sci. Rep.}\ }\textbf {\bibinfo {volume} {7}},\ \bibinfo {pages}
  {16835} (\bibinfo {year} {2017})}\BibitemShut {NoStop}%
\bibitem [{\citenamefont {Yu}\ \emph {et~al.}(2016)\citenamefont {Yu},
  \citenamefont {Qiu}, \citenamefont {Wu}, \citenamefont {Yoon}, \citenamefont
  {Deorani}, \citenamefont {Besbas}, \citenamefont {Manchon},\ and\
  \citenamefont {Yang}}]{yu16SCiRep}%
  \BibitemOpen
  \bibfield  {author} {\bibinfo {author} {\bibfnamefont {J.}~\bibnamefont
  {Yu}}, \bibinfo {author} {\bibfnamefont {X.}~\bibnamefont {Qiu}}, \bibinfo
  {author} {\bibfnamefont {Y.}~\bibnamefont {Wu}}, \bibinfo {author}
  {\bibfnamefont {J.}~\bibnamefont {Yoon}}, \bibinfo {author} {\bibfnamefont
  {P.}~\bibnamefont {Deorani}}, \bibinfo {author} {\bibfnamefont {J.~M.}\
  \bibnamefont {Besbas}}, \bibinfo {author} {\bibfnamefont {A.}~\bibnamefont
  {Manchon}}, \ and\ \bibinfo {author} {\bibfnamefont {H.}~\bibnamefont
  {Yang}},\ }\bibfield  {title} {\enquote {\bibinfo {title} {Spin orbit torques
  and {Dzyaloshinskii-Moriya} interaction in dual-interfaced {Co-Ni}
  multilayers},}\ }\href {\doibase 10.1038/srep32629} {\bibfield  {journal}
  {\bibinfo  {journal} {Sci. Rep.}\ }\textbf {\bibinfo {volume} {6}},\ \bibinfo
  {pages} {32629} (\bibinfo {year} {2016})}\BibitemShut {NoStop}%
\bibitem [{\citenamefont {Zhang}\ \emph {et~al.}(2017)\citenamefont {Zhang},
  \citenamefont {Cao}, \citenamefont {Qiao}, \citenamefont {Tang},
  \citenamefont {Cao}, \citenamefont {Zhao}, \citenamefont {Eimer},
  \citenamefont {Si}, \citenamefont {Lei}, \citenamefont {Wang}, \citenamefont
  {Lin}, \citenamefont {Zhang}, \citenamefont {Wu},\ and\ \citenamefont
  {Zhao}}]{zhang17APL}%
  \BibitemOpen
  \bibfield  {author} {\bibinfo {author} {\bibfnamefont {B.}~\bibnamefont
  {Zhang}}, \bibinfo {author} {\bibfnamefont {A.}~\bibnamefont {Cao}}, \bibinfo
  {author} {\bibfnamefont {J.}~\bibnamefont {Qiao}}, \bibinfo {author}
  {\bibfnamefont {M.}~\bibnamefont {Tang}}, \bibinfo {author} {\bibfnamefont
  {K.}~\bibnamefont {Cao}}, \bibinfo {author} {\bibfnamefont {X.}~\bibnamefont
  {Zhao}}, \bibinfo {author} {\bibfnamefont {S.}~\bibnamefont {Eimer}},
  \bibinfo {author} {\bibfnamefont {Z.}~\bibnamefont {Si}}, \bibinfo {author}
  {\bibfnamefont {N.}~\bibnamefont {Lei}}, \bibinfo {author} {\bibfnamefont
  {Z.}~\bibnamefont {Wang}}, \bibinfo {author} {\bibfnamefont {X.}~\bibnamefont
  {Lin}}, \bibinfo {author} {\bibfnamefont {Z.}~\bibnamefont {Zhang}}, \bibinfo
  {author} {\bibfnamefont {M.}~\bibnamefont {Wu}}, \ and\ \bibinfo {author}
  {\bibfnamefont {W.}~\bibnamefont {Zhao}},\ }\bibfield  {title} {\enquote
  {\bibinfo {title} {Influence of heavy metal materials on magnetic properties
  of {Pt/Co/heavy metal} tri-layered structures},}\ }\href@noop {} {\bibfield
  {journal} {\bibinfo  {journal} {Appl. Phys. Lett.}\ }\textbf {\bibinfo
  {volume} {110}},\ \bibinfo {pages} {012405} (\bibinfo {year}
  {2017})}\BibitemShut {NoStop}%
\bibitem [{\citenamefont {Sato}\ \emph {et~al.}(2012)\citenamefont {Sato},
  \citenamefont {Yamanouchi}, \citenamefont {Ikeda}, \citenamefont {Fukami},
  \citenamefont {Matsukura},\ and\ \citenamefont {Ohno}}]{sato12APL}%
  \BibitemOpen
  \bibfield  {author} {\bibinfo {author} {\bibfnamefont {H.}~\bibnamefont
  {Sato}}, \bibinfo {author} {\bibfnamefont {M.}~\bibnamefont {Yamanouchi}},
  \bibinfo {author} {\bibfnamefont {S.}~\bibnamefont {Ikeda}}, \bibinfo
  {author} {\bibfnamefont {S.}~\bibnamefont {Fukami}}, \bibinfo {author}
  {\bibfnamefont {F.}~\bibnamefont {Matsukura}}, \ and\ \bibinfo {author}
  {\bibfnamefont {H.}~\bibnamefont {Ohno}},\ }\bibfield  {title} {\enquote
  {\bibinfo {title} {Perpendicular-anisotropy {CoFeB-MgO} magnetic tunnel
  junctions with a {MgO/CoFeB/Ta/CoFeB/MgO} recording structure},}\ }\href@noop
  {} {\bibfield  {journal} {\bibinfo  {journal} {Appl. Phys. Lett.}\ }\textbf
  {\bibinfo {volume} {101}},\ \bibinfo {pages} {022414} (\bibinfo {year}
  {2012})}\BibitemShut {NoStop}%
\bibitem [{\citenamefont {Jang}\ \emph {et~al.}(2010)\citenamefont {Jang},
  \citenamefont {Lim},\ and\ \citenamefont {Lee}}]{Jang10JAP}%
  \BibitemOpen
  \bibfield  {author} {\bibinfo {author} {\bibfnamefont {S.~Y.}\ \bibnamefont
  {Jang}}, \bibinfo {author} {\bibfnamefont {S.~H.}\ \bibnamefont {Lim}}, \
  and\ \bibinfo {author} {\bibfnamefont {S.~R.}\ \bibnamefont {Lee}},\
  }\bibfield  {title} {\enquote {\bibinfo {title} {Magnetic dead layer in
  amorphous {CoFeB} layers with various top and bottom structures},}\
  }\href@noop {} {\bibfield  {journal} {\bibinfo  {journal} {J. Appl. Phys.}\
  }\textbf {\bibinfo {volume} {107}},\ \bibinfo {pages} {09C707} (\bibinfo
  {year} {2010})}\BibitemShut {NoStop}%
\bibitem [{\citenamefont {Oguz}\ \emph {et~al.}(2008)\citenamefont {Oguz},
  \citenamefont {Jivrajka}, \citenamefont {Venkatesan}, \citenamefont {Feng},\
  and\ \citenamefont {Coey}}]{oguz08JAP}%
  \BibitemOpen
  \bibfield  {author} {\bibinfo {author} {\bibfnamefont {K.}~\bibnamefont
  {Oguz}}, \bibinfo {author} {\bibfnamefont {P.}~\bibnamefont {Jivrajka}},
  \bibinfo {author} {\bibfnamefont {M.}~\bibnamefont {Venkatesan}}, \bibinfo
  {author} {\bibfnamefont {G.}~\bibnamefont {Feng}}, \ and\ \bibinfo {author}
  {\bibfnamefont {J.~M.~D.}\ \bibnamefont {Coey}},\ }\bibfield  {title}
  {\enquote {\bibinfo {title} {Magnetic dead layers in sputtered
  $\text{Co}_{40}\text{Fe}_{40}\text{B}_{20}$ films},}\ }\href@noop {}
  {\bibfield  {journal} {\bibinfo  {journal} {J. Appl. Phys.}\ }\textbf
  {\bibinfo {volume} {103}},\ \bibinfo {pages} {07B526} (\bibinfo {year}
  {2008})}\BibitemShut {NoStop}%
\bibitem [{\citenamefont {Wang}\ \emph {et~al.}(2006)\citenamefont {Wang},
  \citenamefont {Chen}, \citenamefont {Yang}, \citenamefont {Shen},
  \citenamefont {Park}, \citenamefont {Kao},\ and\ \citenamefont
  {Tsai}}]{wang06JAP}%
  \BibitemOpen
  \bibfield  {author} {\bibinfo {author} {\bibfnamefont {Y.-H.}\ \bibnamefont
  {Wang}}, \bibinfo {author} {\bibfnamefont {W.-C.}\ \bibnamefont {Chen}},
  \bibinfo {author} {\bibfnamefont {S.-Y.}\ \bibnamefont {Yang}}, \bibinfo
  {author} {\bibfnamefont {K.-H.}\ \bibnamefont {Shen}}, \bibinfo {author}
  {\bibfnamefont {C.}~\bibnamefont {Park}}, \bibinfo {author} {\bibfnamefont
  {M.-J.}\ \bibnamefont {Kao}}, \ and\ \bibinfo {author} {\bibfnamefont
  {M.-J.}\ \bibnamefont {Tsai}},\ }\bibfield  {title} {\enquote {\bibinfo
  {title} {Interfacial and annealing effects on magnetic properties of {CoFeB}
  thin films},}\ }\href@noop {} {\bibfield  {journal} {\bibinfo  {journal} {J.
  Appl. Phys.}\ }\textbf {\bibinfo {volume} {99}},\ \bibinfo {pages} {08M307}
  (\bibinfo {year} {2006})}\BibitemShut {NoStop}%
\bibitem [{\citenamefont {Ingvarsson}\ \emph {et~al.}(2002)\citenamefont
  {Ingvarsson}, \citenamefont {Xiao}, \citenamefont {Parkin},\ and\
  \citenamefont {Gallagher}}]{ingvarsson02JMMM}%
  \BibitemOpen
  \bibfield  {author} {\bibinfo {author} {\bibfnamefont {S.}~\bibnamefont
  {Ingvarsson}}, \bibinfo {author} {\bibfnamefont {G.}~\bibnamefont {Xiao}},
  \bibinfo {author} {\bibfnamefont {S.~S.~P.}\ \bibnamefont {Parkin}}, \ and\
  \bibinfo {author} {\bibfnamefont {W.~J.}\ \bibnamefont {Gallagher}},\
  }\bibfield  {title} {\enquote {\bibinfo {title} {Thickness-dependent magnetic
  properties of {Ni\textsubscript{81}Fe\textsubscript{19}},
  {Co\textsubscript{90}Fe\textsubscript{10}} and
  {Ni\textsubscript{65}Fe\textsubscript{15}Co\textsubscript{20}} thin films},}\
  }\href {\doibase https://doi.org/10.1016/S0304-8853(02)00577-2} {\bibfield
  {journal} {\bibinfo  {journal} {J. Magn. Magn. Mater.}\ }\textbf {\bibinfo
  {volume} {251}},\ \bibinfo {pages} {202} (\bibinfo {year}
  {2002})}\BibitemShut {NoStop}%
\bibitem [{\citenamefont {Freimuth}\ \emph {et~al.}(2014)\citenamefont
  {Freimuth}, \citenamefont {Bl{\"u}gel},\ and\ \citenamefont
  {Mokrousov}}]{Freimuth14JPhysCM}%
  \BibitemOpen
  \bibfield  {author} {\bibinfo {author} {\bibfnamefont {F.}~\bibnamefont
  {Freimuth}}, \bibinfo {author} {\bibfnamefont {S.}~\bibnamefont
  {Bl{\"u}gel}}, \ and\ \bibinfo {author} {\bibfnamefont {Y.}~\bibnamefont
  {Mokrousov}},\ }\bibfield  {title} {\enquote {\bibinfo {title} {Berry phase
  theory of {Dzyaloshinskii-Moriya} interaction and spin-orbit torques},}\
  }\href@noop {} {\bibfield  {journal} {\bibinfo  {journal} {J. Phys.: Condens.
  Matt.}\ }\textbf {\bibinfo {volume} {26}},\ \bibinfo {pages} {104202}
  (\bibinfo {year} {2014})}\BibitemShut {NoStop}%
\bibitem [{\citenamefont {Chikazumi}(1997)}]{chikazumi1997}%
  \BibitemOpen
  \bibfield  {author} {\bibinfo {author} {\bibfnamefont {S.}~\bibnamefont
  {Chikazumi}},\ }\href@noop {} {\emph {\bibinfo {title} {Physics of
  Ferromagnetism}}},\ \bibinfo {edition} {2nd}\ ed.\ (\bibinfo  {publisher}
  {Oxford University Press},\ \bibinfo {address} {Oxford},\ \bibinfo {year}
  {1997})\BibitemShut {NoStop}%
\end{thebibliography}%

\end{document}